\documentclass[pdflatex,sn-basic,Numbered]{sn-jnl}

\usepackage{graphicx}%
\usepackage{multirow}%
\usepackage{amsmath,amssymb,amsfonts}%
\usepackage{amsthm}%
\usepackage{mathrsfs}%
\usepackage[title]{appendix}%

\usepackage{multirow}%
\usepackage[table]{xcolor}
\usepackage{array}

\usepackage{textcomp}%
\usepackage{manyfoot}%
\usepackage{booktabs}%
\usepackage{algorithm}%
\usepackage{algorithmicx}%
\usepackage{algpseudocode}%
\usepackage{listings}%
\usepackage{todonotes}%
\usepackage{pdflscape}%
\usepackage{tabularx}%
\usepackage{makecell}%
\usepackage{xargs}                      
\usepackage[pdftex,dvipsnames]{xcolor}
\usepackage{indentfirst}
\usepackage{tikz}
\usepackage[most]{tcolorbox}
\usepackage[T1]{fontenc}
\usepackage[utf8]{inputenc}
\usepackage{enumitem}
\usepackage{caption}
\usepackage{comment}
\usepackage{longtable}

\definecolor{Gray}{gray}{0.92}

\newcolumntype{a}{>{\columncolor{Gray}}c}

\usepackage{array}
\newcolumntype{P}[1]{>{\centering\arraybackslash}p{#1}}
\newtcolorbox{colorb}{
enhanced,
boxrule=0pt,frame hidden,
borderline west={2pt}{0pt}{green!50!black},
colback=green!05!white,
sharp corners,
top = 2.25pt,
bottom=2.25pt,
left=5pt
}
\newtcolorbox{colora}{
enhanced,
boxrule=0pt,frame hidden,
borderline west={2pt}{0pt}{gray!50!black},
colback=gray!05!white,
sharp corners,
top = 2.25pt,
bottom=2.25pt,
left=5pt
}
\theoremstyle{thmstyleone}%
\theoremstyle{thmstyletwo}%

\theoremstyle{thmstylethree}%

\begin{document}

\title{Supporting Viewpoint Coordination for Compliance by Design with an Artifact Model for Regulatory Requirements Engineering}

\author*[1,2]{\fnm{Oleksandr} \sur{Kosenkov}}\email{oleksandr.kosenkov@bth.se}

\author[1]{\fnm{Davide} \sur{Fucci}}

\author[1]{\fnm{Michael} \sur{Unterkalmsteiner}}

\author[1,2]{\fnm{Daniel} \sur{Mendez}}

\affil*[1]{\orgname{Blekinge Institute of Technology}, \orgaddress{\city{Karlskrona}, \country{Sweden}}}

\affil[2]{\orgname{fortiss GmbH}, \orgaddress{\city{Munich}, \country{Germany}}}

\abstract{
Regulatory compliance has become an essential software requirement. Legal demands for compliance by design (CbD) call for requirements engineering (RE) approaches that derive requirements and support compliance implementation in subsequent software engineering (SE) processes such as software design.

The complexity of the legal knowledge demanded in regulatory RE makes it important to involve legal experts and ensure the relevant specifications are valid and usable for them. Regulatory requirements must support compliance implementation in software design. However, there is a lack of approaches bridging legal and SE viewpoints, and interaction challenges between the two persist.

To address engineering–legal coordination in specification of regulatory requirements, we propose the Artifact Model for Regulatory Requirements Engineering (AM4RRE), operationalized through templates. AM4RRE was positively evaluated by most of the 37 practitioners for coordination in regulatory RE and for supporting requirements quality. Results from a quasi-experiment with 34 of them in 10 groups indicated that template-guided coordination was associated with, on average, shorter total specification time when executed first and higher requirements quality in most groups in which it varied.

Existing RE research and practice provide little attention and support for viewpoint coordination in regulatory RE for CbD. Artifact-oriented approaches such as AM4RRE have the potential to guide coordination in RE for better RE results and contribute to concretizing the role and contribution of RE to CbD.
}




\maketitle

\section{Introduction}\label{sec:intro}
In recent years, countries worldwide have enacted multiple sophisticated regulations affecting software engineering (SE)~\cite{state2023legislation,osborne2025digital} across domains. Consequently, regulatory compliance has emerged as a cross-cutting software quality requirement that demands legally verifiable conformance. Recently introduced compliance by design (CbD) obligations demand integrating regulatory requirements into software design rather than addressing them as an afterthought.
To address the challenge of systematically accounting for regulations and supporting compliance implementation in SE, regulatory requirements engineering (RE) approaches must both derive requirements from regulations and support their implementation throughout software engineering processes~\cite{hjerppe2019general,kempe2021regulatory}. In practice, awareness and adoption of regulatory RE approaches remain limited~\cite{klymenko2022understanding,kosenkov2025privacy}. We argue this is because two conditions tied to core RE challenges are overlooked.

First, processing the legal knowledge required to implement compliance, along with SE expertise~\cite{martin2018methods,kosenkov2025systematic}, is more complex than other types of domain knowledge addressed in RE. Legal knowledge and specifications are required in RE and downstream processes to demonstrate compliance~\cite{breaux2008analyzing,syed2010emerging} and in legal disputes~\cite{boella2014critical}.
Applying legal knowledge involves specific legal \textit{roles}, \textit{processes}, and \textit{goals} that together form a distinct viewpoint. For example, legal interpretation concretizes abstract regulatory norms into case-specific demands and supports compliance demonstrability through regulation-to-implementation traceability, which differs from the traceability typically required in SE~\cite{breaux2008analyzing,kosenkov2025privacy}. Even when legal experts are available, the semantic gap between engineering and legal professionals~\cite{sleimi2021automated, bednar2019engineering} makes it difficult to explicate, specify, and reuse legal knowledge~\cite{kosenkov2024regulatory} or to achieve shared understanding. Still, existing RE approaches provide limited support for either enabling engineers with basic legal knowledge to complete legal specifications~\cite{massey2014evaluating} or meaningfully involving legal experts.

The second overlooked practical condition is the role of regulatory RE in supporting downstream compliance implementation by coordinating interdependent legal and SE viewpoints~\cite{bobkowska2010efficient, boella2014critical, horstmann2024those}. Regulatory requirements are often a ``natural afterthought'' because determining regulations’ applicability and selecting implementation options~\cite{ghanavati2015impact} depend on existing SE specifications~\cite{ingolfo2013choosing}. Engineering-legal decision-making must therefore address how regulations impact non-regulatory requirements~\cite{massey2010evaluating} and system design~\cite{kosenkov2025privacy}, while balancing legal constraints, business objectives, and technical feasibility~\cite{massey2014evaluating,kosenkov2024regulatory}. CbD obligations emphasize such coordination to keep interpretations and implementation aligned. Although engineering-legal interaction already occurs in practice, it remains largely unstructured~\cite{klymenko2022understanding}, complicated by semantic gaps, differences in methods~\cite{kasauli2020charting}, and ways of working~\cite{kosenkov2024regulatory}. Practitioners, therefore, would benefit from approaches that support systematic specification and coordination of legal and SE viewpoints in regulatory RE for CbD. Existing research offers little practical support for this.

To address this gap, we propose the Artifact Model for Regulatory Requirements Engineering (AM4RRE\footnote{AM4RRE is pronounced like amphora—an ancient ceramic container for water or wine with two large handles. Some amphorae were so large and heavy that two people had to carry them collaboratively.}), which supports regulatory RE for CbD by guiding specification and coordination activities of legal, managerial, RE, and software architecture viewpoints, and by guiding the specification of regulatory requirements grounded in these viewpoints. Our proposed approach follows artifact orientation, which guides RE activities through a blueprint of artifacts and their dependencies rather than through prescribed activities. AM4RRE makes two primary contributions to the existing state of regulatory RE research and practice, which are:
\begin{itemize}[nosep]
    \item it offers a systematic way to capture and make explicit specific knowledge, processes, goals, and context of the roles involved in regulatory RE through viewpoint modeling. In particular, the legal viewpoint model incorporates legal concepts used in legal interpretation to help legal experts complete the legal specifications required in regulatory RE. This is intended to facilitate their involvement and make the resulting specifications useful, rather than detached from legal practice.
    \item it supports engineering-legal coordination by making relationships between viewpoint concerns explicit and enabling these concerns to be reflected in resulting regulatory requirements specifications. In particular, it supports the inclusion of information about regulatory demands directly relevant to software design and the specification of new or affected software design components.
\end{itemize}

The evaluation of AM4RRE presented in this study focuses on these two contributions---usefulness for completing SE and legal viewpoint specifications, and support for guiding coordination between viewpoints. For the evaluation, we tailored the AM4RRE artifact model to the General Data Protection Regulation (GDPR) and the Cyber Resilience Act (CRA), both of which require CbD, and we operationalized it through a series of templates. We evaluated AM4RRE using subjective feature analysis with screening~\cite{kitchenham1996desmet} with 40 practitioners (3 of whom withdrew) and a quasi-experiment~\cite{wohlin2012experimentation} with 34 of these practitioners in 10 groups.

Practitioners found the coordination driven by AM4RRE templates to be aligned with regulatory RE practices and positively perceived the usefulness of AM4RRE to support the quality of regulatory requirements, coordination in regulatory RE, compliance implementation in software design, and achievement of viewpoint-specific goals. The quasi-experiment results support the feature analysis by indicating an association between using AM4RRE-based templates and, compared with free-form coordination, shorter average total requirements-specification time when guided specification was executed first, higher completeness in all groups in which completeness varied, higher consistency in eight out of ten groups, and higher correctness in nine out of ten groups.

Our results further suggest that RE can help address the well-known domain knowledge challenge in SE and, on that basis, contributes to coordinating domain experts and SE roles in downstream SE processes such as software design. This contributes to further understanding of both the domain knowledge and the intermediating role of RE between the software application domain and SE processes.

The paper is structured as follows. Sections~\ref{sec:background} and~\ref{sec:related} introduce background and related work. Section~\ref{sec:am4rre} presents AM4RRE and the regulatory requirements templates. Section~\ref{sec:methodology} describes our methodology, and Sections~\ref{sec:results} and~\ref{sec:discussion} report and discuss our key results. Section~\ref{sec:threats} discusses the main threats to validity and how we mitigate them, before outlining conclusions and future work in Section~\ref{sec:conclusion}.

\section{Background}\label{sec:background}
This study focuses on RE for \textit{regulatory compliance of software systems} -- a state of verifiable implementation of obligatory, pre-defined, legally binding norms specified in applicable regulations~\cite{kosenkov2025systematic}. Because regulations (e.g., GDPR) usually require demonstrating compliance, we suggest that both software engineers and legal experts should be able to verify compliance implementation.
Recent EU regulations introduce the obligation of \textit{compliance by design (CbD)}, requiring regulatory compliance to be implemented in software design rather than being addressed as an afterthought. CbD has different regulation-specific facets: \textit{data protection- or privacy by design (PbD)}, required by Article 25 of the GDPR, and \textit{security by design (SbD)}, de facto required under Article 13 of the Cyber Resilience Act (CRA)~\cite{ec2024state,ec2024seventh}.
\textit{Regulatory RE for CbD} is a specialized area of RE that translates regulations into requirements and supports regulatory compliance implementation in software design.

Regulatory RE for CbD involves multiple \textit{viewpoints} -- complementary and interdependent perspectives that contribute to regulatory RE and compliance implementation and are characterized by specific syntax and specification concepts, roles with their knowledge and qualifications, processes, and goals pursued in regulatory RE~\cite{leite1996viewpoints}. The legal viewpoint in regulatory RE is characterized by the demand for legal knowledge to execute a \textit{legal interpretation}, that is, to concretize abstract regulatory norms into concrete context-specific demands and system-specific requirements. We define \emph{legal knowledge} as knowledge about legal concepts, their relationships, and how they are applied in legal processes. \emph{Legal concepts} are ``stereotypical'' semantic elements of regulatory acts that carry the meaning relevant to legal interpretation and that eventually determine interpretation outcomes~\cite{kosenkov2024developing}.

Regulatory RE requires actionable mechanisms that address divergent interpretations of specifications~\cite{glinz2015shared} and differing methods and tools~\cite{wohlrab2019boundary} between the disciplines. We define \textit{coordination of viewpoints in regulatory RE for CbD} as the explication, identification, specification, and exchange of information between legal and SE viewpoints to specify regulatory requirements and implement regulatory compliance in software design. \textit{Specification transparency} is the quality of a specification underpinning coordination by allowing one to find and understand the required information. Because substantial research on interdisciplinary coordination in RE is lacking, our study focuses on measuring specification time and requirements specification quality as outcomes rather than the coordination process. \textit{Correctness of regulatory requirements} refers to the extent to which the requirements accurately reflect applicable regulations and support their implementation; \textit{internal completeness} means that all required elements are specified; and \textit{consistency} is the absence of conflicts between regulatory requirements and their source specifications (viewpoint specifications under guided or case descriptions under free-form coordination).

This study applies artifact-oriented regulatory RE, which guides RE through the content of RE results rather than RE activities~\cite{mendez2010meta}. We use the state-of-the-art Artifact Model for Domain-independent Requirements Engineering (AMDiRE)~\cite{mendez2015artefact}, composed of role, process, and an artifact model as the backbone, as the basis for developing our AM4RRE approach. In the context of artifact orientation, we use the terms \textit{regulatory requirement specifications} and regulatory requirements as synonyms.

\section{Related work}\label{sec:related}
Next, we provide an overview of the studies in research tracks related to our study.

\paragraph{Regulatory RE and compliance by design research}
In regulatory RE studies, approaches to complex legal knowledge range from generic methods such as grounded theory or deontic logic (e.g.,~\cite{breaux2008analyzing,kiyavitskaya2007extracting}) to custom conceptual models (e.g.,~\cite{siena2012capturing,sleimi2021automated}) and adapted legal concepts (e.g., Hohfeldian model of jural relations~\cite{ghanavati2014legal}).
However, existing approaches do not support the semantic analysis and legal interpretation that legal experts perform, preventing both meaningful involvement and expert review of the results~\cite{kosenkov2024developing}.
Many existing regulatory RE approaches focus solely on modeling regulations (e.g.,~\cite{breaux2008analyzing,sleimi2021automated}) and do not account, in the RE process, for the system or context to which the regulation applies. Approaches that account for such context usually suggest using the same engineering notations for both regulations and requirements (e.g., a goal-oriented requirements language~\cite{ghanavati2014legal}). This constrains the expressiveness of regulation models and the execution of legal interpretation.
Many studies assume that compliance should be addressed only in the RE process~\cite{engiel2017tool} and ignore the demand for CbD. Studies focusing on CbD emphasize the importance of engineering-legal collaboration and specifying both viewpoints~\cite{sion2020dpmf,boltz2026enabling}; however, they do not suggest systematic ways to analyze different regulations~\cite{kosenkov2025privacy}.
Some empirical studies (e.g.,~\cite{massey2014evaluating}) suggest that RE tools, rather than replacing lawyers, should guide software engineers in legal interpretation and in exchanging information with legal experts.
Recent studies on automated regulatory RE methods, including AI-based methods, also recognize the need to involve legal experts~\cite{sleimi2021automated,singhal2026code}.
Only a few studies considered such involvement, and they either proposed a high-level vision of the activities legal experts should perform~\cite{ayala2024gdpr,engiel2017tool} or focused on information shared between engineers and legal experts~\cite{bobkowska2010efficient}, neither of which allows meaningful engineering-legal interaction.

\paragraph{Shared understanding}
Shared understanding is essential in supporting RE~\cite{glinz2015shared} and downstream  activities~\cite{werner2020lack}. Studies in this area point to legal knowledge as a challenge to shared understanding~\cite{glinz2015shared,werner2020lack}. However, no approaches specifically address shared understanding in regulatory RE, and available studies mainly focus on shared understanding between RE and downstream SE processes~\cite{fricker2007handshaking}, while addressing aspects at a higher level (e.g.,~\cite{quinones2024guide}). This makes it difficult to adapt them for interdisciplinary engineering-legal shared understanding. These studies also mention artifacts as an important but auxiliary means of establishing shared understanding~\cite{fricker2007handshaking,glinz2015shared}.

\paragraph{Viewpoints in RE and SE}
Existing RE studies suggest using viewpoints as a means for organizing RE processes that involve multiple heterogeneous stakeholders, large volumes of viewpoint-specific information, and potential conflicts and inconsistencies between stakeholders~\cite{leite1996viewpoints,kotonya1996requirements,easterbrook2005viewpoints,sommerville1997viewpoints}. Purposes of viewpoint application include facilitating a deeper understanding of the problem domain, improving backward traceability, addressing problem-domain complexity, and explicitly capturing differences between viewpoints~\cite{easterbrook2005viewpoints}. In this research track, the legal viewpoint is defined as the indirect viewpoint that generates requirements constraining the SE viewpoints~\cite{kotonya1996requirements}. Still, there is a lack of studies that apply viewpoints to regulatory RE specifically. Further, viewpoint applications remain at an abstract level (e.g.,~\cite{kotonya1996requirements}) and are only loosely adaptable to regulatory RE. In addition, to the best of our knowledge, viewpoints have seen limited application in RE practice because of the tendency to express requirements in viewpoint-specific working notations and the limited capacity of viewpoint models to reflect organizational problems alongside technical ones~\cite{sommerville1997viewpoints}.

\paragraph{Coordination in requirements and software engineering}
Some SE studies focus on coordination and interdependencies across SE teams that apply different specialized practices (e.g., agile vs. waterfall), belong to different disciplines (e.g., hardware vs. software), or involve stakeholders with conflicting interests~\cite{toffolon2000framework}. In this area, boundary objects (artifacts in which knowledge is manifested~\cite{wohlrab2019boundary}) were identified as coordination mechanisms in large-scale agile software development~\cite{kasauli2020charting}, facilitating interaction, knowledge sharing, and integration of work across SE teams~\cite{wohlrab2019boundary}.
Studies in this research track identified regulations as one type of boundary object used to ensure common understanding and compliance~\cite{kasauli2020charting}. However, these studies focus on interactions among engineers (e.g.,~\cite{shenouda2024using}), rather than interdisciplinary engineering-legal interaction; thus, they neither specify a legal viewpoint nor support legal interpretation.

\paragraph{Research gap}
Existing studies recognize legal knowledge as a barrier to shared understanding, identify a distinct legal viewpoint, and characterize regulations as boundary objects for coordination across SE teams. However, there is a lack of approaches that support the specification of legal viewpoints (with or without legal experts), enable legal interpretation, and meet other legal goals required in practice. Existing methods do not explicitly and transparently capture each viewpoint's implicit knowledge as domain-specific concepts and goals, so they do little to help roles understand other viewpoints or coordinate. The literature also lacks methods for specifying regulatory requirements that consider downstream artifacts, as regulatory RE for CbD requires. As a result, stakeholders usually have to capture additional information separate from the engineering artifacts, creating unnecessary overhead.
Finally, to the best of our knowledge, we still lack studies considering approaches under realistic conditions involving legal experts and software engineering roles at the same time. Earlier versions of AM4RRE~\cite{kosenkov2021vision,kosenkov2024developing} aimed to enable engineers to support legal interpretation using corresponding legal concepts. However, recent empirical studies identified additional legal concepts relevant to CbD~\cite{kosenkov2025privacy}, potentially diverging viewpoint-specific goals~\cite{kosenkov2025privacy,kosenkov2026towards} and views on challenges to regulatory RE~\cite{kosenkov2025systematic}. This made it clear that an effective coordination between legal experts and software engineers is essential. Since no approach accommodates viewpoint differences and implements more structured coordination, we address this gap with the contributions presented here.

\section{Artifact Model for Regulatory Requirements Engineering (AM4RRE)}\label{sec:am4rre}

In this section, we describe the Artifact Model for Regulatory Requirements Engineering (AM4RRE), its development, tailoring, operationalization with templates, and AM4RRE-based templates for regulatory requirements specifications for CbD. We focus only on the aspects relevant to the study at hand.

\paragraph*{AM4RRE in a nutshell}
AM4RRE is an artifact-oriented approach that guides all roles involved in regulatory RE for CbD by means of a central artifact model and the dependencies between the artifacts it covers. By focusing on what should be specified rather than how it should be specified, it avoids prescribing complex processes and supports consistency and shared terminology in central artifacts.
Artifact orientation provides several benefits~\cite{mendez2015artefact,mendez2010meta}, including a closer focus on deliverable quality, greater flexibility (it can drive RE as the core approach, complement activity-based approaches, and integrate into existing SE process models), and better management of artifact availability and dependencies. We chose it as the basis for AM4RRE because it suits the distinctive features of regulatory RE.

\begin{figure}
    \centering
    \includegraphics[width=1\linewidth]{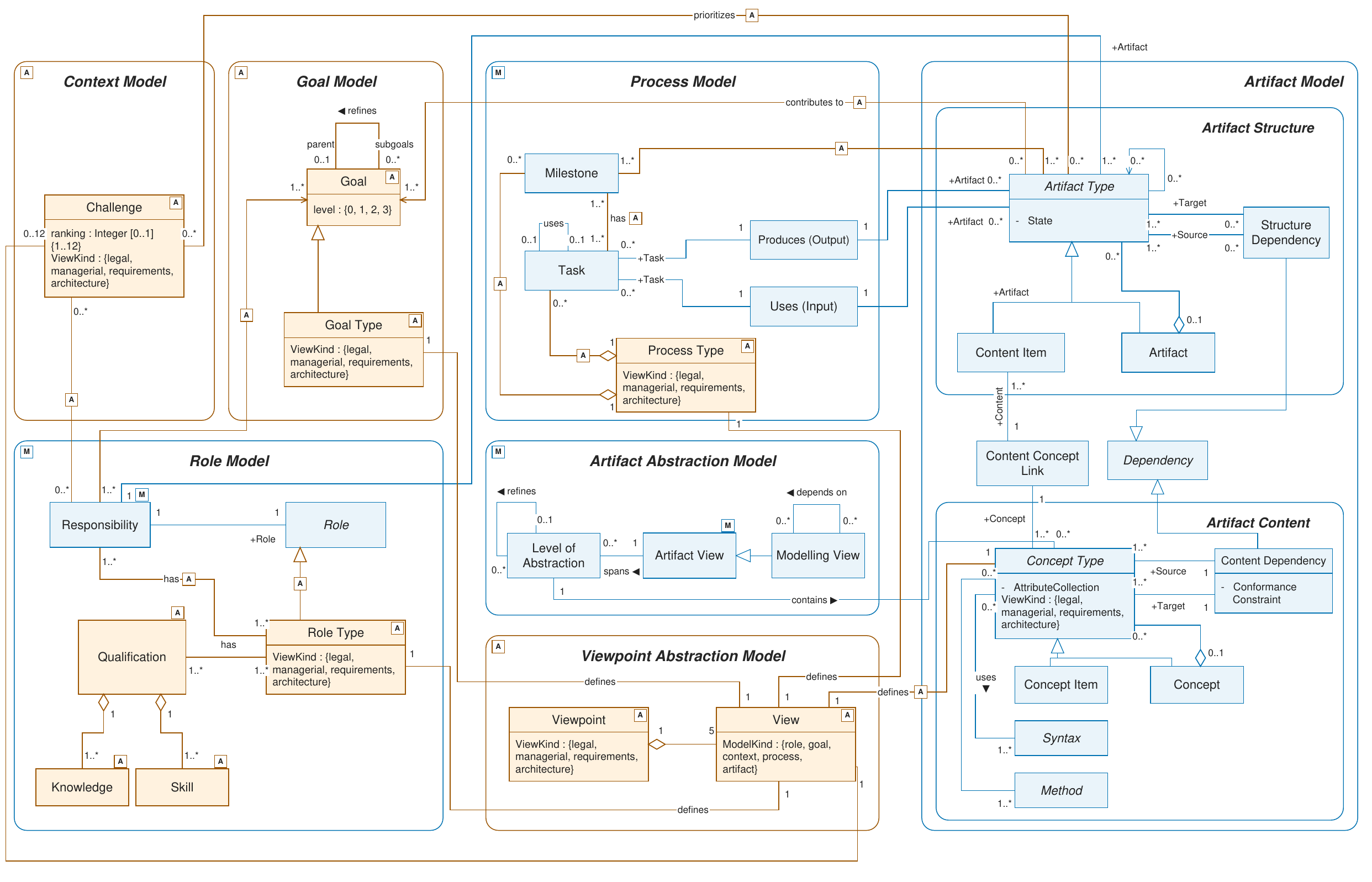}
    \caption{\small AM4RRE meta-model based on the AMDiRE meta-model~\cite{mendez2010meta}. Added elements are yellow (marked ``A'' ); original elements are blue (marked ``M'' if modified).}
    \label{fig:am4rre-meta}
\end{figure}

Figure~\ref{fig:am4rre-overview} provides a high-level overview of AM4RRE; Figure~\ref{fig:am4rre-meta} complements it with the AM4RRE meta-model, showing extensions of the original AMDiRE components to accommodate regulatory RE. AM4RRE components are as follows:

\begin{figure}
    \centering
    \includegraphics[width=1\linewidth]{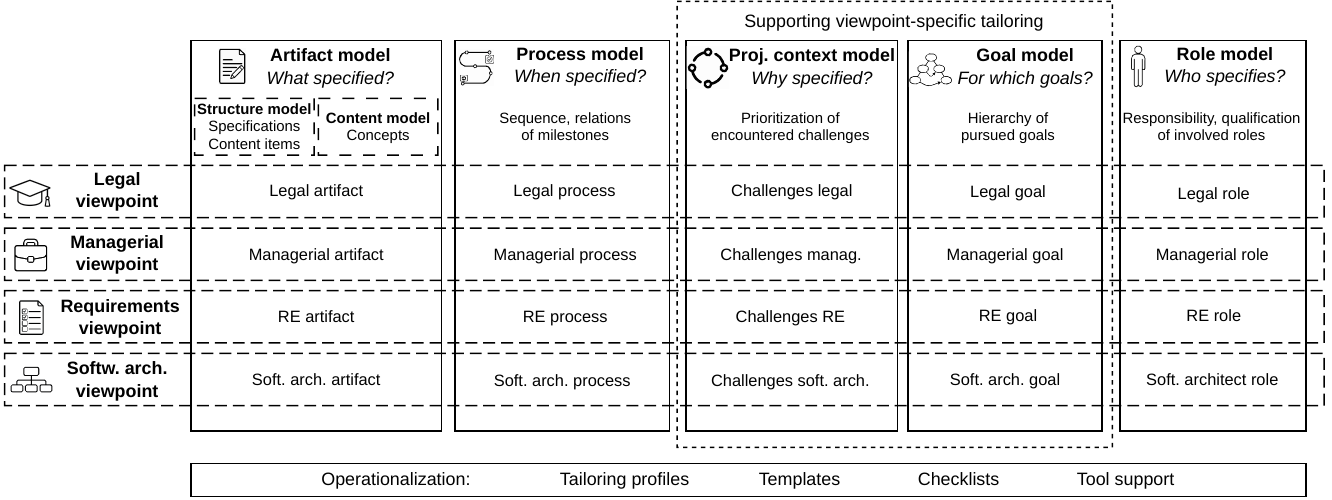}
    \caption{\small Overview of the core AM4RRE components and associated viewpoints, intentionally kept at an abstract level.}
    \label{fig:am4rre-overview}
\end{figure}

    \textbf{Artifact model} (answering the question \textit{What is specified?}) is the backbone of AM4RRE and defines the artifacts and information specified in regulatory RE for CbD. It contains two artifact views, each a separate (sub)model: a structure model and an underlying content model. The structure model organizes related concepts into artifacts and uses a common specification syntax to form so-called content items (the smallest semantically coherent aggregation of information with the same purpose and meaning) and specifications (groups of content items serving the goals of one role and viewpoint). It can be compared to documents and underlying sections with associated responsibilities captured in the role model. The detailed content view, in turn, abstracts from the modeling concepts through a content model and, thus, supports consistency and seamless modeling.
    
    \textbf{Role model} (\textit{Who specifies?}) describes the roles involved, their responsibilities for artifacts, and their required qualifications.
    
    \textbf{Process model} (\textit{When specified?}) establishes the milestones at which corresponding roles complete, review, or accept artifacts.
    
    \textbf{Goal model} (\textit{For which goals?}) specifies the goals relevant in regulatory RE for CbD. Goals are organized into different hierarchical levels, with the root goal (level 0) supporting CbD implementation itself; each goal can be supported by a non-exhaustive set of subgoals on the next level. Goals guide project-specific decisions.
    
    \textbf{Project context model} (\textit{Why specified?}) captures the project context by prioritizing challenges to regulatory RE for CbD from each role's perspective.

To manage concerns of different roles and enable their involvement, we organize elements of each model related to a particular role into \textit{views} (e.g., legal artifact model view). Views across models related to the same role constitute one of the four \textit{viewpoints}: legal, managerial, requirements, and software architecture. For brevity in this paper and the quasi-experiment materials, we occasionally refer to corresponding views as models (e.g., legal artifact model).

The version of AM4RRE we present extends the state-of-the-art Artifact Model for Domain-independent Requirements Engineering (AMDiRE) for non-regulatory RE~\cite{mendez2015artefact} using the data of recent empirical studies as follows:
\begin{itemize}[nosep]
    \item extends the artifact model with legal concepts and artifacts identified in~\cite{kosenkov2024developing,kosenkov2025privacy};
    \item introduces views and viewpoints (see Figure~\ref{fig:am4rre-meta} for comparison with AMDiRE);
    \item adds the legal expert role and changes other roles in the role model~\cite{kosenkov2025systematic,klymenko2022understanding,kosenkov2024regulatory};
    \item adds legal interpretation milestones in the process model based on~\cite{kosenkov2024developing,kosenkov2024regulatory};
    \item introduces the project context model prioritizing challenges identified in~\cite{kosenkov2025systematic};
    \item introduces the goal model based on~\cite{kosenkov2025privacy,kosenkov2026towards}.
\end{itemize}

\begin{figure}
   \centering
   \includegraphics[width=0.85\linewidth]{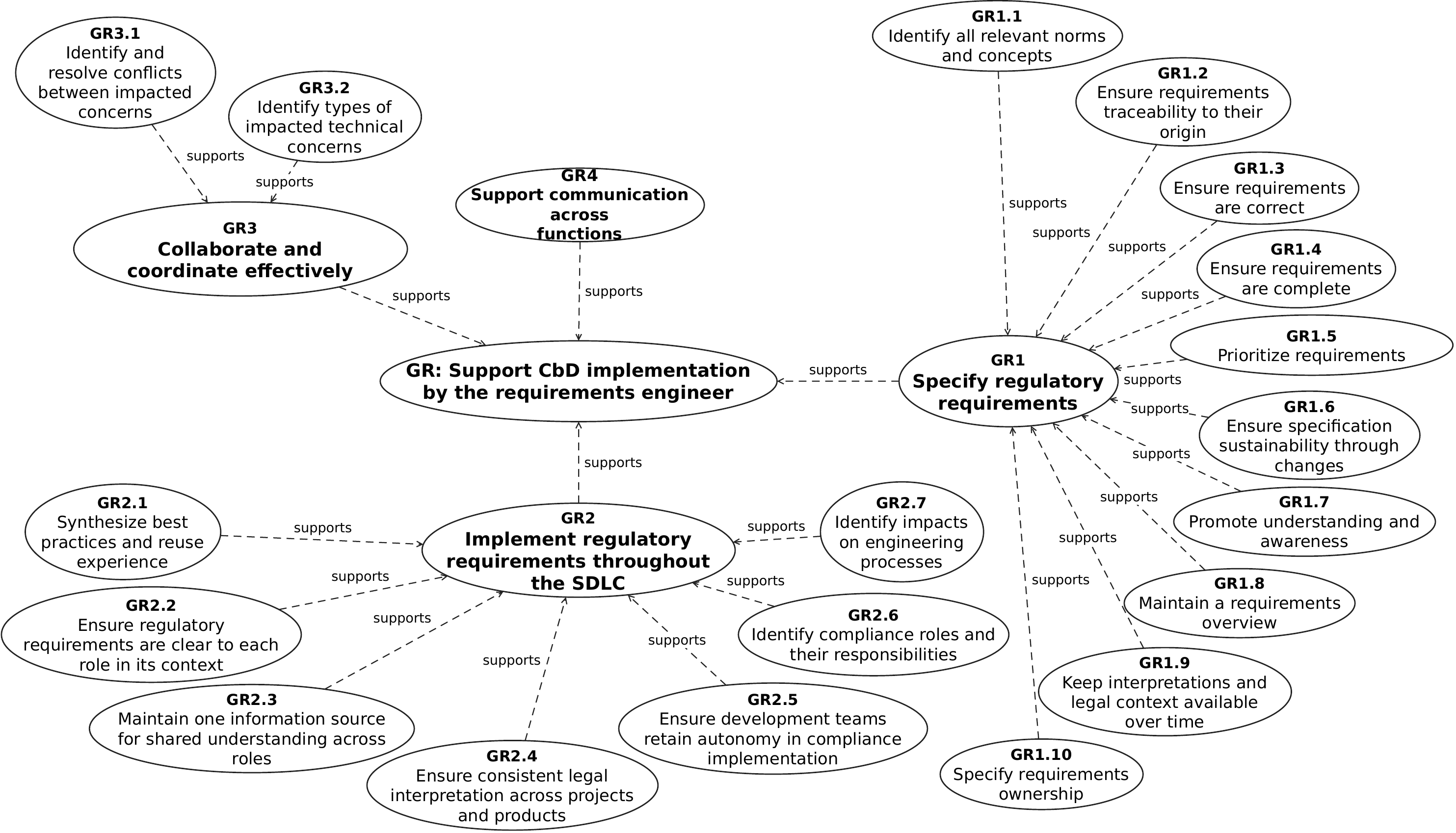}
   \caption{\small RE view of the goal model. The root goal is not shown; the first-level goal is denoted with a view ID, the second-level goals as whole numbers following the view ID (e.g., GR1), and third-level goals as a fractional part (e.g., GR1.1).}
   \label{fig:mg-re}
\end{figure}

\begin{figure}
    \centering
    \includegraphics[width=1\linewidth]{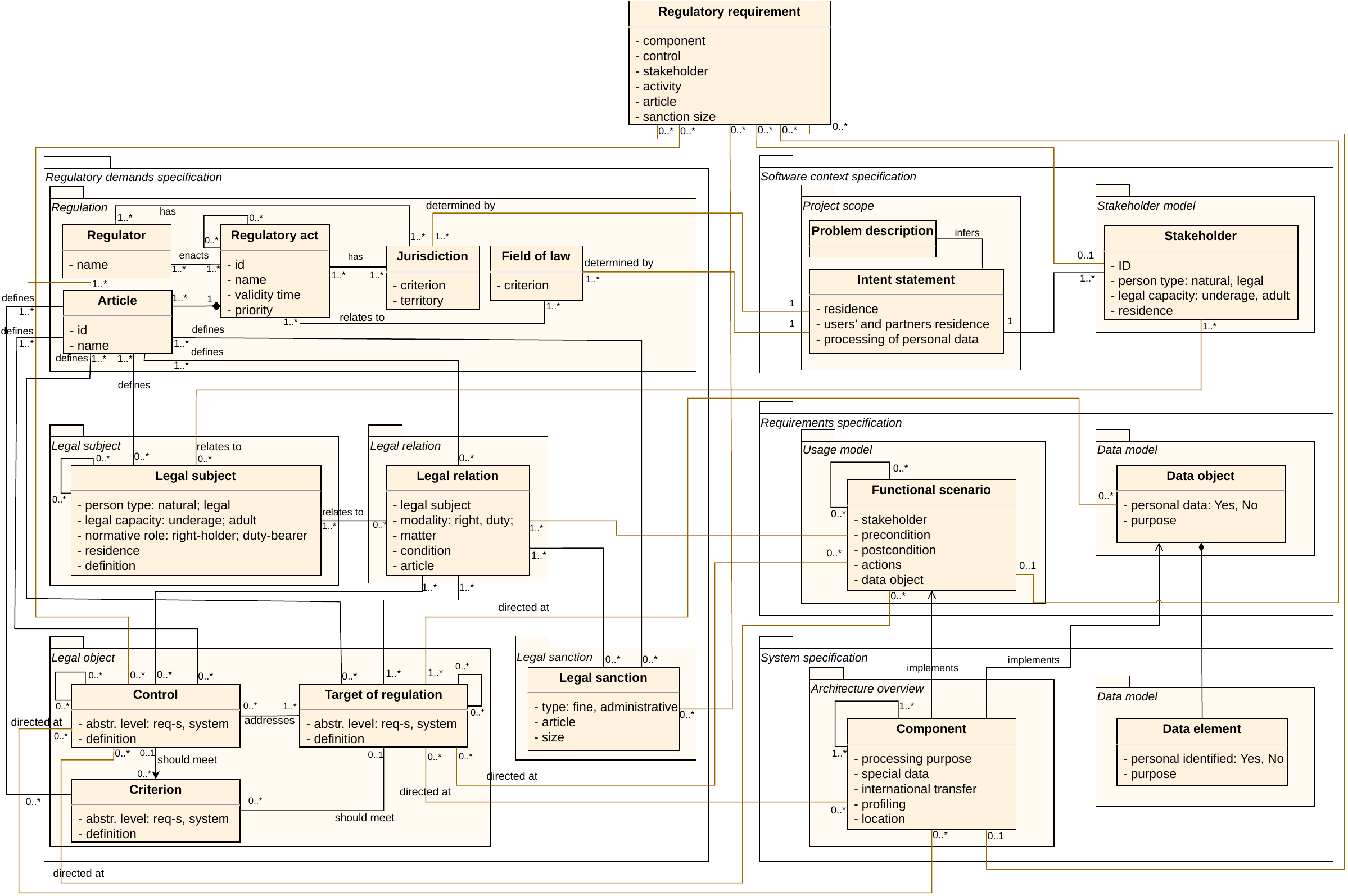}
    \caption{\small Content model tailored to the GDPR and used for template development. A UML class diagram depicts viewpoint-specific concepts (content view); a UML package diagram depicts the structure view overlaid with content items and specifications (regulatory demands, software context, requirements, and system specifications).}
    \label{fig:artifact-model}
\end{figure}

Our study extends the previous version of AM4RRE~\cite{kosenkov2024developing} by integrating legal concepts from~\cite{kosenkov2025privacy}, merging two legal specifications into one, adding project context and goal models, and revising role and process models using empirical data~\cite{kosenkov2024regulatory}. Most importantly, we evaluate AM4RRE operationalized through templates.

\paragraph*{Viewpoint specification and coordination}
AM4RRE guides \textit{viewpoint specification} by defining baseline elements for each view (see above). For the legal viewpoint, for example, these include regulation-independent concepts and artifacts for a legal interpretation, legal experts' responsibilities and qualifications, interpretation milestones and tasks, legal experts' goals in regulatory RE for CbD, and regulatory RE challenges to prioritize from their perspective. Together, these views capture the legal viewpoint's distinctive features, help explicate legal specifications and align them with legal practice, and provide SE roles with a structured overview that may guide specification completion.

AM4RRE supports \textit{viewpoint coordination} by enabling the involved roles to (1) identify concepts required for a legal interpretation using the legal viewpoint tailored to a regulation, as in the GDPR example, (2) identify existing or new SE concepts that correspond to the required legal concepts, guided by their properties and definitions, and (3) integrate the required concepts into the means used for SE specification completion (e.g., via templates) with the preservation of traceability to the relevant legal concepts. Relations between concepts in different views guide the regulatory RE process. Finally, to ensure that the resulting specifications (project-level artifacts) address concerns from different viewpoints, the artifact model captures concepts that constitute the content of regulatory requirements specifications and belong to different viewpoints. AM4RRE supports the following forms of engineering-legal coordination. The first is when the model is tailored to a domain and/or regulation, e.g., before a project starts or when a unit adapts its methodologies. Here, team members define the reference model to be used in projects. Next, the model can be tailored to the specific project setting. Here, legal experts and SE roles responsible for specific artifacts establish relations and context-specific equivalences between legal and SE concepts. During project execution, engineering-legal coordination occurs when legal experts and corresponding SE roles use the operationalized, tailored model to specify the corresponding legal concepts embedded in the SE specifications, and directly interact to specify regulatory requirements according to the tailored template.

\paragraph{AM4RRE tailoring}\label{sec:tailoring}
Before AM4RRE can guide regulatory RE for CbD, it must be adapted to specific regulation and organizational or project contexts. We denote this adaptation as tailoring, which includes some of the instantiation activities. Tailoring can be applied at the organizational, project, or both levels in three ways: [T1] regulation-specific, [T2] project-specific, and [T3] optional goal-driven tailoring. Next, we provide an overview of each of them.

%\begin{enumerate}[nosep]
    \textbf{T1.} Regulation-specific tailoring of the artifact model identifies the legal concepts that constitute the artifact content model for implementing a specific regulation. It includes the following steps: [T1.1] identification of the instances of legal concepts from the AM4RRE legal artifact content model in the selected regulation (e.g., by annotations). [T1.2] identification of the relevant attributes of concept instances according to the regulation. First, existing attributes of corresponding concepts in the non-tailored AM4RRE are checked and transferred to the tailored instance if they are relevant. If there is no corresponding attribute, it should be added to the tailored AM4RRE instance. [T1.3] removal of concepts and content items not relevant to the corresponding regulation. [T1.4] ad hoc addition of new legal concepts required for legal interpretation if a person possessing legal knowledge considers them important. Experts with the required legal knowledge should conduct this tailoring independently.

    \textbf{T2.}  Project-specific tailoring adapts models constituting AM4RRE to organizational or project context and establishes relations between the legal and SE concepts of the artifact content model. It is conducted in the following steps. [T2.1]: for each legal concept and its attribute in the tailored legal artifact model, semantically related concepts in SE artifact content models are identified by coordinating and negotiating between the legal and SE roles responsible for the corresponding relevant concepts (tailoring-time coordination). [T2.2]: checking that definitions of legal and SE concepts between which a relation was established are compatible; If a concept or an attribute is missing from the SE specification, it is added to the most suitable SE content item and concept and defined according to the corresponding legal concept definition. The relations established during this tailoring further guide the roles toward specific artifacts, content items, and concepts in regulatory RE. It must be performed for each organizational or project context and regulation, by the responsible SE roles, with legal experts validating the relations.
    
    \textbf{T3.} Goal-driven tailoring is optional and is used to scope regulatory RE toward the achievement of specific goals. For this, relevant goals are selected using the goal model, and relationships between goals and artifacts in the artifact model are used to determine which concepts need to be specified and prioritized to achieve the corresponding goals (e.g., effective management of the non-compliance goal requires specifying the legal sanction concept). This type of tailoring is outside the scope of this study.
%\end{enumerate}

Figure~\ref{fig:artifact-model} already shows the artifact model tailored to the GDPR, and the following example illustrates this tailoring.

\textbf{Example.} During the GDPR-specific tailoring, the legal expert of a company identifies the data subject as a GDPR-specific instance of <<legal subject>> and adds the attribute \textit{residence} to the legal concept <<legal subject>> because, according to the GDPR, the residence of data subjects is one of the factors determining GDPR applicability (we abstract from additional legal intricacies regarding GDPR applicability). Next, during project-specific tailoring, the legal expert identifies the managerial concept <<stakeholder>> as a specialization of the legal concept <<legal subject>>. The legal expert, in coordination with the managerial role, adds the attribute \textit{residence} to the <<stakeholder>> managerial concept, as required for legal interpretation. After that, the legal expert and manager verify that attributes are coherently defined from both views and that residence and person type attributes will be understandable to specifiers (tailoring-time coordination). After the tailoring, the model is operationalized through templates. Next, project roles use the templates individually, specifying the corresponding attributes that support identifying the stakeholder instances to which GDPR applies. Once specifications are available, the involved roles collaboratively specify requirements using the tailored regulatory requirements specification template. Here, the relations between concept instances drive specification-time coordination: systematic, targeted discussion of the applicability of legal concepts, followed by specification of regulatory requirements.

\paragraph{AM4RRE operationalization}
In this study, we operationalize AM4RRE using specification templates. AM4RRE-based templates are intended to facilitate the capture of viewpoint-specific information required for regulatory RE, guide specifiers with or without viewpoint-specific knowledge, and support the exchange of information between the viewpoints. We developed specification templates based on the artifact model, and each template usually corresponds to one content item. The process model determines the sequence of template specifications and their dependencies.

\begin{figure}[!htb]
    \centering
    \begin{minipage}{.45\textwidth}
        \centering
        \frame{\includegraphics[width=1\linewidth]{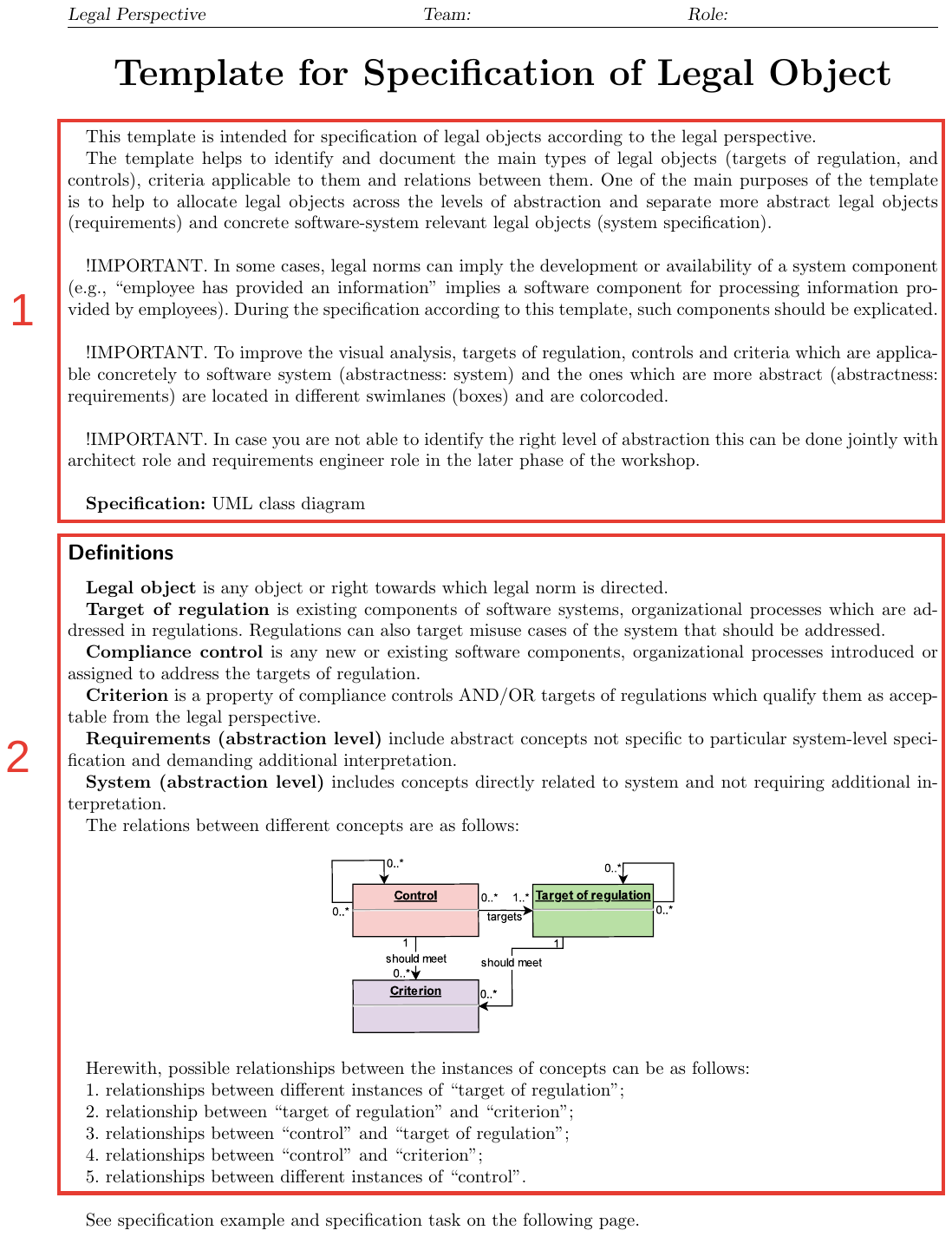}}
    \end{minipage}%
    \quad
    \begin{minipage}{.45\textwidth}
        \centering
        \frame{\includegraphics[width=1\linewidth]{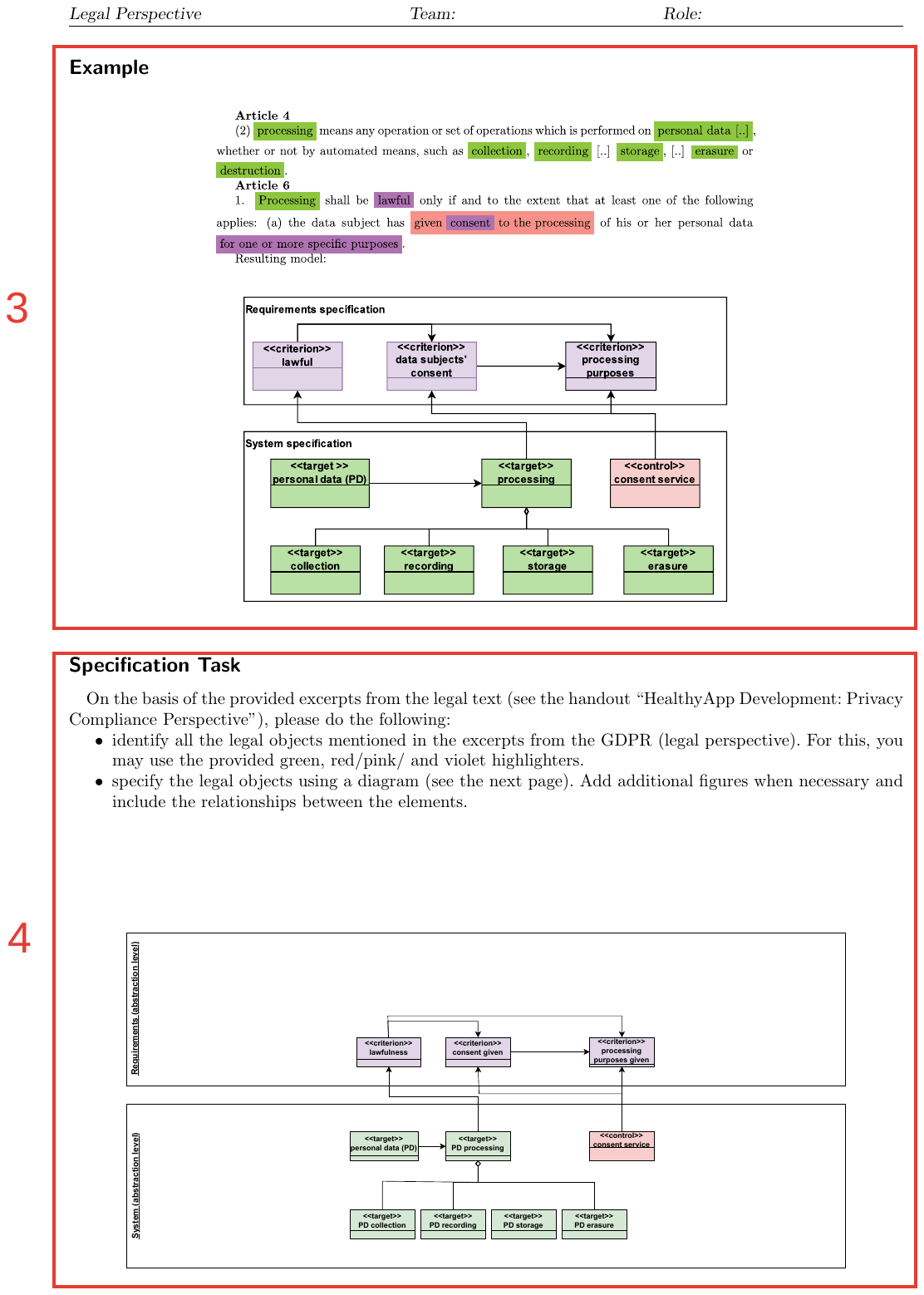}}
    \end{minipage}
    \caption{Example of an AM4RRE-based template: legal object specification (legal viewpoint). Numbers 1 to 4 mark the four template sections.}\label{fig:template}
\end{figure}

AM4RRE-based templates developed for this study are composed of the following sections (see Figure~\ref{fig:template}): (1) template description including its purpose, references to the viewpoint-specific case description, relations to other templates, a short description of the syntax and elements; (2) definitions of the concepts used; (3) example of specification (for SE templates, examples were based on systems and regulations unrelated to the case, and in legal templates, examples were based on the GDPR or CRA); (4) specification task and form or figures with some basic information (e.g., UML component diagram figures) for the specification.

\paragraph{Regulatory requirements template}
Finally, the AM4RRE content model also covers the concepts of different views used for the requirements specification itself. The non-tailored template for specification of regulatory requirements for compliance by design is as follows:
\\

$\underset{\text{Target of regulation / control}}{\underline{\hspace{3.5cm}}}$ should implement
$\underset{\text{control / matter of legal relation}}{\underline{\hspace{3.86cm}}}$ 
involving $\underset{\text{legal subject(s)}}{\underline{\hspace{1.75cm}}}$ for compliance with  $\underset{\text{sanction article}}{\underline{\hspace{1.8cm}}}$ under the threat of $\underset{\text{sanction size}}{\underline{\hspace{1.5cm}}}$
\\

The template follows the content model and resembles structured natural-language requirement patterns, extended with regulatory elements. The template should be tailored as described above. Tailoring allows for the inclusion of only the relevant information and the optimization of the specification process (e.g., if the prioritization of regulatory requirements based on potential sanctions is not required, specification of sanctions can be excluded using the goal-driven tailoring). The template tailored for a specific regulation is constructed from the SE concepts to which legal concepts were related during project-specific tailoring (T2). For example, a regulatory requirements template tailored for GDPR has the following structure:

$\underset{\text{Component / control}}{\underline{\hspace{2.5cm}}}$ should implement
$\underset{\text{actions / control}}{\underline{\hspace{2.5cm}}}$ 
involving $\underset{\text{stakeholder(s)}}{\underline{\hspace{2cm}}}$ for compliance with  $\underset{\text{GDPR article}}{\underline{\hspace{2cm}}}$ under the threat of $\underset{\text{size of the sanction}}{\underline{\hspace{2.25cm}}}$ 
\\

The template was constructed from the concepts belonging to different viewpoints to address the essential regulatory RE challenges, such as capturing the relevant information across all four viewpoints, ensuring traceability of compliance implementation to software design specification, addressing compliance implementation in both existing and new software components, and allowing the initial prioritization using the information about sanctions.

\begin{landscape}
    \begin{figure}
        \centering
        \includegraphics[width=1\linewidth]{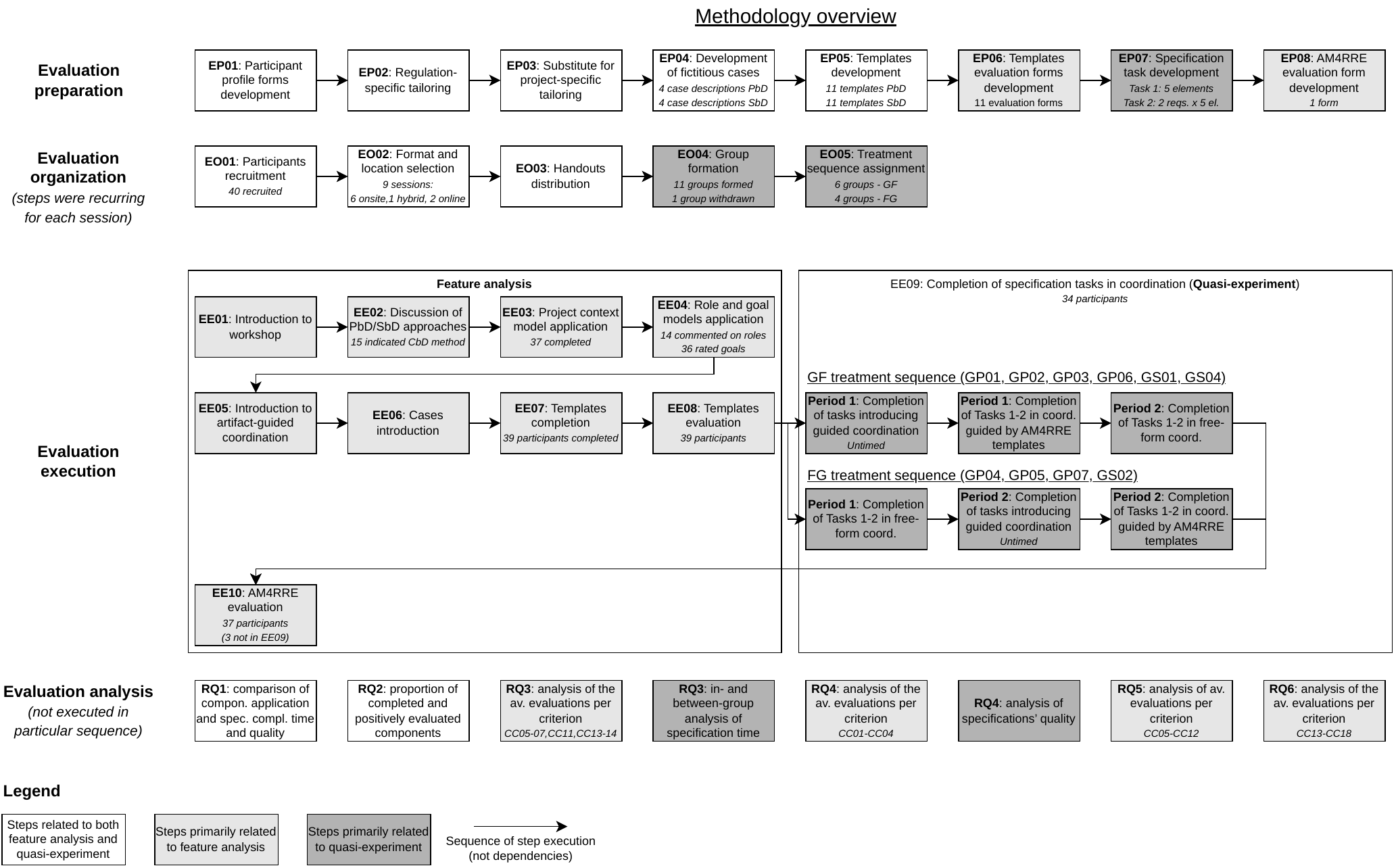}
        \caption{\small Schematic visual overview of the methodology applied in the study.}
        \label{fig:methodology}
    \end{figure}
\end{landscape}

\section{Methodology}\label{sec:methodology}
We conducted this study to \textbf{evaluate AM4RRE in systematically guiding practitioners in specifying and coordinating legal and SE viewpoints during regulatory RE for CbD}. The following research questions guided our study:

\begin{enumerate}[leftmargin=28pt,nosep]
    \item[RQ1:] To what extent can AM4RRE be applied in regulatory RE for privacy by design (PbD) according to GDPR and for security by design (SbD) according to CRA?
    \item[RQ2:] To what extent can practitioners apply AM4RRE in regulatory RE for CbD?
    \item[RQ3:] To what extent can AM4RRE support coordination among legal, managerial, RE, and software architecture viewpoints?
    \item[RQ4:] To what extent can AM4RRE support the quality of regulatory requirements?
    \item[RQ5:] To what extent can AM4RRE support the regulatory RE process?
    \item[RQ6:] To what extent can AM4RRE support achieving the goals of different viewpoints in regulatory RE?
\end{enumerate}

For the AM4RRE evaluation, we used a mixed-methods approach that combines feature analysis as the main data source~\cite {kitchenham1996desmet} with a crossover quasi-experiment as an auxiliary source to complement the feature analysis results~\cite{vegas2015crossover,wohlin2012experimentation}. To implement the feature analysis and quasi-experiment, we created fictitious cases focused on implementing PbD under GDPR and SbD under CRA. We simulated the work and role coordination from four viewpoints: legal, managerial, RE, and software architecture.

For the feature analysis, each participant reviewed and applied project context, role, and goal models and completed viewpoint-specific AM4RRE templates to specify the information required for coordination, using the corresponding case descriptions. Based on this exposure to AM4RRE, participants individually evaluated the usefulness of viewpoint-specific templates using 9 criteria and, after the quasi-experiment, evaluated AM4RRE against 18 further criteria. 

In the two-period crossover quasi-experiment, we explored the use of completed AM4RRE-based templates as a treatment in guided coordination between roles (G), compared with free-form coordination without completed templates (F). We allocated 34 participants to ten groups. Each group completed the same requirements specification tasks twice, once in Period 1 (P1) and once in Period 2 (P2). They used completed AM4RRE-based templates after untimed introductory tasks (G) or case descriptions only (F). Groups followed two treatment sequences: G in P1 followed by F in P2 (GF treatment sequence), or F in P1 followed by G in P2 (FG treatment sequence). We analyzed specification time and the quality of specified regulatory requirements. We selected a quasi-experimental design because we could not randomly assign group composition or fully randomize treatment assignment due to the limited availability of legal experts and other participants. For several reasons, including the lack of systematic CbD methods for comparison and potential carry-over effects, our study aims not to infer causal relations but to examine the association between applying AM4RRE-based templates and the time and quality of regulatory requirements specification. Hence, the hypotheses and evaluation goals we formulate in this study are exploratory rather than confirmatory. We use feature analysis to answer RQ1-6, and the quasi-experiment as an additional data source to answer RQ1-4. We formulate the following exploratory evaluation goals for feature analysis and hypotheses for the quasi-experiment in connection with selected RQs requiring them.

\textit{Evaluation goal 1 (RQ2). Each template can be fully completed by more than 50\% of participants using the guidance embedded within the templates.}
We assessed this goal by analyzing participants' completed templates and evaluations.

\textit{Evaluation goal 2 (RQ2). Participants can specify at least 50\% of the elements assigned in specification tasks using AM4RRE-based templates}.
To assess this goal, we analyzed the specifications each group completed using AM4RRE-based templates.

\textit{Hypothesis 1 for RQ3: Total specification time is shorter in the groups that use AM4RRE-guided coordination in Period 1 (GF sequence) than the total time in the groups that apply free-form coordination in Period 1 (FG sequence): 
\(T_{\mathrm{Total,GF}} = T_{\mathrm{P1,guided}} + T_{\mathrm{P2,free}} < T_{\mathrm{Total,FG}} = T_{\mathrm{P1,free}} + T_{\mathrm{P2,guided}}\bigr)\).}
We suggest that coordination guided by AM4RRE templates, when executed first, can be associated with a shorter total specification times.

\textit{Hypothesis 2 (RQ3): The difference in specification time between Period 1 and Period 2 is larger in groups with FG treatment sequence than in groups with GF treatment sequence: \(\Delta T_{\mathrm{FG}} = \bigl(T_{\mathrm{P1,free}} - T_{\mathrm{P2,guided}}\bigr) > \Delta T_{\mathrm{GF}} = \bigl(T_{\mathrm{P1,guided}} - T_{\mathrm{P2,free}}\bigr)\).}

\textit{Hypothesis 3 (RQ4): In each group, specification under coordination guided by AM4RRE templates is associated with higher specification completeness (Cm), consistency (Cn), and correctness (Cr) than specification under free-form coordination: \((Cm_{\mathrm{P,guided}} > Cm_{\mathrm{P,free}}\bigr)\); \((Cn_{\mathrm{P,guided}} > Cn_{\mathrm{P,free}}\bigr)\); \((Cr_{\mathrm{P,guided}} > Cr_{\mathrm{P,free}}\bigr)\)}.

We report the steps of the methodology following its four phases: (1) evaluation materials preparation (EP), (2) evaluation organization (EO), (3) evaluation execution (EE), and (4) analysis of the evaluation results (EA). Figure~\ref{fig:methodology} provides an overview of the methodology, and Table~\ref{tab:handouts} presents the handout materials.

\begin{table*}[ht]
\tiny
\renewcommand\theadfont{\tiny}
\setlength{\tabcolsep}{2pt}
\centering
\rowcolors{3}{gray!15}{white}
\begin{tabular}{lllllllllll}

\thead{\tiny{ID}} & \thead{\tiny{Name}} & \thead{\tiny{Role(s)}} & \thead{\tiny{Output of}} & \thead{\tiny{Input to}} & \thead{\tiny{Completion method}} & \thead{\tiny{Pp.}}\\ \hline

\multicolumn{7}{l}{\textit{Handout Set I}}  \\

\textit{FD} & Demographic data & Each role & EP01 & EE01 & Free text, option selection & 1 \\

\textit{FC01} & Form for spec. of challenges & Each role & EP01 & EE03 & Free text & 1 \\

\multicolumn{7}{l}{\textit{Handout Set II}}  \\
\textit{FC02} & Form for ranking challenges & Each role & EP03 & EE03 & Ranking & 1 \\

\textit{FR} & Form with roles descriptions & Each role & EP03 & EE04 & Free-text spec. & 2 \\

\textit{GL} & Form with goals for legal role & Legal & EP03 & EE04 & Rating & 2 \\

\textit{GM} & Form with goals for managerial role & Manager. & EP03 & EE04 & Rating & 2 \\

\textit{GR} & Form with goals for RE role & RE & EP03 & EE04 & Rating & 2 \\

\textit{GA} & Form with goals for architect role & Arch. & EP03 & EE04 & Rating & 2 \\

\multicolumn{7}{l}{\textit{Handout Set III}}  \\
\textit{CL\_PB/SB} & Viewpoint-specific case description & Legal & EP04 & EE06 & Not applicable & 1 \\

\textit{CM\_PB/SB} & Viewpoint-specific case description & Manager. & EP04 & EE06 & Not applicable & 1 \\

\textit{CR\_PB/SB} & Viewpoint-specific case description & RE & EP04 & EE06 & Not applicable & 1 \\

\textit{CA\_PB/SB} & Viewpoint-specific case description & Arch. & EP04 & EE06 & Not applicable & 1 \\

\textit{TAct\_LP/LS} & Template for regulatory act spec. & Legal & EP05 & EE07 & Guided text. & 1 \\

\textit{TSubj\_LP/LS} & Template for legal subject spec. & Legal & EP05 & EE07 & Guided use case diagram & 2 \\

\textit{TRel\_LP/LS} & Template for regulatory relations spec. & Legal & EP05 & EE07 & Guided text. & 1 \\

\textit{TObj\_LP/LS} & Template for legal object spec. & Legal & EP05 & EE07 & G. class diag., text highligh. & 3 \\

\textit{TSanc\_LP/LS} & Template for regulatory sanctions spec. & Legal & EP05 & EE07 & Guided text. & 1 \\

\textit{TScop\_MP/MS} & Template for project scope spec. & Manager. & EP05 & EE07 & Guided text. & 1 \\

\textit{TStak\_MP/MS} & Template for stakeholder model spec. & Manager. & EP05 & EE07 & Guided case diagram & 2 \\

\textit{TUsag\_RP/RS} & Template for usage model spec. & RE & EP05 & EE07 & Guided text. & 2 \\

\textit{TData01\_RP/RS} & Template for data model spec. & RE & EP05 & EE07 & Guided class diag. & 1 \\

\textit{TArch\_AP/AS} & Template for archit. overview spec. & Arch. & EP05 & EE07 & Guided component diag. & 2 \\

\textit{TData02\_AP/AS} & Template for data model spec. & Arch. & EP05 & EE07 & Guided class diag. & 1 \\

\multicolumn{7}{l}{\textit{Handout Set IV}}  \\
\textit{SAct\_LP/LS} & Eval. form for regulatory act \textit{TAct\_} & Legal & EP06 & EE08 & Rating, free text eval. & 1 \\

\textit{SSubj\_LP/LS} & Eval. form. for legal subject \textit{TSubj\_} & Legal & EP06 & EE08 & Rating, free text eval. & 2 \\

\textit{SRel\_LP/LS} & Eval. form. for legal relations \textit{TRel\_} & Legal & EP06 & EE08 & Rating, free text eval. & 1 \\

\textit{SObj\_LP/LS} & Eval. form. for legal object \textit{TObj\_} & Legal & EP06 & EE08 & Rating, free text eval. & 3 \\

\textit{SSanc\_LP/LS} & Eval. form. for legal sanctions \textit{TSanc\_} & Legal & EP06 & EE08 & Rating, free text eval. & 1 \\

\textit{SScop\_MP/MS} & Eval. form. for project scope \textit{TScop\_} & Manager. & EP06 & EE08 & Rating, free text eval. & 1 \\

\textit{SStak\_MP/MS} & Eval. form. for stakeh. model \textit{TStak\_} & Manager. & EP06 & EE08 & Rating, free text eval. & 2 \\

\textit{SUsag\_RP/RS} & Eval. form. for usage model \textit{TUsag\_} & RE & EP06 & EE08 & Rating, free text eval. & 2 \\

\textit{SData01\_RP/RS} & Eval. form. for data model \textit{TData01\_} & RE & EP06 & EE08 & Rating, free text eval. & 1 \\

\textit{SArch\_AP/AS} & Eval. form. for archit. overview \textit{TArch\_} & Arch. & EP06 & EE08 & Rating, free text eval. & 2 \\

\textit{SData02\_AP/AS} & Eval. form. for data model \textit{TData02\_} & Arch. & EP06 & EE08 & Rating, free text eval. & 1 \\

\multicolumn{7}{l}{\textit{Handout Set V (order depended on the assigned sequence of specification tasks execution)}}  \\
\textit{CF} & Tasks for free form coordination & Jointly & EP07 & EE09 & Free text., requir. template & 1 \\

\multicolumn{7}{l}{\textit{Handout Set VI (order depended on the assigned sequence of specification tasks execution)}}  \\
\textit{CG01\_P/S} & Intro. tasks for guided coordination & Jointly & EP07 & EE09 & Free text. & 1 \\

\textit{CG02} & Tasks for guided coordination & Jointly & EP07 & EE09 & Free text., requir. template & 1 \\

\multicolumn{7}{l}{\textit{Handout Set VII}}  \\
\textit{EF} & Evaluation form & Each role & EP08 & EE10 & Rating, free text. & 1 \\
\hline
Total pages &  &  &  &  &  &\textbf{55} \\

\hline
\end{tabular}
\caption{\small Overview of evaluation handouts. The ID suffix (e.g., \_LP) encodes viewpoint (L - legal, M -manag., R - requir., A - archit.), and case (P - PbD, S - SbD). ``Each role''--every role received the same handout; ``Jointly''--joint completion by group participants. ``Guided completion''--specification instructions were given.}
\label{tab:handouts}
\end{table*}

\subsection{Evaluation materials preparation (EP)}\label{sec:ep}
We report the development of the evaluation materials (see Table~\ref{tab:handouts}).

\textbf{EP01: Development of participant profile forms} resulted in two handouts: \textit{FD}, which collected data on participants' organizations, positions, work experience, and use of regulatory RE methods for CbD; and \textit{FC01}, which collected free-form descriptions of regulatory RE challenges known to practitioners.

\textbf{EP02: Regulation-specific tailoring} was conducted by the first author using GDPR or CRA norms commonly examined in prior studies\footnote{We selected excerpts from GDPR articles 4(1),(2),(4),(7),(8),(11),(15); 6(1)(a); 8(1),(2); 9(1),(2); 13(1)(c),(f); 13(2)(b),(f); 15(1)(a),(b),(c),(h); 15(3); 83(4)(a); 83(5)(a),(b) and the following CRA norms recital (67), Article 3(1),(2),(10),(13),(14),(17),(23),(40); 13(1),(5); 20(4); Annex I (2)(d),(e),(g),(h).} and  resulted in five tailored content items of the AM4RRE legal artifact model (see Figure~\ref{fig:artifact-model}).

\textbf{EP03: Substitute for project-specific tailoring} used the original AMDiRE artifact model to link SE concepts in six content items---two each from managerial, requirements, and architecture specifications---to legal concepts in five content items tailored in \textit{EP02}, adding new SE concepts where needed. Next, we developed handouts for the remaining untailored AM4RRE models: a ranking of challenges for the project context model, role descriptions for the role model, and four goal model handouts.

\textbf{EP04: Development of fictitious cases} produced four viewpoint-specific case descriptions per case with compliance concerns and minimal overlap (e.g., legal descriptions contained only norm excerpts).

\textbf{EP05: Template development.}\label{sec:ep05}
Using the tailored artifact model, we developed one template for each of the 11 selected content items (Table~\ref{tab:handouts}), covering viewpoint-specific information and information required by related items in the legal viewpoints.

\textbf{EP06: Development of template evaluation forms (feature analysis).}
Evaluation forms included example specifications and the following nine criteria for evaluation using a five-point Likert scale and space for comments:
\begin{enumerate}[leftmargin=45pt,nosep]
    \item[TC01] (\textit{RK} in Table ~\ref{tab:templates-average}) capturing role-specific knowledge (e.g., legal knowledge);
    \item[TC02] (\textit{Expl.}) explicating and explaining the role-specific knowledge that is tacit;
    \item[TC03] (\textit{Tran.}) specifying information in a way transparent to other roles;
    \item[TC04] (\textit{Comm.}) specifying information to use specifications for communication;
    \item[TC05] (\textit{Cm.}) capturing all the required information;
    \item[TC06] (\textit{Trac.}) tracing the information between different roles and specifications;
    \item[TC07] (\textit{Cn.}) specifying information in a consistent way;
    \item[TC08] (\textit{Eas.}) ease of information specification;
    \item[TC09] (\textit{Comp.}) guiding specification by persons without role-specific knowledge / compensating for the absence of the role-specific knowledge.
\end{enumerate} 

During the quasi-experiment, we also substituted example specifications for completed templates when the corresponding role was absent from a group.

\textbf{EP07: Specification task development (quasi-experiment).}
Handouts for both guided (\textit{CG02}) and free-form (\textit{CF}) coordination described two tasks: (1) specifying two existing regulated software components, two new system components or controls (e.g., processes) to implement compliance, and one relationship between existing and new components; and (2) specifying two regulatory requirements using the regulatory requirements template (Section~\ref{sec:am4rre}). The guided-coordination handout (CG02) included requirement templates with placeholders; the free-form coordination handout included an example template and a free-text form to allow flexibility. For free-form coordination, we instructed participants to use only case descriptions, whereas during guided coordination, participants used their completed templates and consulted case descriptions as needed.
To introduce AM4RRE-guided coordination and its templates, we developed five instructional tasks (\textit{CG01}), distinct from the main tasks (see open dataset for their description).
PbD and SbD handouts differed only in task descriptions and wording.

\textbf{EP08 Development of AM4RRE evaluation form (feature analysis).}
The form contained 18 criteria in three groups corresponding to the RQs they address.

\textbf{Requirements quality (\textit{Qual.} in Table~\ref{tab:evaluation}) (RQ4)}
\begin{enumerate}[leftmargin=45pt,nosep]
    \item[CC01] (\textit{Cm.} in Table~\ref{tab:evaluation}) completeness of requirements deriving from GDPR / CRA;
    \item[CC02] (\textit{Cn.1}) consistency between requirements deriving from GDPR / CRA;
    \item[CC03] (\textit{Cn.2}) consistency between GDPR/CRA and other requirements;
    \item[CC04] (\textit{Cr.}) correctness of requirements deriving from GDPR / CRA.
\end{enumerate}

\textbf{Regulatory RE process (\textit{RE}) (RQ3, RQ5)}
\begin{enumerate}[leftmargin=45pt,nosep]
    \item[CC05] (\textit{Coord.}) facilitating coordination for regulatory requirements specification;
    \item[CC06] (\textit{LK}) capturing legal domain knowledge;
    \item[CC07] (\textit{RK}) capturing other role-specific knowledge (business, requirements, etc.);
    \item[CC08] (\textit{Comp.}) in case of the absence of legal knowledge, compensation for the absence of it (guiding a non-specialist in specification);
    \item[CC09] (\textit{Trac.}) facilitate traceability between specifications and roles;
    \item[CC10] (\textit{Cn3}) facilitate consistency between roles (when coordinating);
    \item[CC11] (\textit{Tran.}) assure transparency of specifications to other roles;
    \item[CC12] (\textit{Inter.}) support interpretation of regulations.
\end{enumerate}

\textbf{Achievement of role-specific goals (\textit{Goal}) (RQ3, RQ6)}
\begin{enumerate}[leftmargin=45pt,nosep]
    \item[CC13] (\textit{SA}) facilitating GDPR / CRA compliance in software architecture;
    \item[CC14] (\textit{SDLC}) facilitating GDPR / CRA compliance throughout the SDLC;
    \item[CC15] (\textit{Decis.}) enabling decision-making about compliance;
    \item[CC16] (\textit{Doc.}) documenting the required information;
    \item[CC17] (\textit{Gov.}) facilitating compliance governance;
    \item[CC18] (\textit{Bus.}) addressing business concerns in CbD implementation.
\end{enumerate}
\vspace{1em}
We used the following five-point Likert scale for rating goals (\textit{EP03}), template evaluation (\textit{EP06}), and AM4RRE evaluation (\textit{EP08}): -2--not useful, -1--potentially not useful, 0--no assessment, +1--potentially useful, +2--useful.

\subsection{Evaluation organization (EO)}\label{sec:eo}
Next, we report participant recruitment, evaluation format, and organization.

\textbf{EO01: Participants recruitment}\label{sec:eo01} using purposive sampling~\cite{baltes2022sampling} gained 40 participants (28 in evaluation for PbD, 12 -- in SbD) from 23 companies (see Table~\ref{tab:participants}) with GDPR or CRA experience in SE to fill four roles: legal expert (9-PbD, 4-SbD), software architect (8-PbD, 4-SbD), requirements engineer (6-PbD, 3-SbD), and manager (5-PbD, 1-SbD). Eligibility required one year of GDPR or CRA experience, and we counted relevant training or regulatory exposure as fulfilling this requirement. Of the 40 participants, 37 completed the final AM4RRE evaluation\footnote{P3.1R, S9.3R, S10.3A withdrew due to unforeseen circumstances. We kept data from the activities they completed: for all three — known CbD methods, free-form challenge descriptions, challenge ranking, and goal ratings; for P3.1R — evaluation of one template; for S9.3R — evaluation of two RE templates.}. Of these 37, 34 completed the quasi-experiment in ten groups, as P15.L was not assigned to any group, and S7.3L and S8.3M did not complete tasks because two members of GS03 withdrew.

\textbf{EO02: Evaluation format and location.}\label{sec:eo02}
We held sessions on-site using printed handouts, except for GS01 and GS03, which were fully online via Zoom and used Zoom Whiteboard, and GP05, with P16.5L joining remotely and three participants attending on-site (we printed completed digital templates for on-site use and scanned completed paper templates to Zoom Whiteboard).

\textbf{EO03: Handouts distribution} was organized in sets (see Table~\ref{tab:handouts}) to prevent bias from later materials (e.g., templates and forms with specification instances).

\textbf{EO04: Group formation (quasi-experiment)}\label{sec:eo04} aimed to represent all four viewpoints per group, which we achieved in GP04-GP07. GP01 included two architects and no RE role, and GP07 included two legal experts. GP03, GS01, and GS02 lacked managerial roles. Managerial and RE roles were absent in GP02 and GS04. In these groups, specification instances from the evaluation forms replaced absent roles. Most groups participated in separate evaluation sessions. GP02, GP03, and GP04 worked in parallel during one shared evaluation session.

\textbf{EO05: Treatment sequence assignment (quasi-experiment).}\label{sec:eo05}
Each group completed the specification tasks under both coordination conditions: guided coordination (G) and free-form coordination (F). Each group followed either the GF or FG treatment sequence. To accommodate practitioners' limited availability and avoid prolonged free-form coordination, we avoided the FG treatment sequence for groups with (1) missing roles, (2) no prior collaboration, (3) limited GDPR/CRA experience or (4) CbD knowledge, or (5) participants from small and medium organizations.

\begin{landscape}
\begin{table*}[ht]
\tiny
\renewcommand\theadfont{\tiny}
\setlength{\tabcolsep}{3.75pt}
\centering
\rowcolors{3}{gray!15}{white}
\begin{tabular}{ccccccccccccccc}

\thead{\tiny{Group}} & \thead{\tiny{}} & \thead{\tiny{Part.}} & \thead{\tiny{}} & \thead{\tiny{}} & \thead{\tiny{}} & \thead{\tiny{Org.}} & \thead{\tiny{Org.}} & \thead{\tiny{}} & \thead{\tiny{Wk.}} & \thead{\tiny{Reg.}} & \thead{\tiny{}}  & \thead{\tiny{}} & \thead{\tiny{}}  & \thead{\tiny{Addit}}\\

\thead{\tiny{ID}} & \thead{\tiny{Durat.}} & \thead{\tiny{ID}} & \thead{\tiny{Role}} & \thead{\tiny{Format}} & \thead{\tiny{Position}} & \thead{\tiny{ID}} & \thead{\tiny{Size}} & \thead{\tiny{Industry}} & \thead{\tiny{Exp.}} & \thead{\tiny{Exp.}} & \thead{\tiny{CbD}}  & \thead{\tiny{FA}} & \thead{\tiny{QE}}  & \thead{\tiny{Handout}}\\ \hline
GP01 &  & P1.1L & Legal expert & Onsite & Product Security Lead & C1 & >250 & IT & 8 & 3 & No & Yes & Yes & - \\

GP01 &  & P2.1M & Manager & Onsite & Security consultant & C2 & <10 & IT, Retail, Fin., Tel. & 20 & 8 & No & Yes & Yes & - \\ 

GP01 &  & P3.1R & RE & Onsite & Business architecture analyst & C3 & >250 & IT & 3 & 1 & No & No & No & - \\

GP01 &  & P4.1A & Software architect & Onsite & Product Security Engineer & C4 & >250 & Finances & 7 & 2 & No & Yes & Yes & - \\

GP01 & \multirow{-5}{*}{6h:25m} & P5.1A & Software architect & Onsite & Information Security Architect & C5 & >250 & Healthcare & 10 & 2 & Yes & Yes & Yes & TObj\_L \\

\midrule

GP02 &  & P6.2L & Legal expert & Onsite & Privacy Operations Manager & C6 & >250 & IT & 7 & 7 & Yes & Yes & Yes & - \\

GP02 &  & P7.2A & Software architect & Onsite & Product Security Engineer & C6 & >250 & IT & 5 & 5 & No & Yes & Yes & -\\

\cmidrule{1-1}
\cmidrule{3-15}

GP03 &  & P8.3L & Legal expert & Onsite & GRC Analyst & C6 & >250 & IT & 3 & 1 & No & Yes & Yes & - \\

GP03 &  & P9.3R & RE & Onsite & Senior GRC Analyst & C6 & >250 & IT & 4 & 3 & Yes & Yes & Yes & - \\

GP03 &  & P10.3A & Software architect & Onsite & Product Security Engineer & C6 & >250 & IT & 2 & 2 & No & Yes & Yes & - \\

\cmidrule{1-1}
\cmidrule{3-15}

GP04 &  & P11.4L & Legal expert & Onsite & ISO System Manager & C6 & >250 & IT & 9 & 7 & No & Yes & Yes & - \\

GP04 &  & P12.4M & Manager & Onsite & Vice President Engineering & C6 & >250 & IT & 6 & 7 & Yes & Yes & Yes & -\\

GP04 &  & P13.4R & RE & Onsite & Cybersecurity Analyst & C6 & >250 & IT & 4 & 1 & No & Yes & Yes & - \\

GP04 &  & P14.4A & Software architect & Onsite & Senior Software Engineer & C6 & >250 & IT & 10 & 1 & No & Yes & Yes & - \\

\cmidrule{1-1}
\cmidrule{3-15}

NA & \multirow{-11}{*}{6h:40m} & P15.L & Legal expert & Onsite & GRC Analyst & C6 & >250 & IT & 3 & 1 & No & Yes & No & -\\

\midrule
GP05 &  & P16.5L & Legal expert & Online & Corp. Data Protection Manager & C7 & >250 & Logistics & 10 & 10 & NA & Yes & Yes & - \\

GP05 &  & P17.5M & Manager & Onsite & Web Analyst & C8 & 50-249 & Telecommunications & 9 & 5 & No & Yes & Yes & - \\

GP05 &  & P18.5R & RE & Onsite & Quality\&Compliance manager & C9 & <10 & Aerospace & 3 & 1 & No & Yes & Yes & - \\

GP05 & \multirow{-4}{*}{5h:45m} & P19.5A & Software architect & Onsite & IT architect & C10 & >250 & Automotive & 10 & 9 & Yes & Yes & Yes & - \\

\midrule
GP06 &  & P20.6L & Legal expert & Onsite & Privacy manager & C11 & >250 & Retail & 7 & 7 & No & Yes & Yes & -\\

GP06 &  & P21.6M & Manager & Onsite & Technical manager & C12 & 50-249 & IT & 10 & 1 & No & Yes & Yes & - \\

GP06 &  & P22.6R & RE & Onsite & IT Manager & C13 & >250 & Automotive, IT, Fin. & 20 & 6 & No & Yes & Yes & - \\

GP06 & \multirow{-4}{*}{5h:15m} & P23.6A & Software architect & Onsite & Cloud Security Architect & C14 & >250 & IT & 3 & 5 & Yes & Yes & Yes & GL; TObj\_L \\

\midrule

GS01 &  & S1.1L & Legal expert & Online & Lawyer & C15 & <10 & NA & 3-5 & 1 & Yes & Yes & Yes & - \\

GS01 &  & S2.1R & RE & Online & External CIO & C16 & 50-249 & Finances & 25 & 0 & Yes & Yes & Yes & - \\

GS01 & \multirow{-3}{*}{6h:00m} & S3.1A & Software architect & Online & Managing Software Architect & C17 & <10 & Automotive, IT & 10 & 0.5 & Yes & Yes & Yes & - \\

\midrule
GS02 &  & S4.2L & Legal expert & Onsite & Cybersecurity Professional & C18 & NA & Manufacturing & 7 & 1 & Yes & Yes & Yes & - \\

GS02 &  & S5.2R & RE & Onsite & Product manager & C19 & >250 & Telecommunications & 20 & 1 & No & Yes & Yes & TObj\_L \\

GS02 & \multirow{-3}{*}{4h:10m} & S6.2A & Software architect & Onsite & Cyber Threat Intel. Expert & C20 & >250 & Automotive & 10 & 0.5 & Yes & Yes & Yes & - \\

\midrule

GS03 &  & S7.3L & Legal expert & Online & Information Security Officer & C21 & >250 & IT, Finances & 7 & 1 & Yes & Yes & No & - \\

GS03 &  & S8.3M & Manager & Online & Deputy CTO Security & C21 & >250 & IT, Finances & 15 & 1.5 & Yes & Yes & No & - \\

GS03 &  & S9.3R & RE & Online & Head of Product Security & C21 & >250 & IT, Finances & 15 & 0.5 & Yes & Part. & No & - \\

GS03 & \multirow{-4}{*}{4h:30m} & S10.3A & Software architect & Online & Security compliance manager & C21 & >250 & IT, Finances & 15 & 1 & No & Part. & No & - \\

\midrule
GP07 &  & P24.7L & Legal expert & Onsite & DPO Assoc. Gener. Counsel & C6 & >250 & IT & 25 & 7 & No & Yes & Yes & - \\
GP07 &  & P25.7L & Legal expert & Onsite & Solution Architect & C6 & >250 & IT & 14 & 1 & No & Yes & Yes & - \\
GP07 &  & P26.7M & Manager & Onsite & VP Product Development & C6 & >250 & IT & 20 & 6 & No & Yes & Yes & -\\
GP07 &  & P27.7R & RE & Onsite & Head of Development & C6 & >250 & IT & 12 & 1 & No & Yes & Yes & - \\
GP07 & \multirow{-5}{*}{5h:30m} & P28.7A & Software architect & Onsite & Product manager & C6 & >250 & IT & 4 & 1 & No & Yes & Yes & GL; TObj\_L \\

\midrule

GS04 &  & S11.4L & Legal expert & Onsite & Data Protection Officer & C22 & >250 & IT & 6 & 2 & Yes & Yes & Yes & - \\

GS04 & \multirow{-2}{*}{4h:40m} & S12.4A & Software architect & Onsite & Solution Architect & C23 & >250 & IT & 15 & 1 & No & Yes & Yes & - \\

\hline
\end{tabular}
\caption{\small Data about participants and groups. Columns are as follows: (1) group ID with prefix indicating PbD (e.g., GP01) or SbD (e.g., GS01); (2) participant ID (e.g., P1.1L) in a format: case identifier, unique participant number (e.g., P1), group number, role/viewpoint identifier (L for legal, M--manager, R--RE, A--software architect); (3) the role the participant represented; (4) participation format, (5) the participant's position; (6) company ID; (7) company size; (8) company industry; (9) experience in current or related position (years); (10) experience with GDPR or CRA ; (11) knowledge/use of method for PbD/SbD, (12) fully completed feature analysis, (12) fully completed quasi-experiment, (13) ID(s) of additionally filled-in handouts.}
\label{tab:participants}
\end{table*}
\end{landscape}

\subsection{Evaluation execution (EE)}
The first author conducted nine evaluation sessions from October 2025 to March 2026, averaging 5h:26m (see Table~\ref{tab:participants}). In addition to comments in handouts, we collected qualitative data by taking notes on observations, comments, and discussions between participants. We report each evaluation step with approximate duration.

\textbf{EE01: Introduction (5-10 min)} covered the evaluation overview, completion of the demographics forms \textit{FD}, and participant introductions.

\textbf{EE02: Recap and discussion of PbD/SbD approaches (10-15 min)} covered the PbD/SbD methods known to or used by participants.

\textbf{EE03: Application of project context model (feature analysis) (15-20 min)} included specifying the regulatory RE challenge in free form using \textit{FC01} and, after a joint discussion, ranking 12 predefined challenges using \textit{FC02}.

\textbf{EE04: Role and goal models application (feature analysis) (15-20 min)} involved reviewing the role model (\textit{FR} handout), selecting a role, and rating all of its viewpoint-specific goals and second-level goals of other viewpoints. In GP01, only the goals of the assigned viewpoints were rated. Additionally, P7.2A, P23.6A, P28.7A, S6.2A rated the legal, and P20.6L - software architecture viewpoints goals.

\textbf{EE05: Introduction to artifact-guided coordination (10 min)} by the session lead covered artifact-guided coordination, without the templates.

\textbf{EE06: Case introduction (10 min)} using viewpoint-specific descriptions.

\textbf{EE07: Viewpoint-specific templates completion (feature analysis) (90-120 min)} was done in the assigned order using case descriptions. P5.1A, P23.6A, P28.7A, and S5.2R additionally completed the legal object template \textit{TObj\_L}.

\textbf{EE08: Templates evaluation (feature analysis) (30-40 min)} involved participants reviewing specification instances, evaluating templates without changing completed templates, and returning evaluation forms upon completion.

\textbf{EE09: Completion of specification tasks in coordination (quasi-experiment) (40-80 min)} required participants to use completed templates and case descriptions when necessary (e.g., the TObj template required highlighting concepts in the case description) for AM4RRE-guided coordination and case descriptions with any auxiliary materials for free-form coordination.
Before each Period, groups had time to read, ask questions, or discuss the tasks. Task timing began and ended on participants' explicit signals. Immediately before performing the guided coordination tasks, participants completed untimed tasks (\textit{CG01}) introducing template-guided coordination. Specification instances from the evaluation handouts substituted for absent roles. The lead could introduce the instances or point to information in them in an ad hoc way, but only when questions were raised and without intervening in task completion. Groups with multiple completed templates per viewpoint (GP01, GP07) used all available specifications, each completed only by a participant in the corresponding role. In GP01, handouts provided only minimal guidance for organization (e.g., assigning someone to record results) and for introductory and specification tasks (\textit{CG01}, \textit{CG02}), with the session lead providing guidance ad hoc. This organizational guidance was also embedded in handouts for later groups.

\textbf{EE10: AM4RRE evaluation (10 min) (feature analysis).}\label{sec:ee10}
Participants evaluated AM4RRE as a whole against 18 criteria using handout \textit{EF}.

\subsection{Evaluation analysis (EA)}
To answer RQ1, we compared results for PbD and SbD cases: role model (free-form comments); goal model (goal ratings); AM4RRE evaluation (average evaluation per criterion); specification time; and requirements quality. For evaluations, a difference of 1 point or more ($\geq$ 25\%) was considered substantial.

For RQ2, we used the following applicability criteria. We considered the project context model applicable if rankings were heterogeneous (low Kendall's W) and participants confirmed model usefulness to capture heterogeneity. The role model -- in the absence of substantially critical written comments. The goal model -- if over 50\% of all ratings were positive (+1 or +2), and over 50\% of goals had an average rating above 0. And the templates -- if over 50\% of participants completed them fully, over 50\% of evaluations across nine criteria were positive, and the average evaluation for each criterion was above 0. For the quasi-experiment, we considered AM4RRE applicable for specification if more than 50\% of specification elements were completed. We chose the 50\% threshold to ensure that at least one of the two regulatory requirements is available for evaluation.

For RQ3-6, we analyzed the number of evaluations and average evaluations per criterion in feature analysis, focusing on outlier groups or participants. For RQ3, we analyzed the results across six criteria contributing to coordination (\textit{CC05-07}, \textit{CC11}, \textit{CC13-14}). For RQ4-6, we focused on criteria for requirements quality (CC01-CC04), RE process (CC05-CC12), and viewpoint goals support (CC13-CC18), respectively.

To answer RQ1-4, we also analyzed specification time and quality in the quasi-experiment by comparing outcomes between groups following the GF and FG treatment sequences, and between Period 1 and Period 2 within each group. We assessed specification quality for internal completeness (Task 2), consistency, and correctness of each element in Tasks 1 and 2 (see definitions in Section~\ref{sec:background}). Completely missing specification elements (separate elements in Task 1 and requirements in Task 2) were not assessed for quality and were reported for RQ2. We included any additional requirements or specification components in the quality assessment. For completeness, we assigned 0.2 for each of the five requirement element types, following the regulatory requirements template (see Section~\ref{sec:am4rre}). For consistency, we assigned 1 for full agreement with the completed templates (guided) or the case descriptions (free-form), and 0.5 when the relevant component was present but specified inconsistently. For correctness, we assigned 1 if the element contributed to compliance implementation according to GDPR or CRA norms in the legal case description, and 0.5 if it was partially correct or if the specification concerned norms outside the scope of the case description. If participants used legal instead of SE concepts and these were consistent and correct, we assigned 0.5. Specification quality was assessed in two rounds involving independent referees: a legal expert with experience in both GDPR and CRA for correctness, and an undergraduate computer science student with privacy and security knowledge for consistency. First, the first author and referees used case descriptions, assessment rules reported above, completed templates, and requirements specifications to assess correctness independently (referee was blinded for coordination type) and consistency (not blinded). Afterward, we discussed and resolved disagreements to finalize the assessment.

\section{Results}\label{sec:results}
Before presenting the results for each RQ, we summarize the regulatory RE or CbD methods and the challenges reported in the profile forms.

\subparagraph{State of practice and challenges}
In \textit{FD}, 15 of 40 participants reported using or knowing a regulatory RE approach for CbD (6 for PbD, 9 for SbD), with the most frequently mentioned threat modeling, data protection impact assessments, standards, and technical approaches (e.g., zero-trust architecture). While some approaches involved interaction between roles, none specifically focused on it. This suggests that approaches used or known to participants are unlikely to affect the quasi-experiment substantially.
To analyze the 177 challenges to regulatory RE for CbD reported by 35 participants in the \textit{FC01} handout, we classified them into 12 challenge categories used for ranking challenges in the \textit{FC02} handout. Participants most often mentioned conflict with existing SE practices, compliance enforcement, interaction between experts, organizational challenges, and resource intensity (see Table~\ref{tab:challenges01}).

\begin{table*}[ht!]
\tiny
\renewcommand\theadfont{\tiny}
\setlength{\tabcolsep}{2pt}
\centering
\rowcolors{3}{gray!15}{white}
\begin{tabular}{cp{2.2cm}p{5.1cm}p{2.5cm}c|cc}

\tiny{Ch.} & \tiny{Challenge} & \tiny{Challenge} & \tiny{Challenge} & \tiny{Ment} & \tiny{A.rk.} & \tiny{S.rk.} \\
\tiny{ID} & \tiny{Name} & \tiny{Description} & \tiny{Example} & \tiny{FC01} & \tiny{FC02} & \tiny{FC02} \\ \hline

Ch01 & \textbf{Abstractness of regulations} &     Regulations are often too abstract to be directly applicable in SE and therefore require interpretation and translation into concrete software requirements. & "lack of clear rules on what to implement exactly (S11.4L)" & 7 & \textbf{5.46} & \textbf{202} \\ \hline

Ch02 & \textbf{Conflicts/changes to existing SE practices} & Regulations may directly or indirectly conflict with existing software engineering practices or necessitate changes to them. & "balance it out out with all other requirements (S3.1A)" & \textbf{41} & 6.03 & 217\\ \hline

Ch03 & \textbf{Demand for knowledge/expertise} & Implementing regulations requires legal domain knowledge to translate regulatory texts; additionally, SE expertise, business domain knowledge, and privacy and security expertise are required for compliance implementation. & "team only knows one part not end-to-end (P16.5L)" & 16 & \textbf{4.95} & \textbf{183} \\ \hline

Ch04 & \textbf{Complexity of regulations processing and implementation} & The implementation of regulatory compliance is complex, primarily due to the inherent complexity of the regulations themselves and the complexity of modern software systems. & "a big amount of new features and how to cover all of it (P1.1L)" & 14 & 6.22 & 230 \\ \hline

Ch05 & \textbf{Interaction between different experts} & Implementing regulatory compliance is interdisciplinary and requires interaction among experts from various domains (e.g., law or compliance, software technology, business, privacy, and security). These experts are also highly interdependent. & "teams working in silos (P23.6A)" & \textbf{21} & \textbf{5.32} & \textbf{197}\\ \hline

Ch06 & \textbf{Resource intensity of processing and implementation of regulations} & Ensuring regulatory compliance in software systems can require significant financial, technical, human, time, and other resources. These resources are needed not only for processing regulations and implementing technical solutions but also for auxiliary activities such as documentation and audits. & "Make estimations regarding time, costs, quality (magic triangle) or avoid them (S3.1A)" & \textbf{17} & \textbf{5.14} & \textbf{185} \\ \hline

Ch07 & \textbf{Provability challenges} & Implementation of regulatory compliance must always be supported by evidence such as documentation. Producing such evidence is difficult due to the complexity of the required information and the multiplicity of sources that must be considered. & NA & NA & 9.32 & 345 \\ \hline

Ch08 & \textbf{Enforcement challenges} & It is often challenging to ensure that regulations are implemented as intended (e.g., ensuring that engineers follow the required processes). & "Some changes don't go through the change management procedures (S7.3L)" & \textbf{28} & 6.97 & 251 \\ \hline

Ch09 & \textbf{Dynamics of software context and regulations} & Both regulations and the business domains of software systems can change rapidly, creating multiple barriers to maintaining compliance. & "unstable nature of project goals (P16.5L)" & 3 & 6.69 & 234 \\ \hline

Ch10 & \textbf{Multiplicity of regulations} & Multiple regulations often need to be complied with simultaneously within the same software system. & "[s]ame feature must comply with different regulations (P24.7L)" & 5 & 7.81 & 289 \\ \hline

Ch11 & \textbf{Dynamics of software systems} & Software systems can evolve rapidly, especially in agile software development environments or in the case of AI-based systems, dynamic in nature. & "continuous changes (P17.5M)" & 5 & 7.44 & 268 \\ \hline

Ch12 & \textbf{Organizational challenges} & Regulatory compliance in software systems requires changes in both the organization of software development and the overall organizational structure (e.g., planning, strategy). & "lack of structural approach to PbD (4) (P6.2L)" & \textbf{20} & \textbf{5.92} & \textbf{213} \\

\hline

\hline
\end{tabular}
\caption{\small Overview of challenges to regulatory RE mentioned in free-form in FC01 and ranked in FC02 (in the order presented to participants in FC02). Columns: challenge ID, name, description, example of challenge from FC01, number of mentions in FC01, average ranking in FC02, sum of ranks in FC02.}
\label{tab:challenges01}
\end{table*}

\subsection{RQ1: To what extent can AM4RRE be applied in regulatory RE for PbD and SbD?}\label{sec:rq1}
We compared goal ratings, template evaluation, and coordination results in PbD and SbD cases to assess AM4RRE applicability in both.

\subparagraph{Project context model (step EE03)} The comparison of challenge rankings did not identify a substantial difference--the same four challenges appeared in the top five ranking positions, and agreement according to Kendall's W was 0.15 in PbD and SbD.

\subparagraph{Role model (step EE04)} Qualitative analysis of participants' written comments did not identify a substantial difference between PbD and SbD evaluation sessions. SbD participants made 27 suggestions and PbD participants 21, although fewer people took part in SbD sessions.

\subparagraph{Goal models (step EE04)}
The proportion of positive ratings for different goal models differed by 4 to 7 percentage points (see Table~\ref{tab:goals-summary}).
Average usefulness ratings (see open dataset) differed by at least one point between PbD and SbD for 18 of 74 goals across four viewpoints (24\%). Eight of these were management-viewpoint goals, with only one participant in the SbD sessions. For 11 of the 18 goals, the average usefulness was positive, indicating agreement on their usefulness but to different degrees. For the remaining seven goals, the average rating either in PbD or SbD was 0 (no assessment). Only one of 74 goals (GR1.2)
had a positive average rating in PbD, and a negative average rating in SbD evaluation, indicating considerable disagreement.

\subparagraph{Templates evaluation (step EE08)}
Of 99 average evaluations of 11 templates against nine evaluation criteria, 16 (16.16\%) across eight templates differed by at least one point between PbD and SbD (see Table~\ref{tab:templates-average})—six of these concerned the managerial-viewpoint templates, which were evaluated by only one participant assigned to the SbD case. From these 16, six had opposite signs in PbD and SbD, indicating considerable disagreement. Of 198 total average evaluations, one average evaluation had opposite signs, but differed by less than 1 point.

\subparagraph{AM4RRE evaluation (step EE10)}
The total average evaluation of AM4RRE was 0.21 lower in SbD (1.02) than in PbD (1.23) (see Table~\ref{tab:evaluation}). In none of the average per-criterion evaluations was the difference between PbD and SbD 1 or higher.

\subparagraph{Specification time and requirements quality (EE09)} The analysis of specification time and quality (see Table~\ref{tab:coordination}) indicated no considerable differences between PbD and SbD cases.

\subparagraph{Summary}
Application results and feedback on the project context, role, and goal models indicated no substantial differences between PbD and SbD contexts. Template evaluations were mixed across cases, and AM4RRE evaluations were lower in the SbD case. The differences reflected disagreement about the degree of usefulness, while the overall evaluation was positive. These findings provide preliminary evidence that AM4RRE is useful in both PbD and SbD sessions.

\subsection{RQ2: To what extent can practitioners apply AM4RRE in regulatory RE for CbD?}\label{sec:rq2}
We report our analysis and observations on applying the project context, role, and goal models, and AM4RRE-based templates.

\subparagraph{Project context model (step EE03)}
Of 37 participants who ranked the challenges\footnote{Including P3.1R, S9.3R, S10.3A}, 32 provided complete rankings; two omitted at least one rank, and four assigned duplicate ranks. Based on average and summed ranks, the top-ranked challenges were abstractness of regulations, demand for domain knowledge, expert interaction, resource intensity, and organizational challenges (see Table~\ref{tab:challenges01}). However, average ranks ranged from 4.95 to 9.32, and Kendall's W was 0.12, indicating differences in prioritization. Further analysis of average ranks by participant group (roles, company affiliation) also showed no clearly dominant challenges; however, RE roles showed higher agreement (Kendall's W = 0.39) than managerial roles (0.23), architect roles (0.18), and legal roles (0.13). In a written comment, P16.5L suggested that challenge rankings strongly depend on company size, sector, culture, and other context factors.
Oral comments from some participants (e.g., in groups GP02, GP03, and GP04) suggested that the ranking of challenges could contribute to better mutual understanding and prioritization of challenges across roles when discussed collectively.

\subparagraph{Role model (step EE04)}
Participants suggested 48 changes to the role models: 18 for requirements engineers, 13 for software architects, nine for legal experts, and eight for managerial roles. Suggestions primarily focused on additional role-specific knowledge, such as risk management (managerial role); cross-domain and technical knowledge (requirements engineer); and product and business knowledge (software architect role). Suggestions regarding role skills included, for example, communication and strong critical thinking (legal expert); communication and prioritization (requirements engineer); and basic compliance review (managerial role). According to P20.6L, legal experts need a basic understanding of the business domain and only consult and validate, rather than ensure compliance.

\begin{table*}[ht!]
\tiny
\renewcommand\theadfont{\tiny}
\setlength{\tabcolsep}{2.5pt}
\centering
\rowcolors{3}{gray!15}{white}
\begin{tabular}{cccccc|ccccc|c|c}

 &  &  &  &  &  & \multicolumn{5}{c}{Ratings count} &  & \tiny{Total}\\
 
ID & Viewpoint & Viewp. Goals & Other Goals & Part. & Total Rates & -2 & -1 & 0 &  +1 & +2 & Posit. & Posit. \\
\hline

GL PbD &  &  &  & 10 & 271 & 5 & 13 & 19 & 89 & 145 & 86.35\% \\
GL SbD & \multirow{-2}{*}{Legal} & \multirow{-2}{*}{21} & \multirow{-2}{*}{11} & 5 & 158 & 2 & 13 & 13 & 62 & 68 &  82.28\% & \multirow{-2}{*}{84.85\%} \\
\hline

GM PbD &  &  &  & 5 & 129 & 3 & 10 & 13 & 41 & 62 & 79.84\% & \\
GM SbD & \multirow{-2}{*}{Manag.} & \multirow{-2}{*}{19} & \multirow{-2}{*}{11} & 1 & 29 & 0 & 0 & 7 & 12 & 10 & 75.86\% & \multirow{-2}{*}{79.11\%} \\
\hline

GR PbD &  &  & & 5 & 153 & 6 & 10 & 11 & 51 & 75 & 82.35\% & \\
GR SbD & \multirow{-2}{*}{RE} & \multirow{-2}{*}{23} & \multirow{-2}{*}{12} & 3 & 101 & 1 & 5 & 16 & 26 & 53 & 78.22\% & \multirow{-2}{*}{80.71\%}\\
\hline

GA PbD &  &  &  & 7 & 139 & 4 & 6 & 19 & 50 & 60 & 79.12\% & \\
GA SbD & \multirow{-2}{*}{Archit.} & \multirow{-2}{*}{12} & \multirow{-2}{*}{14} & 4 & 102 & 0 & 4 & 10 & 47 & 41 & 86.27\% & \multirow{-2}{*}{82.16\%}\\
\hline
Total & - & 74 & - & 36 & 1082 & 21 & 61 & 108 & 378 & 514 & - & 82.44\%\\

\hline
\end{tabular}
\caption{\small Summary of the goals rating: goal model and case, viewpoint, number of viewpoint-specific goals, number of second-level goals of other viewpoints, number of participants who completed rating, number of ratings per evaluation scale.}
\label{tab:goals-summary}
\end{table*}

\subparagraph{Goal model (step EE04)}
Of all goal ratings (see Table~\ref{tab:goals-summary}), 82.44\% were positive on average. Of the 74 rated goals in four viewpoints, one had a negative average (GA2.2--isolation from business logic), and three had an average of 0--no assessment (GM4.5--ability to choose between compliance solutions, and GR1.2--traceability to the origin, GR2.5--assuring development team autonomy). For the remaining goals, the average rating was above zero. Participants' oral comments indicated that nearly all goals seemed useful but would need prioritizing in practice.

\subparagraph{Templates (EE07-EE08)}
Of the 11 templates, five were fully completed by all participants (TRel\_L, TScop\_M, TUsag\_R, TData01\_R, TData02\_A), and five by 67\% to 92\% of them. TObj\_L was the only template that was completed by fewer than 50\% of participants--41.18\% of participants completed it fully and 41.18\% completed it partially.

\begin{table*}[ht!]
\tiny
\renewcommand\theadfont{\tiny}
\setlength{\tabcolsep}{3.75pt}
\centering
\rowcolors{3}{gray!15}{white}
\begin{tabular}{lllcc|ccccc|c}

Templ. & Templ. & Partic. & Total &  & \multicolumn{5}{c}{Evaluations number} \\
\tiny{ID} & \tiny{Name} & \tiny{Viewp.} & \tiny{number} & \tiny{Eval.} & \tiny{-2} & \tiny{-1} & \tiny{0} &  \tiny{+1} & \tiny{+2} & \tiny{Posit.} \\
\hline

\textit{TAct\_L} & Regulatory act spec. templ. & Legal & 12 & 105 &  6 & 10 & 14 & 37 & 38 & 71.43\% \\

\textit{TSubj\_L} & Legal subject spec. templ. & Legal & 12 & 105 & 1 & 5 & 29 & 30 & 40 & 66.70\% \\

\textit{TRel\_L} & Regulatory relations spec. templ. & Legal & 12 & 106 & 0 & 12 & 12 & 30 & 52 & 77.36\% \\

\textit{TObj\_L} & Legal object spec. templ. & Legal & 17 & 150 & 14 & 23 & 38 & 47 & 28 & 50.00\% \\

\textit{TSanc\_L} & Regulatory sanctions spec. templ. & Legal & 12 & 104 & 3 & 5 & 16 & 29 & 51 & 76.92\% \\

\textit{TScop\_M} & Project scope spec. templ. & Manager. & 6 & 52 & 4 & 5 & 13 & 21 & 9 & 57.69\% \\

\textit{TStak\_M} & Stakeholder model spec. templ. & Manager. & 6 & 53 & 0 & 9 & 10 & 18 & 16 & 64.15\% \\ 

\textit{TUsag\_R} & Usage model spec. templ. & RE & 9 & 80 & 0 & 2 & 12 & 31 & 35 & 82.50\% \\

\textit{TData01\_R} & Data model spec. templ. & Requirem. & 8 & 71 & 2 & 2 & 14 & 23 & 30 & 74.65\% \\

\textit{TArch\_A} & Architecture overview spec. templ. & Architect. & 11 & 97 & 0 & 6 & 16 & 49 & 26 & 77.32\% \\

\textit{TData02\_A} & Data model spec. templ. & Architect. & 10 & 88 & 5 & 3 & 12 & 34 & 34 & 77.27\%\\

\hline\hline

Total &  &  &  & 1011 & 35 & 82 & 186 & 349 & 359 & 70.03\%\\

\hline
\end{tabular}
\caption{Summary of template evaluation}
\label{tab:templates-summary}
\end{table*}

From a total of 1011 evaluations of templates by 39 participants, according to one of the 9 criteria (see Table~\ref{tab:templates-summary}), 70.03\% of evaluations were positive (min 50.00\% for TObj\_L and max 82.50\% for TUsag\_R). Of all evaluations 35 evaluations (3.46\%) identified templates as ``not useful'' (-2) (min 0\% for \textit{TRel\_L}, max 9.33\% for \textit{TObj\_L}), 82 evaluations (8.11\%) were ``potentially not useful'' (-1) (min 2.50\% for \textit{TUsag\_R}, max 16.98\% for \textit{TStak\_M}), 186 evaluations (18.40\%) were ``no assessment'' (0) (min 11.32\% for \textit{TRel\_L}, max 27.62\% for \textit{TSubj\_L}), 349 evaluations (34.52\%) were ``potentially useful'' (+1) (min 27.88\% for \textit{TSanc\_L}, max 50.52\% for \textit{TArch\_A}) and 359 evaluations (35.51\%) were ``useful'' (min 17.31\% for \textit{TScop\_M}, max 49.06\% for \textit{TRel\_L}).
Of 99 template-by-criterion average evaluations (see Table~\ref{tab:templates-average}), only two were negative (ease of specification of \textit{TObj}, traceability of \textit{TScop}), three were 0 (for role knowledge explication of \textit{TStak}, capturing role knowledge and completeness of \textit{TScop}); the remaining 94 evaluations were above 0, and 41 were above +1. Positive evaluation was also supported by some of the written comments (see below):

\begin{colorb}
    \small
    \textit{``It is a good overview without security assessment. It can be used as a basis for discussion with other stakeholders.''} - P5.1A on \textit{TArch\_AP}.
\end{colorb}

\begin{table*}[ht]
\tiny
\renewcommand\theadfont{\tiny}
\setlength{\tabcolsep}{3.75pt}
\centering
\rowcolors{3}{gray!15}{white}
\begin{tabular}{ll|lll|lllllllll}

\tiny{Templ.} &  &  &  &  & \tiny{TC01} & \tiny{TC02} & \tiny{TC03} & \tiny{TC04} & \tiny{TC05} & \tiny{TC06} &  \tiny{TC07} & \tiny{TC08} & \tiny{TC09} \\

\tiny{ID} & Summary & FC & PC & NC & \tiny{RK} & \tiny{Expl.} & \tiny{Tran.} & \tiny{Comm.} & \tiny{Cm.} & \tiny{Trac.} &  \tiny{Cn.} & \tiny{Eas.} & \tiny{Comp.} \\
\hline

\textit{TAct\_L} & Total Avg. & 8 & 1 & 3 & \cellcolor{LimeGreen!25}+0.50 & \cellcolor{LimeGreen!25}+0.92 & \cellcolor{Green!45}+1.27 & \cellcolor{LimeGreen!25}+0.83 & \cellcolor{LimeGreen!25}+0.75 & \cellcolor{LimeGreen!25}+0.67 & \cellcolor{Green!45}+1.17 & \cellcolor{LimeGreen!25}+1.00 & \cellcolor{LimeGreen!25}+0.70 \\

\textit{TAct\_LP} & PbD Avg. & 4 & 1 & 3 & \cellcolor{LimeGreen!25}+0.38 & \cellcolor{LimeGreen!25}+0.88 & \cellcolor{Green!45}+1.38 & \cellcolor{LimeGreen!25}+0.88 & \cellcolor{LimeGreen!25}+0.75 & \cellcolor{LimeGreen!25}+0.75 & \cellcolor{Green!45}+1.25 & \cellcolor{LimeGreen!25}+0.63 & \cellcolor{LimeGreen!25}+0.43\\

\textit{TAct\_LS} & SbD Avg. & 4 & 0 & 0 & \cellcolor{LimeGreen!25}+0.75 & \cellcolor{LimeGreen!25}+1.00 & \cellcolor{LimeGreen!25}+1.00 & \cellcolor{LimeGreen!25}+0.75 & \cellcolor{LimeGreen!25}+0.75 & \cellcolor{LimeGreen!25}+0.50 & \cellcolor{LimeGreen!25}+1.00 & \cellcolor{Green!45}+1.75 & \cellcolor{Green!45}+1.33\\

\textit{TAct\_L} & PbD-SbD Diff. &  &  &  & |0.38| & |0.13| & |0.38| & |0.13| & 0.00 & |0.25| & |0.25| & \cellcolor{red!25}|1.13| & |0.90| \\
\hline

\textit{TSubj\_L} & Total Avg. & 9 & 4 & 0 & \cellcolor{LimeGreen!25}+1.00 & \cellcolor{Green!45}+1.08 & \cellcolor{Green!45}+1.33 & \cellcolor{Green!45}+1.09 & \cellcolor{LimeGreen!25}+0.50 & \cellcolor{LimeGreen!25}+1.00 & \cellcolor{LimeGreen!25}+1.00 & \cellcolor{LimeGreen!25}+0.75 & \cellcolor{Green!45}+1.10\\

\textit{TSubj\_LP} & PbD Avg. & 5 & 4 & 0 & \cellcolor{LimeGreen!25}+0.88 & \cellcolor{LimeGreen!25}+0.88 & \cellcolor{Green!45}+1.38 & \cellcolor{Green!45}+1.25 & \cellcolor{LimeGreen!25}+0.25 & \cellcolor{LimeGreen!25}+0.88 & \cellcolor{LimeGreen!25}+0.75 & \cellcolor{LimeGreen!25}+0.25 & \cellcolor{LimeGreen!25}+0.67\\

\textit{TSubj\_LS} & SbD Avg. & 4 & 0 & 0 & \cellcolor{Green!45}+1.25 & \cellcolor{Green!45}+1.50 & \cellcolor{Green!45}+1.25 & \cellcolor{LimeGreen!25}+0.67 & \cellcolor{LimeGreen!25}+1.00 & \cellcolor{Green!45}+1.25 & \cellcolor{Green!45}+1.50 & \cellcolor{Green!45}+1.75 & \cellcolor{Green!45}+1.75\\

\textit{TSubj\_L} & PbD-SbD Diff. & - & - & - & |0.38| & |0.63| & |0.13| & |0.58| & |0.75| & |0.38| & |0.75| & \cellcolor{red!25}|1.50| & \cellcolor{red!25}|1.08|\\
\hline

\textit{TRel\_L} & Total Avg. & 13 & 0 & 0 & \cellcolor{Green!45}+1.58 & \cellcolor{Green!45}+1.64 & \cellcolor{Green!45}+1.50 & \cellcolor{Green!45}+1.42 & \cellcolor{LimeGreen!25}+1.00 & \cellcolor{Green!45}+1.25 & \cellcolor{LimeGreen!25}+0.83 & \cellcolor{LimeGreen!25}+0.36 & \cellcolor{LimeGreen!25}+0.75 \\

\textit{TRel\_LP} & PbD Avg. & 9 & 0 & 0 & \cellcolor{Green!45}+1.50 & \cellcolor{Green!45}+1.43 & \cellcolor{Green!45}+1.38 & \cellcolor{Green!45}+1.13 & \cellcolor{LimeGreen!25}+0.88 & \cellcolor{Green!45}+1.13 & \cellcolor{LimeGreen!25}+0.75 & \cellcolor{LimeGreen!25}+0.25 & \cellcolor{LimeGreen!25}+0.63\\

\textit{TRel\_LS} & SbD Avg. & 4 & 0 & 0 & \cellcolor{Green!45}+1.75 & \cellcolor{Green!45}+2.00 & \cellcolor{Green!45}+1.75 & \cellcolor{Green!45}+2.00 & \cellcolor{Green!45}+1.25 & \cellcolor{Green!45}+1.50 & \cellcolor{LimeGreen!25}+1.00 & \cellcolor{LimeGreen!25}+0.67 & \cellcolor{LimeGreen!25}+1.00\\

\textit{TRel\_L} & PbD-SbD Diff. & - & - & - & |0.25| & |0.57| & |0.38| & |0.88| & |0.38| & |0.38| & |0.25| & |0.42| & |0.38|\\
\hline

\textit{TObj\_L} & Total Avg. & 7 & 7 & 3 & \cellcolor{LimeGreen!25}+0.76 & \cellcolor{LimeGreen!25}+0.59 & \cellcolor{LimeGreen!25}+0.41 & \cellcolor{LimeGreen!25}+0.47 & \cellcolor{LimeGreen!25}+0.33 & \cellcolor{LimeGreen!25}+0.29 & \cellcolor{LimeGreen!25}+0.24 & \cellcolor{orange!55}-0.35 & \cellcolor{LimeGreen!25}+0.38\\

\textit{TObj\_LP} & PbD Avg. & 4 & 6 & 2 & \cellcolor{LimeGreen!25}+1.00 & \cellcolor{LimeGreen!25}+0.83 & \cellcolor{LimeGreen!25}+0.58 & \cellcolor{LimeGreen!25}+0.33 & \cellcolor{LimeGreen!25}+0.64 & \cellcolor{LimeGreen!25}+0.50 & \cellcolor{LimeGreen!25}+0.33 & \cellcolor{orange!55}
-0.08 & \cellcolor{LimeGreen!25}+0.18\\

\textit{TObj\_LS} & SbD Avg. & 3 & 1 & 1 & \cellcolor{LimeGreen!25}+0.20 & \cellcolor{blue!20}0.00 & \cellcolor{blue!20}0.00 & \cellcolor{LimeGreen!25}+0.80 & \cellcolor{orange!55}-0.50 & \cellcolor{orange!55}-0.20 & \cellcolor{blue!20}0.00 & \cellcolor{orange!55}-1.00 & \cellcolor{LimeGreen!25}+0.80\\

\textit{TObj\_L} & PbD-SbD Diff. & - & - & - & |0.80| & |0.83| & |0.58| & |0.47| & \cellcolor{red!25}|1.14| & |0.70| & |0.33| & |0.92| & |0.62|\\
\hline

\textit{TSanc\_L} & Total Avg. & 12 & 0 & 1 & \cellcolor{Green!45}+1.42 & \cellcolor{Green!45}+1.20 & \cellcolor{Green!45}+1.50 & \cellcolor{Green!45}+1.42 & \cellcolor{LimeGreen!25}+0.64 & \cellcolor{LimeGreen!25}+0.92 & \cellcolor{LimeGreen!25}+1.00 & \cellcolor{Green!45}+1.42 & \cellcolor{LimeGreen!25}+0.82\\

\textit{TSanc\_LP} & PbD Avg. & 8 & 0 & 1 & \cellcolor{Green!45}+1.50 & \cellcolor{Green!45}+1.67 & \cellcolor{Green!45}+1.50 & \cellcolor{Green!45}+1.25 & \cellcolor{LimeGreen!25}+0.63 & \cellcolor{LimeGreen!25}+1.00 & \cellcolor{LimeGreen!25}+1.00 & \cellcolor{Green!45}+1.25 & \cellcolor{LimeGreen!25}+0.71\\

\textit{TSanc\_LS} & SbD Avg. & 4 & 0 & 0 & \cellcolor{Green!45}+1.25 & \cellcolor{LimeGreen!25}+0.50 & \cellcolor{Green!45}+1.50 & \cellcolor{Green!45}+1.75 & \cellcolor{LimeGreen!25}+0.67 & \cellcolor{LimeGreen!25}+0.75 & \cellcolor{LimeGreen!25}+1.00 & \cellcolor{Green!45}+1.75 & \cellcolor{LimeGreen!25}+1.00\\

\textit{TSanc\_L} & PbD-SbD Diff. & - & - & - & |0.25| & \cellcolor{red!25}|1.17| & 0.00 & |0.50| & |0.04| & |0.25| & 0.00 & |0.50| & |0.29|\\
\hline

\textit{TScop\_M} & Total Avg. & 6 & 0 & 0 & \cellcolor{blue!20}0.00 & \cellcolor{LimeGreen!25}+0.50 & \cellcolor{Green!45}+1.17 & \cellcolor{Green!45}+1.17 & \cellcolor{blue!20}0.00 & \cellcolor{orange!55}-0.20 & \cellcolor{LimeGreen!25}+0.50 & \cellcolor{LimeGreen!25}+0.50 & \cellcolor{LimeGreen!25}+0.80 \\

\textit{TScop\_MP} & PbD Avg. & 5 & 0 & 0 & \cellcolor{blue!20}0.00 & \cellcolor{LimeGreen!25}+0.80 & \cellcolor{Green!45}+1.40 & \cellcolor{Green!45}+1.60 & \cellcolor{orange!55}-0.20 & \cellcolor{orange!55}-0.25 & \cellcolor{LimeGreen!25}+0.60 & \cellcolor{LimeGreen!25}+0.40 & \cellcolor{LimeGreen!25}+0.75 \\

\textit{TScop\_MS} & SbD Avg. & 1 & 0 & 0 & \cellcolor{blue!20}0.00 & \cellcolor{orange!55}-1.00 & \cellcolor{blue!20}0.00 & \cellcolor{orange!55}-1.00 & \cellcolor{LimeGreen!25}+1.00 & \cellcolor{blue!20}0.00 & \cellcolor{blue!20}0.00 & \cellcolor{LimeGreen!25}+1.00 & \cellcolor{LimeGreen!25}+1.00 \\

\textit{TScop\_M} & PbD-SbD Diff. & - & - & - & 0.00 & \cellcolor{red!25}|1.80| & \cellcolor{red!25}|1.40| & \cellcolor{red!25}|2.60| & \cellcolor{red!25}|1.20| & |0.25| & |0.60| & |0.60| & |0.25|\\
\hline

\textit{TStak\_M} & Total Avg. & 5 & 1 & 0 & \cellcolor{LimeGreen!25}+0.50 & \cellcolor{blue!20}0.00 & \cellcolor{Green!45}+1.50 & \cellcolor{Green!45}+1.33 & \cellcolor{LimeGreen!25}+0.67 & \cellcolor{LimeGreen!25}+0.83 & \cellcolor{LimeGreen!25}+1.00 & \cellcolor{LimeGreen!25}+0.33 & \cellcolor{LimeGreen!25}+0.80 \\

\textit{TStak\_MP} & PbD Avg. & 5 & 0 & 0 & \cellcolor{LimeGreen!25}+0.20 & \cellcolor{blue!20}0.00 & \cellcolor{Green!45}+1.60 & \cellcolor{Green!45}+1.40 & \cellcolor{LimeGreen!25}+0.60 & \cellcolor{LimeGreen!25}+0.80 & \cellcolor{LimeGreen!25}+1.00 & \cellcolor{LimeGreen!25}+0.40 & \cellcolor{LimeGreen!25}+1.00 \\

\textit{TStak\_MS} & SbD Avg. & 0 & 1 & 0 & \cellcolor{Green!45}+2.00 & \cellcolor{blue!20}0.00 & \cellcolor{LimeGreen!25}+1.00 & \cellcolor{LimeGreen!25}+1.00 & \cellcolor{LimeGreen!25}+1.00 & \cellcolor{LimeGreen!25}+1.00 & \cellcolor{LimeGreen!25}+1.00 & \cellcolor{blue!20}0.00 & \cellcolor{blue!20}0.00 \\

\textit{TStak\_M} & PbD-SbD Diff. & - & - & - & \cellcolor{red!25}|1.80| & |0.00| & |0.60| & |0.40| & |0.40| & |0.20| & |0.00| & |0.40| & \cellcolor{red!25}|1.00|\\

\hline

\textit{TUsag\_R} & Total Avg. & 8 & 0 & 0 & \cellcolor{Green!45}+1.67 & \cellcolor{Green!45}+1.11 & \cellcolor{Green!45}+1.89 & \cellcolor{Green!45}+1.33 & \cellcolor{LimeGreen!25}+0.89 & \cellcolor{LimeGreen!25}+0.67 & \cellcolor{Green!45}+1.11 & \cellcolor{Green!45}+1.33 & \cellcolor{Green!45}+1.13 \\

\textit{TUsag\_RP} & PbD Avg. & 5 & 0 & 0 & \cellcolor{Green!45}+1.50 & \cellcolor{LimeGreen!25}+1.00 & \cellcolor{Green!45}+1.83 & \cellcolor{Green!45}+1.50 & \cellcolor{LimeGreen!25}+1.00 & \cellcolor{LimeGreen!25}+0.83 & \cellcolor{Green!45}+1.17 & \cellcolor{Green!45}+1.17 & \cellcolor{LimeGreen!25}+1.00 \\

\textit{TUsag\_RS} & SbD Avg. & 3 & 0 & 0 & \cellcolor{Green!45}+2.00 & \cellcolor{Green!45}+1.33 & \cellcolor{Green!45}+2.00 & \cellcolor{LimeGreen!25}+1.00 & \cellcolor{LimeGreen!25}+0.67 & \cellcolor{LimeGreen!25}+0.33 & \cellcolor{LimeGreen!25}+1.00 & \cellcolor{Green!45}+1.67 & \cellcolor{Green!45}+1.33 \\

\textit{TUsag\_R} & PbD-SbD Diff. & - & - & - & |0.50| & |0.33| & |0.17| & |0.50| & |0.33| & |0.50| & |0.17| & |0.50| & |0.33|\\
\hline

\textit{TData01\_R} & Total Avg. & 8 & 0 & 0 & \cellcolor{Green!45}+1.63 & \cellcolor{Green!45}+1.38 & \cellcolor{Green!45}+1.71 & \cellcolor{Green!45}+1.13 & \cellcolor{LimeGreen!25}+0.63 & \cellcolor{LimeGreen!25}+0.25 & \cellcolor{LimeGreen!25}+1.00 & \cellcolor{LimeGreen!25}+1.00 & \cellcolor{Green!45}+1.13\\

\textit{TData01\_RP} & PbD Avg. & 5 & 0 & 0 & \cellcolor{Green!45}+1.60 & \cellcolor{Green!45}+1.40 & \cellcolor{Green!45}+1.60 & \cellcolor{Green!45}+1.20 & \cellcolor{Green!45}+1.20 & \cellcolor{LimeGreen!25}+1.00 & \cellcolor{Green!45}+1.60 & \cellcolor{LimeGreen!25}+1.00 & \cellcolor{LimeGreen!25}+1.00\\

\textit{TData01\_RS} & SbD Avg. & 3 & 0 & 0 & \cellcolor{Green!45}+1.67 & \cellcolor{Green!45}+1.33 & \cellcolor{Green!45}+2.00 & \cellcolor{LimeGreen!25}+1.00 & \cellcolor{orange!55}-0.33 & \cellcolor{orange!55}-1.00 & \cellcolor{blue!20}0.00 & \cellcolor{LimeGreen!25}+1.00 & \cellcolor{Green!45}+1.33\\

\textit{TData01\_R} & PbD-SbD Diff. & - & - & - & |0.07| & |0.07| & |0.40| & |0.20| & \cellcolor{red!25}|1.53| & \cellcolor{red!25}|2.00| & \cellcolor{red!25}|1.60| & |0.00| & |0.33| \\
\hline

\textit{TArch\_A} & Total Avg. & 10 & 1 & 0 & \cellcolor{Green!45}+1.55 & \cellcolor{Green!45}+1.20 & \cellcolor{Green!45}+1.18 & \cellcolor{LimeGreen!25}+0.82 & \cellcolor{LimeGreen!25}+0.36 & \cellcolor{LimeGreen!25}+0.91 & \cellcolor{Green!45}+1.20 & \cellcolor{LimeGreen!25}+0.91 & \cellcolor{LimeGreen!25}+0.73\\

\textit{TArch\_AP} & PbD Avg. & 7 & 1 & 0 & \cellcolor{Green!45}+1.75 & \cellcolor{Green!45}+1.14 & \cellcolor{Green!45}+1.13 & \cellcolor{LimeGreen!25}+0.63 & \cellcolor{LimeGreen!25}+0.38 & \cellcolor{LimeGreen!25}+0.88 & \cellcolor{LimeGreen!25}+1.00 & \cellcolor{LimeGreen!25}+0.75 & \cellcolor{LimeGreen!25}+0.50\\

\textit{TArch\_AS} & SbD Avg. & 3 & 0 & 0 & \cellcolor{LimeGreen!25}+1.00 & \cellcolor{Green!45}+1.33 & \cellcolor{Green!45}+1.33 & \cellcolor{Green!45}+1.33 & \cellcolor{LimeGreen!25}+0.33 & \cellcolor{LimeGreen!25}+1.00 & \cellcolor{Green!45}+1.67 & \cellcolor{Green!45}+1.33 & \cellcolor{Green!45}+1.33\\

\textit{TArch\_A} & PbD-SbD Diff. & - & - & - & |0.75| & |0.19| & |0.21| & |0.71| & |0.04| & |0.13| & |0.67| & |0.58| & |0.83| \\
\hline

\textit{TData02\_A} & Total Avg. & 11 & 0 & 0 & \cellcolor{Green!45}+1.80 & \cellcolor{Green!45}+1.10 & \cellcolor{LimeGreen!25}+0.70 & \cellcolor{LimeGreen!25}+0.70 & \cellcolor{Green!45}+1.13 & \cellcolor{LimeGreen!25}+0.70 & \cellcolor{Green!45}+1.30 & \cellcolor{Green!45}+1.20 & \cellcolor{LimeGreen!25}+0.50\\

\textit{TData02\_AP} & PbD Avg. & 8 & 0 & 0 & \cellcolor{Green!45}+1.86 & \cellcolor{LimeGreen!25}+1.00 & \cellcolor{LimeGreen!25}+0.43 & \cellcolor{LimeGreen!25}+0.57 & \cellcolor{LimeGreen!25}+1.00 & \cellcolor{LimeGreen!25}+0.86 & \cellcolor{LimeGreen!25}+1.00 & \cellcolor{LimeGreen!25}+1.00 & \cellcolor{LimeGreen!25}+0.14\\

\textit{TData02\_AS} & SbD Avg. & 3 & 0 & 0 & \cellcolor{Green!45}+1.67 & \cellcolor{Green!45}+1.33 & \cellcolor{Green!45}+1.33 & \cellcolor{LimeGreen!25}+1.00 & \cellcolor{Green!45}+1.33 & \cellcolor{LimeGreen!25}+0.33 & \cellcolor{Green!45}+2.00 & \cellcolor{Green!45}+1.67 & \cellcolor{Green!45}+1.33\\

\textit{TData02\_A} & PbD-SbD Diff. & - & - & - &  |0.19| & |0.33| & |0.90| & |0.43| & |0.33| & |0.52| & \cellcolor{red!25}|1.00| & |0.67| & \cellcolor{red!25}|1.19| \\
\hline
\end{tabular}
\caption{\small Number of fully completed templates (FC), partially completed (PC), non-completed (NC) templates and average evaluation of each of 11 templates as per (1) PbD evaluation, (2) SbD evaluation, (3) difference between average evaluation in PbD and SbD evaluation (absolute value), and (4) both PbD and SbD evaluation. For color coding see Table~\ref{tab:evaluation}.}
\label{tab:templates-average}
\end{table*}

Participants' written comments concerned unsuitable specification syntax (for \textit{TArch\_AP}) and format (\textit{TData02\_AP}), potentially high complexity (\textit{TSubj\_LP}, \textit{TRel\_LP}), or limited readability of specifications (\textit{TStak\_MP}). Others suggested that specifications need more flexibility (\textit{TSubj\_LP}), that the resulting specifications can be too high-level (\textit{TStak\_MP}, \textit{TScop\_MP}), or that they require additional detail (\textit{TStak\_MP}, \textit{TArch\_AS}). Most participants completed the templates using only the embedded instruction; however, some participants in the legal role (e.g., P6.2L, P20.6L) initially struggled with the idea of completing templates using only regulatory norms and without software project-specific information.

\begin{colorb}
    \small
    \textit{``This is a little too complex [..]. As a model, it can be useful, but only after the different subjects have been identified, as it can ``focus'' on one specific role and illustrate the relationships around it.''} - P24.7L on TRel\_LP.
\end{colorb}

The legal object specification template (\textit{TObj\_L}) was particularly challenging for participants to complete. Of 17 participants, 7 completed it fully (highlighting and modeling concepts), 7 partially (only one of the two), and three not at all. Written comments suggest that some participants understood the template's purpose better after completing it.

\begin{colorb}
    \small
    \textit{``Makes sense after the fact, very hard to wrap the mind around to create from text - method feels ``off'' ''.} - P16.5L regarding TObj\_LP.
\end{colorb}

\begin{colorb}
    \small
    \textit{``Now I understand. It’s just about reading and identifying the various meanings. Not sure this is really useful, but maybe. I am afraid that making it into a scheme would lose meaning and nuance.''} -  P24.7L TObj\_LP.
\end{colorb}

Of the four participants additionally assigned \textit{TObj\_L}, P5.1A highlighted concepts and added elements, P23.6A only highlighted concepts, P28.7A annotated concepts to a limited degree due to colorblindness, and S5.2R completed neither task. S5.2R and P23.6A evaluated the template mainly negatively; P23.6A's written comment stated that the template lacks anything specific to the architecture role. P5.1A and P28.7A evaluated the template mainly positively, with P5.1A pointing to the need for additional training (see quotation below).

\begin{colorb}
    \small
    \textit{``You need practice to find and differentiate between control, criterion, and target of regulation.''} - P5.1A about TObj\_LP.
\end{colorb}

\subparagraph{AM4RRE (EE09-EE10)} When discussing AM4RRE applicability from the legal perspective, P16.5L mentioned that it reminded them of their daily work and P24.7L directly confirmed its compatibility with existing practice.
\begin{colorb}
    \small
    Session lead: \textit{- Can you imagine that this approach could be used in your company?}\\
    P24.7L: \textit{- We did something like this, but in a less structured way.}
\end{colorb}

In the quasi-experiment, under guided coordination, GP06 omitted three, GP07 five, and GS02 one of five specification elements in Task 1, but specified all 10 requirement elements in Task 2, corresponding to 80\%, 67\%, and 93\% completion. The remaining groups specified all required elements. In free-form coordination, GP01 omitted two elements, GP07 one, and GS01 completely omitted one requirement in Task 2. Based on observations, participants discussed but did not finish specifications for unclear reasons. Several groups specified more than the tasks required. In Task 1, nine additional elements were specified in five free-form periods and eight in four guided periods. In Task 2, guided periods produced 24 requirements in total (GP02 and GP03 each specified one extra requirement and GP04 two), and free-form periods produced 20 (GP03 specified one extra, and GS01 omitted one) (see the open dataset).

\begin{colora}
    \textit{Evaluation goal 1} was partially met as 10 of 11 templates, except for TObj\_L, were completed by more than 50\% of participants and got more than 50\% of positive evaluations; for 8 templates, except for TObj\_L, TScop\_M, and TStak\_M, all average evaluations across nine criteria were above 0.

    \textit{Evaluation goal 2} was achieved with all ten groups specifying more than 50\% of the required elements under guided coordination.
    
    \textit{RQ2:} The results indicate that participants applied most AM4RRE components: the project context model was useful for capturing heterogeneous opinions about challenges; the role model received only minor change suggestions; and 70 of 74 goals in the goal model received an average rating above zero, and, on average, 82.44\% of all given rates were positive.
\end{colora}

\begin{landscape}
    \vspace{-1cm}
\begin{table*}[ht]
\tiny
\renewcommand\theadfont{\tiny}
\setlength{\tabcolsep}{1.5pt}
\centering
\begin{tabular}{c|a|a|cccc|a|cccccccc|a|cccccc}

\tiny{Part.} & \tiny{Part.} & \tiny{\textbf{Qual.}} & \tiny{CC01} & \tiny{CC02} & \tiny{CC03} & \tiny{CC04} & \tiny{\textbf{RE}} & \tiny{CC05} & \tiny{CC06} & \tiny{CC07} & \tiny{CC08} & \tiny{CC09} & \tiny{CC10} & \tiny{CC11} & \tiny{CC12} & \tiny{\textbf{Goal.}} & \tiny{CC13} & \tiny{CC14} & \tiny{CC15} & \tiny{CC16} & \tiny{CC17} & \tiny{CC18}
\\

\tiny{ID} & \tiny{Av} & \tiny{\textbf{Av}} & \tiny{Cm.} & \tiny{Cn.1} & \tiny{Cn.2} & \tiny{Cr.} & \tiny{\textbf{Av}} & \tiny{Coord.} & \tiny{LK} & \tiny{RK} & \tiny{Comp.} & \tiny{Trac.} & \tiny{Cn3.} & \tiny{Tran.} & \tiny{Inter.} & \tiny{\textbf{Av}} & \tiny{SA} & \tiny{SE} & \tiny{Dec.} & \tiny{Doc.} & \tiny{Gov.} & \tiny{Bus.}
\\ \hline
P1.1L & +0.94 & \textbf{+0.5} & \cellcolor{LimeGreen!25}+1 & \cellcolor{blue!20}0 & \cellcolor{blue!20}0 & \cellcolor{LimeGreen!25}+1 & \textbf{+1.25} & \cellcolor{Green!45}+2 & \cellcolor{LimeGreen!25}+1 & \cellcolor{blue!20}0 & \cellcolor{LimeGreen!25}+1 & \cellcolor{Green!45}+2 & \cellcolor{LimeGreen!25}+1 & \cellcolor{LimeGreen!25}+1 & \cellcolor{Green!45}+2 & \textbf{+0.80} & \cellcolor{LimeGreen!25}+1 & \cellcolor{orange!55}-1 & \cellcolor{Green!45}+2 & \cellcolor{LimeGreen!25}+1 & NA & \cellcolor{LimeGreen!25}+1 \\

P2.1M & +1.11 & \textbf{+1.5} & \cellcolor{Green!45}+2 & \cellcolor{Green!45}+2 & \cellcolor{LimeGreen!25}+1 & \cellcolor{LimeGreen!25}+1 & \textbf{+1} & \cellcolor{LimeGreen!25}+1 & \cellcolor{Green!45}+2 & \cellcolor{LimeGreen!25}+1 & \cellcolor{blue!20}0 & \cellcolor{LimeGreen!25}+1 & \cellcolor{LimeGreen!25}+1 & \cellcolor{LimeGreen!25}+1 & \cellcolor{LimeGreen!25}+1 & \textbf{+1} & \cellcolor{LimeGreen!25}+1 & \cellcolor{LimeGreen!25}+1 & \cellcolor{LimeGreen!25}+1 & \cellcolor{LimeGreen!25}+1 & \cellcolor{LimeGreen!25}+1 & \cellcolor{LimeGreen!25}+1 \\

P3.1R & NA & \textbf{NA} & NA & NA & NA & NA & NA & NA & NA & NA & NA & NA & NA & NA & NA & \textbf{NA} & NA & NA & NA & NA & NA & NA\\

P4.1A & +1.06 & \textbf{+1.25} & \cellcolor{LimeGreen!25}+1 & \cellcolor{Green!45}+2 & \cellcolor{LimeGreen!25}+1 & \cellcolor{LimeGreen!25}+1 & \textbf{+1.13} & \cellcolor{LimeGreen!25}+1 & \cellcolor{LimeGreen!25}+1 & \cellcolor{LimeGreen!25}+1 & \cellcolor{LimeGreen!25}+1 & \cellcolor{LimeGreen!25}+1 & \cellcolor{LimeGreen!25}+1 & \cellcolor{Green!45}+2 & \cellcolor{LimeGreen!25}+1 & \textbf{+0.83} & \cellcolor{LimeGreen!25}+1 & \cellcolor{blue!20}0 & \cellcolor{LimeGreen!25}+1 & \cellcolor{LimeGreen!25}+1 & \cellcolor{LimeGreen!25}+1 & \cellcolor{LimeGreen!25}+1 \\

P5.1A & +1.28 & \textbf{+1} & \cellcolor{LimeGreen!25}+1 & \cellcolor{LimeGreen!25}+1 & \cellcolor{LimeGreen!25}+1 & \cellcolor{LimeGreen!25}+1 & \textbf{+1.13} & \cellcolor{Green!45}+2 & \cellcolor{LimeGreen!25}+1 & \cellcolor{LimeGreen!25}+1 & \cellcolor{LimeGreen!25}+1 & \cellcolor{LimeGreen!25}+1 & \cellcolor{LimeGreen!25}+1 & \cellcolor{LimeGreen!25}+1 & \cellcolor{LimeGreen!25}+1 & \textbf{+1.67} & \cellcolor{Green!45}+2 & \cellcolor{Green!45}+2 & \cellcolor{Green!45}+2 & \cellcolor{LimeGreen!25}+1 & \cellcolor{LimeGreen!25}+1 & \cellcolor{Green!45}+2 \\
\hline

\rowcolor{gray!15}\textbf{GP01 Av} & +1.10 & \textbf{+1.06} & \textbf{+1.25} & \textbf{+1.25} & \textbf{+0.75} & \textbf{+1.00} & \textbf{+1.13} & \textbf{+1.50} & \textbf{+1.25} & \textbf{+0.75} & \textbf{+0.75} & \textbf{+1.25} & \textbf{+1.00} & \textbf{+1.25} & \textbf{+1.25} & \textbf{+1.08} & \textbf{+1.25} & \textbf{+0.50} & \textbf{+1.50} & \textbf{+1.00} & \textbf{+1.00} & \textbf{+1.25}\\
\hline
P6.2L & +1.83 & \textbf{+2} & \cellcolor{Green!45}+2 & \cellcolor{Green!45}+2 & \cellcolor{Green!45}+2 & \cellcolor{Green!45}+2 & \textbf{+2} & \cellcolor{Green!45}+2 & \cellcolor{Green!45}+2 & \cellcolor{Green!45}+2 & \cellcolor{Green!45}+2 & \cellcolor{Green!45}+2 & \cellcolor{Green!45}+2 & \cellcolor{Green!45}+2 & \cellcolor{Green!45}+2 & \textbf{+1.5} & \cellcolor{LimeGreen!25}+1 & \cellcolor{LimeGreen!25}+1 & \cellcolor{Green!45}+2 & \cellcolor{Green!45}+2 & \cellcolor{Green!45}+2 & \cellcolor{LimeGreen!25}+1 \\

P7.2A & +1.72 & \textbf{+2} & \cellcolor{Green!45}+2 & \cellcolor{Green!45}+2 & \cellcolor{Green!45}+2 & \cellcolor{Green!45}+2 & \textbf{+1.5} & \cellcolor{Green!45}+2 & \cellcolor{Green!45}+2 & \cellcolor{Green!45}+2 & \cellcolor{Green!45}+2 & \cellcolor{LimeGreen!25}+1 & \cellcolor{LimeGreen!25}+1 & \cellcolor{blue!20}0 & \cellcolor{Green!45}+2 & \textbf{+1.83} & \cellcolor{Green!45}+2 & \cellcolor{LimeGreen!25}+1 & \cellcolor{Green!45}+2 & \cellcolor{Green!45}+2 & \cellcolor{Green!45}+2 & \cellcolor{Green!45}+2 \\
\hline
\rowcolor{gray!15}\textbf{GP02 Av} & +1.78 & \textbf{+2.00} & \textbf{+2.00} & \textbf{+2.00} & \textbf{+2.00} & \textbf{+2.00} & \textbf{+1.75} & \textbf{+2.00} & \textbf{+2.00} & \textbf{+2.00} & \textbf{+2.00} & \textbf{+1.50} & \textbf{+1.50} & \textbf{+1.00} & \textbf{+2.00} & \textbf{+1.67} & \textbf{+1.50} & \textbf{+1.00} & \textbf{+2.00} & \textbf{+2.00} & \textbf{+2.00} & \textbf{+1.50} \\
\hline
P8.3L & +1.67 & \textbf{+1.75} & \cellcolor{Green!45}+2 & \cellcolor{Green!45}+2 & \cellcolor{LimeGreen!25}+1 & \cellcolor{Green!45}+2 & \textbf{+1.38} & \cellcolor{LimeGreen!25}+1 & \cellcolor{Green!45}+2 & \cellcolor{LimeGreen!25}+1 & \cellcolor{Green!45}+2 & \cellcolor{blue!20}0 & \cellcolor{LimeGreen!25}+1 & \cellcolor{Green!45}+2 & \cellcolor{Green!45}+2 & \textbf{+2} & \cellcolor{Green!45}+2 & \cellcolor{Green!45}+2 & \cellcolor{Green!45}+2 & \cellcolor{Green!45}+2 & \cellcolor{Green!45}+2 & \cellcolor{Green!45}+2 \\
P9.3R & +1.72 & \textbf{+1.75} & \cellcolor{Green!45}+2 & \cellcolor{Green!45}+2 & \cellcolor{LimeGreen!25}+1 & \cellcolor{Green!45}+2 & \textbf{+1.75} & \cellcolor{Green!45}+2 & \cellcolor{Green!45}+2 & \cellcolor{Green!45}+2 & \cellcolor{Green!45}+2 & \cellcolor{LimeGreen!25}+1 & \cellcolor{LimeGreen!25}+1 & \cellcolor{Green!45}+2 & \cellcolor{Green!45}+2 & \textbf{+1.67} & \cellcolor{LimeGreen!25}+1 & \cellcolor{LimeGreen!25}+1 & \cellcolor{Green!45}+2 & \cellcolor{Green!45}+2 & \cellcolor{Green!45}+2 & \cellcolor{Green!45}+2 \\
P10.3A & +1.61 & \textbf{+2} & \cellcolor{Green!45}+2 & \cellcolor{Green!45}+2 & \cellcolor{Green!45}+2 & \cellcolor{Green!45}+2 & \textbf{+1.5} & \cellcolor{Green!45}+2 & \cellcolor{Green!45}+2 & \cellcolor{Green!45}+2 & \cellcolor{LimeGreen!25}+1 & \cellcolor{LimeGreen!25}+1 & \cellcolor{LimeGreen!25}+1 & \cellcolor{Green!45}+2 & \cellcolor{LimeGreen!25}+1 & \textbf{+1.5} & \cellcolor{LimeGreen!25}+1 & \cellcolor{LimeGreen!25}+1 & \cellcolor{Green!45}+2 & \cellcolor{LimeGreen!25}+1 & \cellcolor{Green!45}+2 & \cellcolor{Green!45}+2 \\
\hline
\rowcolor{gray!15}\textbf{GP03 Av} & +1.67 & \textbf{+1.83} & \textbf{+2.00} & \textbf{+2.00} & \textbf{+1.33} & \textbf{+2.00} & \textbf{+1.54} & \textbf{+1.67} & \textbf{+2.00} & \textbf{+1.67} & \textbf{+1.67} & \textbf{+0.67} & \textbf{+1.00} & \textbf{+2.00} & \textbf{+1.67} & \textbf{+1.72} & \textbf{+1.33} & \textbf{+1.33} & \textbf{+2.00} & \textbf{+1.67} & \textbf{+2.00} & \textbf{+2.00}\\
\hline
P11.4L & +0.94 & \textbf{+1} & \cellcolor{LimeGreen!25}+1 & \cellcolor{LimeGreen!25}+1 & \cellcolor{LimeGreen!25}+1 & \cellcolor{LimeGreen!25}+1 & \textbf{+1} & \cellcolor{LimeGreen!25}+1 & \cellcolor{blue!20}0 & \cellcolor{blue!20}0 & \cellcolor{LimeGreen!25}+1 & \cellcolor{Green!45}+2 & \cellcolor{Green!45}+2 & \cellcolor{LimeGreen!25}+1 & \cellcolor{LimeGreen!25}+1 & \textbf{+0.83} & \cellcolor{LimeGreen!25}+1 & \cellcolor{LimeGreen!25}+1 & \cellcolor{blue!20}0 & \cellcolor{LimeGreen!25}+1 & \cellcolor{LimeGreen!25}+1 & \cellcolor{LimeGreen!25}+1 \\
P12.4M & +0.72 & \textbf{+0.5} & \cellcolor{LimeGreen!25}+1 & \cellcolor{LimeGreen!25}+1 & \cellcolor{blue!20}0 & \cellcolor{blue!20}0 & \textbf{+0.88} & \cellcolor{Green!45}+2 & \cellcolor{LimeGreen!25}+1 & \cellcolor{LimeGreen!25}+1 & \cellcolor{LimeGreen!25}+1 & \cellcolor{LimeGreen!25}+1 & \cellcolor{blue!20}0 & \cellcolor{LimeGreen!25}+1 & \cellcolor{blue!20}0 & \textbf{+0.67} & \cellcolor{LimeGreen!25}+1 & \cellcolor{blue!20}0 & \cellcolor{LimeGreen!25}+1 & \cellcolor{LimeGreen!25}+1 & \cellcolor{LimeGreen!25}+1 & \cellcolor{blue!20}0 \\
P13.4R & +1.67 & \textbf{+2} & \cellcolor{Green!45}+2 & \cellcolor{Green!45}+2 & \cellcolor{Green!45}+2 & \cellcolor{Green!45}+2 & \textbf{+1.5} & \cellcolor{Green!45}+2 & \cellcolor{Green!45}+2 & \cellcolor{LimeGreen!25}+1 & \cellcolor{Green!45}+2 & \cellcolor{LimeGreen!25}+1 & \cellcolor{LimeGreen!25}+1 & \cellcolor{LimeGreen!25}+1 & \cellcolor{Green!45}+2 & \textbf{+1.67} & \cellcolor{blue!20}0 & \cellcolor{Green!45}+2 & \cellcolor{Green!45}+2 & \cellcolor{Green!45}+2 & \cellcolor{Green!45}+2 & \cellcolor{Green!45}+2 \\

P14.4A & +1.5 & \textbf{+2} & \cellcolor{Green!45}+2 & \cellcolor{Green!45}+2 & \cellcolor{Green!45}+2 & \cellcolor{Green!45}+2 & \textbf{+1.25} & \cellcolor{Green!45}+2 & \cellcolor{blue!20}0 & \cellcolor{Green!45}+2 & \cellcolor{blue!20}0 & \cellcolor{LimeGreen!25}+1 & \cellcolor{LimeGreen!25}+1 & \cellcolor{Green!45}+2 & \cellcolor{Green!45}+2 & \textbf{+1.5} & \cellcolor{Green!45}+2 & \cellcolor{blue!20}0 & \cellcolor{LimeGreen!25}+1 & \cellcolor{Green!45}+2 & \cellcolor{Green!45}+2 & \cellcolor{Green!45}+2 \\
\hline
\rowcolor{gray!15}\textbf{GP04 Av} & +1.21 & \textbf{+1.38} & \textbf{+1.50} & \textbf{+1.50} & \textbf{+1.25} & \textbf{+1.25} & \textbf{+1.16} & \textbf{+1.75} & \textbf{+0.75} & \textbf{+1.00} & \textbf{+1.00} & \textbf{+1.25} & \textbf{+1.00} & \textbf{+1.25} & \textbf{+1.25} & \textbf{+1.17} & \textbf{+1.00} & \textbf{+0.75} & \textbf{+1.00} & \textbf{+1.50} & \textbf{+1.50} & \textbf{+1.25} \\
\hline
P15.L & +1.65 & \textbf{+2} & \cellcolor{Green!45}+2 & \cellcolor{Green!45}+2 & \cellcolor{Green!45}+2 & \cellcolor{Green!45}+2 & \textbf{+1.14} & \cellcolor{Green!45}+2 & NA & \cellcolor{LimeGreen!25}+1 & \cellcolor{blue!20}0 & \cellcolor{blue!20}0 & \cellcolor{LimeGreen!25}+1 & \cellcolor{Green!45}+2 & \cellcolor{Green!45}+2 & \textbf{+2} & \cellcolor{Green!45}+2 & \cellcolor{Green!45}+2 & \cellcolor{Green!45}+2 & \cellcolor{Green!45}+2 & \cellcolor{Green!45}+2 & \cellcolor{Green!45}+2 \\
\hline
P16.5L & +1.06 & \textbf{+1.25} & \cellcolor{LimeGreen!25}+1 & \cellcolor{Green!45}+2 & \cellcolor{LimeGreen!25}+1 & \cellcolor{LimeGreen!25}+1 & \textbf{+0.75} & \cellcolor{LimeGreen!25}+1 & \cellcolor{LimeGreen!25}+1 & \cellcolor{LimeGreen!25}+1 & \cellcolor{blue!20}0 & \cellcolor{Green!45}+2 & \cellcolor{LimeGreen!25}+1 & \cellcolor{LimeGreen!25}+1 & \cellcolor{orange!55}-1 & \textbf{+1.33} & \cellcolor{LimeGreen!25}+1 & \cellcolor{LimeGreen!25}+1 & \cellcolor{Green!45}+2 & \cellcolor{Green!45}+2 & \cellcolor{blue!20}0 & \cellcolor{Green!45}+2 \\

P17.5M & +1.33 & \textbf{+0.75} & \cellcolor{blue!20}0 & \cellcolor{LimeGreen!25}+1 & \cellcolor{LimeGreen!25}+1 & \cellcolor{LimeGreen!25}+1 & \textbf{+1.63} & \cellcolor{Green!45}+2 & \cellcolor{Green!45}+2 & \cellcolor{Green!45}+2 & \cellcolor{LimeGreen!25}+1 & \cellcolor{LimeGreen!25}+1 & \cellcolor{LimeGreen!25}+1 & \cellcolor{Green!45}+2 & \cellcolor{Green!45}+2 & \textbf{+1.33} & \cellcolor{blue!20}0 & \cellcolor{LimeGreen!25}+1 & \cellcolor{Green!45}+2 & \cellcolor{LimeGreen!25}+1 & \cellcolor{Green!45}+2 & \cellcolor{Green!45}+2 \\
P18.5R & +1.06 & \textbf{+1.25} & \cellcolor{LimeGreen!25}+1 & \cellcolor{LimeGreen!25}+1 & \cellcolor{LimeGreen!25}+1 & \cellcolor{Green!45}+2 & \textbf{+1.14} & \cellcolor{blue!20}0 & NA & \cellcolor{LimeGreen!25}+1 & \cellcolor{LimeGreen!25}+1 & \cellcolor{Green!45}+2 & \cellcolor{LimeGreen!25}+1 & \cellcolor{Green!45}+2 & \cellcolor{LimeGreen!25}+1 & \textbf{+0.83} & \cellcolor{LimeGreen!25}+1 & \cellcolor{LimeGreen!25}+1 & \cellcolor{LimeGreen!25}+1 & \cellcolor{blue!20}0 & \cellcolor{LimeGreen!25}+1 & \cellcolor{LimeGreen!25}+1 \\
P19.5A & +0.89 & \textbf{+0.75} & \cellcolor{blue!20}0 & \cellcolor{LimeGreen!25}+1 & \cellcolor{LimeGreen!25}+1 & \cellcolor{LimeGreen!25}+1 & \textbf{+1} & \cellcolor{Green!45}+2 & \cellcolor{LimeGreen!25}+1 & \cellcolor{LimeGreen!25}+1 & \cellcolor{orange!55}-1 & \cellcolor{Green!45}+2 & \cellcolor{LimeGreen!25}+1 & \cellcolor{LimeGreen!25}+1 & \cellcolor{LimeGreen!25}+1 & \textbf{+0.83} & \cellcolor{LimeGreen!25}+1 & \cellcolor{LimeGreen!25}+1 & \cellcolor{LimeGreen!25}+1 & \cellcolor{LimeGreen!25}+1 & \cellcolor{LimeGreen!25}+1 & \cellcolor{blue!20}0 \\
\hline
\rowcolor{gray!15}\textbf{GP05 Av} & +1.08 & \textbf{+1.00} & \textbf{+0.50} & \textbf{+1.25} & \textbf{+1.00} & \textbf{+1.25} & \textbf{+1.13} & \textbf{+1.25} & \textbf{+1.33} & \textbf{+1.25} & \textbf{+0.25} & \textbf{+1.75} & \textbf{+1.00} & \textbf{+1.50} & \textbf{+0.75} & \textbf{+1.08} & \textbf{+0.75} & \textbf{+1.00} & \textbf{+1.50} & \textbf{+1.00} & \textbf{+1.00} & \textbf{+1.25} \\
\hline
P20.6L & +0.39 & \textbf{+0.25} & \cellcolor{LimeGreen!25}+1 & \cellcolor{blue!20}0 & \cellcolor{orange!55}-1 & \cellcolor{LimeGreen!25}+1 & \cellcolor{blue!20}\textbf{0} & \cellcolor{LimeGreen!25}+1 & \cellcolor{orange!55}-1 & \cellcolor{blue!20}0 & \cellcolor{orange!55}-1 & \cellcolor{blue!20}0 & \cellcolor{blue!20}0 & \cellcolor{LimeGreen!25}+1 & \cellcolor{blue!20}0 & \textbf{+1} & \cellcolor{LimeGreen!25}+1 & \cellcolor{blue!20}0 & \cellcolor{LimeGreen!25}+1 & \cellcolor{Green!45}+2 & \cellcolor{LimeGreen!25}+1 & \cellcolor{LimeGreen!25}+1 \\
%%Some of the steps are heavily overloaded. - P20.6L

P21.6M & +1.71 & \textbf{+2} & \cellcolor{Green!45}+2 & \cellcolor{Green!45}+2 & \cellcolor{Green!45}+2 & \cellcolor{Green!45}+2 & \textbf{+1.71} & \cellcolor{Green!45}+2 & \cellcolor{Green!45}+2 & \cellcolor{Green!45}+2 & NA & \cellcolor{LimeGreen!25}+1 & \cellcolor{LimeGreen!25}+1 & \cellcolor{Green!45}+2 & \cellcolor{Green!45}+2 & \textbf{+1.5} & \cellcolor{Green!45}+2 & \cellcolor{LimeGreen!25}+1 & \cellcolor{LimeGreen!25}+1 & \cellcolor{Green!45}+2 & \cellcolor{Green!45}+2 & \cellcolor{LimeGreen!25}+1 \\
P22.6R & +1.22 & \textbf{+0.75} & \cellcolor{LimeGreen!25}+1 & \cellcolor{blue!20}0 & \cellcolor{LimeGreen!25}+1 & \cellcolor{LimeGreen!25}+1 & \textbf{+1.38} & \cellcolor{LimeGreen!25}+1 & \cellcolor{Green!45}+2 & \cellcolor{LimeGreen!25}+1 & \cellcolor{LimeGreen!25}+1 & \cellcolor{Green!45}+2 & \cellcolor{LimeGreen!25}+1 & \cellcolor{Green!45}+2 & \cellcolor{LimeGreen!25}+1 & \textbf{+1.33} & \cellcolor{LimeGreen!25}+1 & \cellcolor{LimeGreen!25}+1 & \cellcolor{LimeGreen!25}+1 & \cellcolor{Green!45}+2 & \cellcolor{LimeGreen!25}+1 & \cellcolor{Green!45}+2 \\
P23.6A & \cellcolor{red!50}-0.39 & \textbf{+0.5} & \cellcolor{LimeGreen!25}+1 & \cellcolor{blue!20}0 & \cellcolor{blue!20}0 & \cellcolor{LimeGreen!25}+1 & \textbf{-0.5} & \cellcolor{red!50}-2 & \cellcolor{LimeGreen!25}+1 & \cellcolor{red!50}-2 & \cellcolor{red!50}-2 & \cellcolor{orange!55}-1 & \cellcolor{LimeGreen!25}+1 & \cellcolor{orange!55}-1 & \cellcolor{Green!45}+2 & \textbf{-0.83} & \cellcolor{LimeGreen!25}+1 & \cellcolor{blue!20}0 & \cellcolor{orange!55}-1 & \cellcolor{red!50}-2 & \cellcolor{red!50}-2 & \cellcolor{orange!55}-1 \\

\hline
\rowcolor{gray!15}\textbf{GP06 Av} & +0.73 & \textbf{+0.88} & \textbf{+1.25} & \textbf{+0.50} & \textbf{+0.50} & \textbf{+1.25} & \textbf{+0.65} & \textbf{+0.50} & \textbf{+1.00} & \textbf{+0.25} & \cellcolor{red!50}\textbf{-0.67} & \textbf{+0.50} & \textbf{+0.75} & \textbf{+1.00} & \textbf{+1.25} & \textbf{+0.75} & \textbf{+1.25} & \textbf{+0.50} & \textbf{+0.50} & \textbf{+1.00} & \textbf{+0.50} & \textbf{+0.75}\\
\hline
P24.7L & +1.17 & \textbf{+1} & \cellcolor{LimeGreen!25}+1 & \cellcolor{LimeGreen!25}+1 & \cellcolor{LimeGreen!25}+1 & \cellcolor{LimeGreen!25}+1 & \textbf{+1.25} & \cellcolor{Green!45}+2 & \cellcolor{LimeGreen!25}+1 & \cellcolor{LimeGreen!25}+1 & \cellcolor{LimeGreen!25}+1 & \cellcolor{LimeGreen!25}+1 & \cellcolor{LimeGreen!25}+1 & \cellcolor{Green!45}+2 & \cellcolor{LimeGreen!25}+1 & \textbf{+1.17} & \cellcolor{LimeGreen!25}+1 & \cellcolor{LimeGreen!25}+1 & \cellcolor{LimeGreen!25}+1 & \cellcolor{LimeGreen!25}+1 & \cellcolor{Green!45}+2 & \cellcolor{LimeGreen!25}+1 \\
P25.7L & +1.06 & \textbf{+1} & \cellcolor{LimeGreen!25}+1 & \cellcolor{LimeGreen!25}+1 & \cellcolor{blue!20}0 & \cellcolor{Green!45}+2 & \textbf{+1.25} & \cellcolor{Green!45}+2 & \cellcolor{Green!45}+2 & \cellcolor{blue!20}0 & \cellcolor{LimeGreen!25}+1 & \cellcolor{LimeGreen!25}+1 & \cellcolor{LimeGreen!25}+1 & \cellcolor{LimeGreen!25}+1 & \cellcolor{Green!45}+2 & \textbf{+0.83} & \cellcolor{blue!20}0 & \cellcolor{LimeGreen!25}+1 & \cellcolor{LimeGreen!25}+1 & \cellcolor{Green!45}+2 & \cellcolor{LimeGreen!25}+1 & \cellcolor{blue!20}0 \\
P26.7M & +1.67 & \textbf{+1.5} & \cellcolor{Green!45}+2 & \cellcolor{Green!45}+2 & \cellcolor{LimeGreen!25}+1 & \cellcolor{LimeGreen!25}+1 & \textbf{+1.63} & \cellcolor{LimeGreen!25}+1 & \cellcolor{Green!45}+2 & \cellcolor{Green!45}+2 & \cellcolor{Green!45}+2 & \cellcolor{blue!20}0 & \cellcolor{Green!45}+2 & \cellcolor{Green!45}+2 & \cellcolor{Green!45}+2 & \textbf{+1.83} & \cellcolor{Green!45}+2 & \cellcolor{Green!45}+2 & \cellcolor{LimeGreen!25}+1 & \cellcolor{Green!45}+2 & \cellcolor{Green!45}+2 & \cellcolor{Green!45}+2 \\
P27.7R & +1.28 & \textbf{+2} & \cellcolor{Green!45}+2 & \cellcolor{Green!45}+2 & \cellcolor{Green!45}+2 & \cellcolor{Green!45}+2 & \textbf{+1.75} & \cellcolor{Green!45}+2 & \cellcolor{LimeGreen!25}+1 & \cellcolor{Green!45}+2 & \cellcolor{LimeGreen!25}+1 & \cellcolor{Green!45}+2 & \cellcolor{Green!45}+2 & \cellcolor{Green!45}+2 & \cellcolor{Green!45}+2 & \textbf{+0.17} & \cellcolor{LimeGreen!25}+1 & \cellcolor{blue!20}0 & \cellcolor{blue!20}0 & \cellcolor{blue!20}0 & \cellcolor{blue!20}0 & \cellcolor{blue!20}0 \\
P28.7A & +1.22 & \textbf{+2} & \cellcolor{Green!45}+2 & \cellcolor{Green!45}+2 & \cellcolor{Green!45}+2 & \cellcolor{Green!45}+2 & \textbf{+0.63} & \cellcolor{Green!45}+2 & \cellcolor{LimeGreen!25}+1 & \cellcolor{LimeGreen!25}+1 & \cellcolor{blue!20}0 & \cellcolor{blue!20}0 & \cellcolor{LimeGreen!25}+1 & \cellcolor{blue!20}0 & \cellcolor{blue!20}0 & \textbf{+1.5} & \cellcolor{Green!45}+2 & \cellcolor{LimeGreen!25}+1 & \cellcolor{Green!45}+2 & \cellcolor{LimeGreen!25}+1 & \cellcolor{LimeGreen!25}+1 & \cellcolor{Green!45}+2 \\
\hline
\rowcolor{gray!15}\textbf{GP07 Av} & +1.28 & \textbf{+1.50} & \textbf{+1.60} & \textbf{+1.60} & \textbf{+1.20} & \textbf{+1.60} & \textbf{+1.30} & \textbf{+1.80} & \textbf{+1.40} & \textbf{+1.20} & \textbf{+1.00} & \textbf{+0.80} & \textbf{+1.40} & \textbf{+1.40} & \textbf{+1.40} & \textbf{+1.10} & \textbf{+1.20} & \textbf{+1.00} & \textbf{+1.00} & \textbf{+1.20} & \textbf{+1.20} & \textbf{+1.00} \\
\hline
S1.1L & +0.44 & \cellcolor{blue!20}\textbf{0} & \cellcolor{orange!55}-1 & \cellcolor{LimeGreen!25}+1 & \cellcolor{LimeGreen!25}+1 & \cellcolor{orange!55}-1 & \textbf{+1} & \cellcolor{LimeGreen!25}+1 & \cellcolor{LimeGreen!25}+1 & \cellcolor{LimeGreen!25}+1 & \cellcolor{LimeGreen!25}+1 & \cellcolor{LimeGreen!25}+1 & \cellcolor{LimeGreen!25}+1 & \cellcolor{LimeGreen!25}+1 & \cellcolor{LimeGreen!25}+1 & \cellcolor{blue!20}\textbf{0} & \cellcolor{LimeGreen!25}+1 & \cellcolor{orange!55}-1 & \cellcolor{orange!55}-1 & \cellcolor{LimeGreen!25}+1 & \cellcolor{LimeGreen!25}+1 & \cellcolor{orange!55}-1 \\
S2.1R & \cellcolor{blue!20}0 & \textbf{+0.25} & \cellcolor{blue!20}0 & \cellcolor{LimeGreen!25}+1 & \cellcolor{blue!20}0 & \cellcolor{blue!20}0 & \cellcolor{blue!20}\textbf{0} & \cellcolor{orange!55}-1 & \cellcolor{orange!55}-1 & \cellcolor{blue!20}0 & \cellcolor{orange!55}-1 & \cellcolor{LimeGreen!25}+1 & \cellcolor{LimeGreen!25}+1 & \cellcolor{LimeGreen!25}+1 & \cellcolor{blue!20}0 & \cellcolor{orange!55}\textbf{-0.17} & \cellcolor{blue!20}0 & \cellcolor{blue!20}0 & \cellcolor{orange!55}-1 & \cellcolor{LimeGreen!25}+1 & \cellcolor{blue!20}0 & \cellcolor{orange!55}-1 \\
S3.1A & +1.44 & \textbf{+1.75} & \cellcolor{Green!45}+2 & \cellcolor{Green!45}+2 & \cellcolor{LimeGreen!25}+1 & \cellcolor{Green!45}+2 & \textbf{+1.25} & \cellcolor{Green!45}+2 & \cellcolor{LimeGreen!25}+1 & \cellcolor{LimeGreen!25}+1 & \cellcolor{blue!20}0 & \cellcolor{Green!45}+2 & \cellcolor{LimeGreen!25}+1 & \cellcolor{Green!45}+2 & \cellcolor{LimeGreen!25}+1 & \textbf{+1.5} & \cellcolor{LimeGreen!25}+1 & \cellcolor{Green!45}+2 & \cellcolor{LimeGreen!25}+1 & \cellcolor{Green!45}+2 & \cellcolor{Green!45}+2 & \cellcolor{LimeGreen!25}+1 \\
\hline
\rowcolor{gray!15}\textbf{GS01 Av} & +0.63 & \textbf{+0.67} & \textbf{+0.33} & \textbf{+1.33} & \textbf{+0.67} & \textbf{+0.33} & \textbf{+0.75} & \textbf{+0.67} & \textbf{+0.33} & \textbf{+0.67} & \cellcolor{blue!20}\textbf{0} & \textbf{+1.33} & \textbf{+1.00} & \textbf{+1.33} & \textbf{+0.67} & \textbf{+0.44} & \textbf{+0.67} & \textbf{+0.33} & \cellcolor{orange!55}\textbf{-0.33} & \textbf{+1.33} & \textbf{+1.00} & \cellcolor{orange!55}\textbf{-0.33} \\
\hline
S4.2L & +1.06 & \textbf{+1.5} & \cellcolor{Green!45}+2 & \cellcolor{LimeGreen!25}+1 & \cellcolor{LimeGreen!25}+1 & \cellcolor{Green!45}+2 & \textbf{+0.43} & \cellcolor{LimeGreen!25}+1 & NA & \cellcolor{LimeGreen!25}+1 & \cellcolor{LimeGreen!25}+1 & \cellcolor{orange!55}-1 & \cellcolor{orange!55}-1 & \cellcolor{LimeGreen!25}+1 & \cellcolor{LimeGreen!25}+1 & \textbf{+1.60} & \cellcolor{LimeGreen!25}+1 & \cellcolor{Green!45}+2 & \cellcolor{Green!45}+2 & \cellcolor{Green!45}+2 & NA & \cellcolor{LimeGreen!25}+1 \\
S5.2R & +0.88 & \textbf{+0.75} & \cellcolor{LimeGreen!25}+1 & \cellcolor{blue!20}0 & \cellcolor{LimeGreen!25}+1 & \cellcolor{LimeGreen!25}+1 & \textbf{+0.71} & \cellcolor{blue!20}0 & \cellcolor{Green!45}+2 & \cellcolor{LimeGreen!25}+1 & NA & \cellcolor{blue!20}0 & \cellcolor{LimeGreen!25}+1 & \cellcolor{LimeGreen!25}+1 & \cellcolor{blue!20}0 & \textbf{+1.17} & \cellcolor{LimeGreen!25}+1 & \cellcolor{LimeGreen!25}+1 & \cellcolor{Green!45}+2 & \cellcolor{Green!45}+2 & \cellcolor{blue!20}0 & \cellcolor{LimeGreen!25}+1 \\
S6.2A & +1.33 & \textbf{+1} & \cellcolor{LimeGreen!25}+1 & \cellcolor{LimeGreen!25}+1 & \cellcolor{blue!20}0 & \cellcolor{Green!45}+2 & \textbf{+1} & \cellcolor{Green!45}+2 & \cellcolor{LimeGreen!25}+1 & \cellcolor{Green!45}+2 & \cellcolor{blue!20}0 & \cellcolor{LimeGreen!25}+1 & \cellcolor{LimeGreen!25}+1 & \cellcolor{LimeGreen!25}+1 & \cellcolor{blue!20}0 & \textbf{+2} & \cellcolor{Green!45}+2 & \cellcolor{Green!45}+2 & \cellcolor{Green!45}+2 & \cellcolor{Green!45}+2 & \cellcolor{Green!45}+2 & \cellcolor{Green!45}+2 \\
\hline
\rowcolor{gray!15}\textbf{GS02 Av} & +1.09 & \textbf{+1.08} & \textbf{+1.33} & \textbf{+0.67} & \textbf{+0.67} & \textbf{+1.67} & \textbf{+0.71} & \textbf{+1.00} & \textbf{+1.50} & \textbf{+1.33} & \textbf{+0.50} & \cellcolor{blue!20}\textbf{0} & \textbf{+0.33} & \textbf{+1.00} & \textbf{+0.33} & \textbf{+1.59} & \textbf{+1.33} & \textbf{+1.67} & \textbf{+2.00} & \textbf{+2.00} & \textbf{+1.00} & \textbf{+1.33} \\
\hline
S7.3L & +1.11 & \textbf{+1} & \cellcolor{blue!20}0 & \cellcolor{LimeGreen!25}+1 & \cellcolor{LimeGreen!25}+1 & \cellcolor{Green!45}+2 & \textbf{+1} & \cellcolor{LimeGreen!25}+1 & \cellcolor{LimeGreen!25}+1 & \cellcolor{Green!45}+2 & \cellcolor{LimeGreen!25}+1 & \cellcolor{blue!20}0 & \cellcolor{LimeGreen!25}+1 & \cellcolor{LimeGreen!25}+1 & \cellcolor{LimeGreen!25}+1 & \textbf{+1.33} & \cellcolor{Green!45}+2 & \cellcolor{LimeGreen!25}+1 & \cellcolor{LimeGreen!25}+1 & \cellcolor{LimeGreen!25}+1 & \cellcolor{Green!45}+2 & \cellcolor{LimeGreen!25}+1 \\
S8.3M & +0.83 & \textbf{+0.5} & \cellcolor{blue!20}0 & \cellcolor{blue!20}0 & \cellcolor{LimeGreen!25}+1 & \cellcolor{LimeGreen!25}+1 & \textbf{+0.88} & \cellcolor{LimeGreen!25}+1 & \cellcolor{blue!20}0 & \cellcolor{LimeGreen!25}+1 & \cellcolor{blue!20}0 & \cellcolor{LimeGreen!25}+1 & \cellcolor{LimeGreen!25}+1 & \cellcolor{Green!45}+2 & \cellcolor{LimeGreen!25}+1 & \textbf{+1} & \cellcolor{Green!45}+2 & \cellcolor{LimeGreen!25}+1 & \cellcolor{blue!20}0 & \cellcolor{blue!20}0 & \cellcolor{LimeGreen!25}+1 & \cellcolor{Green!45}+2 \\
S9.3R & NA & \textbf{NA} & NA & NA & NA & NA & \textbf{NA} & NA & NA & NA & NA & NA & NA & NA & NA & \textbf{NA} & NA & NA & NA & NA & NA & NA \\
S10.3A & NA & \textbf{NA} & NA & NA & NA & NA & \textbf{NA} & NA & NA & NA & NA & NA & NA & NA & NA & \textbf{NA} & NA & NA & NA & NA & NA & NA \\
\hline
\rowcolor{gray!15}\textbf{GS03 Av} & +0.97 & \textbf{+0.75} & \cellcolor{blue!20}\textbf{0} & \textbf{+0.50} & \textbf{+1.00} & \textbf{+1.50} & \textbf{+0.94} & \textbf{+1.00} & \textbf{+0.50} & \textbf{+1.50} & \textbf{+0.50} & \textbf{+0.50} & \textbf{+1.00} & \textbf{+1.50} & \textbf{+1.00} & \textbf{+1.17} & \textbf{+2.00} & \textbf{+1.00} & \textbf{+0.50} & \textbf{+0.50} & \textbf{+1.50} & \textbf{+1.50} \\
\hline
S11.4L & +1.39 & \textbf{+0.75} & \cellcolor{Green!45}+2 & \cellcolor{orange!55}-1 & \cellcolor{LimeGreen!25}+1 & \cellcolor{LimeGreen!25}+1 & \textbf{+1.63} & \cellcolor{Green!45}+2 & \cellcolor{LimeGreen!25}+1 & \cellcolor{Green!45}+2 & \cellcolor{LimeGreen!25}+1 & \cellcolor{Green!45}+2 & \cellcolor{Green!45}+2 & \cellcolor{Green!45}+2 & \cellcolor{LimeGreen!25}+1 & \textbf{+1.5} & \cellcolor{Green!45}+2 & \cellcolor{LimeGreen!25}+1 & \cellcolor{Green!45}+2 & \cellcolor{Green!45}+2 & \cellcolor{Green!45}+2 & \cellcolor{blue!20}0 \\
S12.4A & +1.67 & \textbf{+1.75} & \cellcolor{LimeGreen!25}+1 & \cellcolor{Green!45}+2 & \cellcolor{Green!45}+2 & \cellcolor{Green!45}+2 & \textbf{+1.63} & \cellcolor{Green!45}+2 & \cellcolor{LimeGreen!25}+1 & \cellcolor{Green!45}+2 & \cellcolor{LimeGreen!25}+1 & \cellcolor{LimeGreen!25}+1 & \cellcolor{Green!45}+2 & \cellcolor{Green!45}+2 & \cellcolor{Green!45}+2 & \textbf{+1.67} & \cellcolor{Green!45}+2 & \cellcolor{Green!45}+2 & \cellcolor{Green!45}+2 & \cellcolor{LimeGreen!25}+1 & \cellcolor{Green!45}+2 & \cellcolor{LimeGreen!25}+1 \\
\hline
\rowcolor{gray!15}\textbf{GS04 Av} & +1.53 & \textbf{+1.25} & \textbf{+1.50} & \textbf{+0.50} & \textbf{+1.50} & \textbf{+1.50} & \textbf{+1.63} & \textbf{+2.00} & \textbf{+1.00} & \textbf{+2.00} & \textbf{+1.00} & \textbf{+1.50} & \textbf{+2.00} & \textbf{+2.00} & \textbf{+1.50} & \textbf{+1.58} & \textbf{+2.00} & \textbf{+1.50} & \textbf{+2.00} & \textbf{+1.50} & \textbf{+2.00} & \textbf{+0.50} \\

\hline\hline
\textbf{Total Av} & +1.17 & \textbf{+1.23} & \textbf{+1.24} & \textbf{+1.24} & \textbf{+1.05} & \textbf{+1.38} & \textbf{+1.12} & \textbf{+1.38} & \textbf{+1.18} & \textbf{+1.14} & \textbf{+0.69} & \textbf{+0.97} & \textbf{+1.05} & \textbf{+1.38} & \textbf{+1.19} & \textbf{+1.19} & \textbf{+1.24} & \textbf{+0.95} & \textbf{+1.22} & \textbf{+1.32} & \textbf{+1.29} & \textbf{+1.11}\\
PbD Av & +1.23 & +1.34 & +1.41 & +1.41 & +1.11 & +1.44 & +1.19 & +1.48 & +1.32 & +1.07 & +0.77 & +1.04 & +1.07 & +1.37 & +1.33 & +1.20 & +1.19 & +0.89 & +1.30 & +1.30 & +1.27 & +1.26 \\

SbD Av & +1.02 & +0.93 & +0.80 & +0.80 & +0.90 & +1.20 & +0.95 & +1.10 & +0.78 & +1.30 & +0.44 & +0.80 & +1.00 & +1.40 & +0.80 & +1.16 & +1.40 & +1.10 & +1.00 & +1.40 & +1.33 & +0.70 \\

Diff. & |0.21| & |0.42| & |0.61| & |0.61| & |0.21| & |0.24| & |0.23| & |0.38| & |0.54| & |0.23| & |0.32| & |0.24| & |0.07| & |0.03| & |0.53| & |0.04| & |0.21| & |0.21| & |0.30| & |0.10| & |0.06| & |0.56| \\
\hline\hline
Legal Av & +1.13 & +1.08 & +1.15 & +1.00 & +0.85 & +1.31 & +1.08 & +1.46 & +1.00 & +0.92 & +0.85 & +0.92 & +1.00 & +1.38 & +1.15 & +1.21 & +1.23 & +0.85 & +1.31 & +1.62 & +1.45 & +0.92 \\
Manag. Av & +1.23 & +1.13 & +1.17 & +1.33 & +1.00 & +1.00 & +1.29 & +1.50 & +1.50 & +1.50 & +0.80 & +0.83 & +1.00 & +1.67 & +1.33 & +1.22 & +1.33 & +1.00 & +1.00 & +1.17 & +1.50 & +1.33 \\
RE Av & +1.12 & +1.25 & +1.29 & +1.14 & +1.14 & +1.43 & +1.16 & +0.86 & +1.33 & +1.14 & +1.00 & +1.29 & +1.14 & +1.57 & +1.14 & +0.95 & +0.71 & +0.86 & +1.00 & +1.29 & +0.86 & +1.00 \\
Arch. Av & +1.21 & +1.45 & +1.36 & +1.55 & +1.27 & +1.64 & +1.05 & +1.55 & +1.09 & +1.18 & +0.27 & +0.91 & +1.09 & +1.09 & +1.18 & +1.27 & +1.55 & +1.09 & +1.36 & +1.09 & +1.27 & +1.27 \\
\hline\hline

\multicolumn{1}{c|}{Ev. num.} & Total &  &  &  &  &  &  &  &  &  &  &  &  &  &  &  &  &  &  &  &  \\
\multicolumn{1}{c|}{+2} & 252 & 61 & 16 & 17 & 10 & 18 & 102 & 21 & 13 & 13 & 6 & 10 & 6 & 18 & 15 & 89 & 13 & 9 & 17 & 18 & 17 & 15 \\
\multicolumn{1}{c|}{+1} & 299 & 64 & 15 & 13 & 20 & 16 & 140 & 12 & 16 & 18 & 17 & 18 & 28 & 16 & 15 & 95 & 20 & 19 & 14 & 15 & 13 & 14 \\
\multicolumn{1}{c|}{0} & 80 & 19 & 5 & 6 & 6 & 2 & 35 & 2 & 3 & 5 & 8 & 7 & 2 & 2 & 6 & 26 & 4 & 7 & 3 & 3 & 4 & 5 \\
\multicolumn{1}{c|}{-1} & 23 & 4 & 1 & 1 & 1 & 1 & 11 & 1 & 2 & 0 & 3 & 2 & 1 & 1 & 1 & 8 & 0 & 2 & 3 & 0 & 0 & 3 \\
\multicolumn{1}{c|}{-2} & 5 & 0 & 0 & 0 & 0 & 0 & 3 & 1 & 0 & 1 & 1 & 0 & 0 & 0 & 0 & 2 & 0 & 0 & 0 & 1 & 1 & 0 \\
\hline
\hline
\end{tabular}
\caption{\small Results of AM4RRE evaluation. Columns show participant IDs and average and criterion-specific ratings. For an overview of the criteria and their groups, see Section~\ref{sec:ep}. Color coding, where $x$ denotes an individual rating or an average rating: dark green, $x > 1$; light green, $0 < x \leq 1$; light blue, $x = 0$; orange, $-1 \leq x < 0$; red, $x < -1$.}
\label{tab:evaluation}
\end{table*}
\end{landscape}

\subsection{RQ3: To what extent can AM4RRE support coordination among the viewpoints?}\label{sec:rq3}
To answer RQ3, we report the feature analysis for coordination-related criteria (see Table~\ref{tab:evaluation}) and specification time in the quasi-experiment (see Table~\ref{tab:coordination}).

AM4RRE was evaluated as useful for \textit{role coordination (CC05)} in 21 of 37 evaluations, with an average evaluation of 1.38. S3.1A suggested that AM4RRE could be a ``middle way'' between overly abstract and overly prescriptive methods (see below).

\begin{colorb}
    \small
    \textit{``Method seems to be a good middle way between principles (Marty Cagan) or legal texts only (without guidance) and procedural work (step by step) which involves a lot of work of preparation.''} - S3.1A in the evaluation of AM4RRE
\end{colorb}

Evaluations varied most among RE and software architect roles and in groups GP06 and GS01. As S2.1R explained in an oral comment, they evaluated AM4RRE as potentially not useful (-1) due to the ``delayed'' and ``insufficient'' involvement of a legal expert and the need for earlier and more extensive legal input.
In an exchange during the session in GP06, P23.6A and P20.6L questioned the idea of coordination in regulatory RE, arguing that close interaction between roles is not required because each role is responsible only within its own competence (see the following quotation).

%coordination is shaped by the understanding of the role and responsibilities
\begin{colorb}
    \small
    \textit{``My role is to support development teams by enabling and facilitating their work. The responsibility for implementing GDPR compliance within systems remains with the engineering and product teams. It would not be an effective use of my limited capacity to assign me to work directly with engineers who do not possess foundational GDPR knowledge or relevant experience. [D]evelopment teams are expected to have the necessary baseline competency in GDPR requirements before engaging in implementation-related collaboration.''} - P20.6L in an oral comment.
\end{colorb}

P23.6A agreed with this criticism, arguing that email and other conventional communication channels should suffice for engineering-legal coordination. Still, P20.6L evaluated AM4RRE-guided coordination as potentially useful (+1). Some participants who evaluated the templates negatively (e.g., P24.7L see Section~\ref{sec:rq2}) still considered AM4RRE as potentially useful for coordination.

For \textit{capturing legal domain knowledge (CC06)}, AM4RRE was most often evaluated as potentially useful (16 of 34 evaluations), with an average evaluation of 1.18. P20.6L and S2.1R evaluated AM4RRE as "potentially not useful" for the reasons reported above. We suggest that evaluations according to this criterion were closely related to some legal-role participants' concerns that modeling regulations can be too ``simplistic'', will result in losing the ``meaning'' and ``nuance'' or will not allow for capturing all the legal knowledge (see Section~\ref{sec:rq2} for comments about legal specification templates). Oral comments of participants indicated that failing to capture the complexity of legal knowledge may constrain regulatory RE (see the following quote).

\begin{colorb}
    \small
    \textit{``This is way too schematic. It may serve as a warning, but is very far from the fine-focus of the GDPR and would not serve in the specific case-analysis.''} - P24.7L in a comment regarding TSanc\_LP template.
\end{colorb}

Participants evaluated AM4RRE's usefulness for \textit{capturing role-specific knowledge (CC07)} as 1.14 on average, and 18 of 37 found it potentially useful. P23.6A, who was critical of the coordination approach overall, rated it “not useful”.

For \textit{transparency of specifications to other roles (CC11)}, 18 of 37 participants evaluated it as useful, with an average evaluation of 1.38. Only P23.6A gave a negative evaluation.

For \textit{facilitating the implementation of regulatory compliance in software architecture (CC13)}, participants assigned an average evaluation of 1.24, with 20 of 37 evaluating AM4RRE as potentially useful. For CC13, average evaluations by requirements engineers (0.71) and software architects (1.55) diverged. The written feedback of participants also indicated that AM4RRE is useful for supporting interdisciplinary structured coordination in regulatory RE for CbD (see quote next).

\begin{colorb}
    \small
    \textit{``If properly refined, I believe this could be a significant leap forward in how to structure communication and reusability of knowledge in the field of PbD, well, really AnythingByDesign.''} - P16.5L in the evaluation of AM4RRE
\end{colorb}

For \textit{facilitating the implementation of regulatory compliance throughout the other SE processes (CC14)}, 19 of 37 participants evaluated AM4RRE as potentially useful, and the overall average evaluation was 0.95. Only two participants assigned a ``potentially not useful'' evaluation: P1.1L and S1.1L.

%\begin{landscape}
\begin{table*}[ht!]
\tiny
\renewcommand\theadfont{\tiny}
\setlength{\tabcolsep}{1.25pt}
\centering
\rowcolors{3}{gray!15}{white}
\begin{tabular}{cccccc|cccc|cccc|c|ccccc}

\textbf{Gr.} & \textbf{Pr.} &  \textbf{Av.} & \textbf{Mx.} &  \textbf{} & \textbf{} & 
\multicolumn{4}{|c|}{Period 1} & \multicolumn{4}{|c|}{Period 2} & \textbf{Total} & \multicolumn{5}{c}{Difference} \\
\textbf{ID} & \textbf{\#} &  \textbf{Ex} & \textbf{Ex} &  \textbf{Cl} & \textbf{Sq} & \textbf{Time} & \textbf{Cm} & \textbf{Cn} & \textbf{Cr} & \textbf{Time} & \textbf{Cm} & \textbf{Cn} & \textbf{Cr} & \textbf{Time} & 
\multicolumn{2}{c}{Time} & \textbf{Cm} & \textbf{Cn} & \textbf{Cr} \\ \hline
GP01 & 4 & 3.75 & 8 & No & GF & 18m:31s & 1 & 0.66 & 0.81 & 8m:49s & 0.80 & 0.61 & 0.57 & 27m:20s & -9m:42s & -52\% & -0.20 & -0.05 & -0.24 \\
GP02 & 2 & 6 & 7 & Yes & GF & 11m:00s & 1 & 0.83 & 0.90 & 8m:05s & 1 & 0.75 & 0.85 & 19m:05s & -2m:55s & -27\% & 0 & -0.08 & -0.05 \\
GP03 & 3 & 2 & 3 & Yes & GF & 11m:02s & 1 & 0.62 & 0.79 & 7m:05s & 1 & 0.50 & 0.64 & 18m:07s & -3m:57s & -36\% & 0 & -0.12 & -0.15 \\
GP04 & 4 & 4 & 7 & Yes & FG & 25m:00s & 1 & 0.77 & 0.70 & 7m:32s & 1 & 0.82 & 0.90 & 32m:32s & -17m:28s & -70\% & 0 & +0.05 & +0.20 \\
GP05 & 4 & 6.25 & 10 & No & FG & 34m:42s & 1 & 0.63 & 0.87 & 27m:21s & 1 & 0.93 & 0.97 & 62m:03s & -7m:21s & -21\% & 0 & +0.30 & +0.10\\
GP06 & 4 & 4.75 & 7 & No & GF & 8m:49s & 1 & 0.50 & 0.71 & 10m:35s & 1 & 0.40 & 0.60 & 19m:24s & +1m:46s & +20\% & 0 & -0.10 & -0.11\\
GP07 & 5 & 3.2 & 7 & Yes & FG & 29m:22s & 0.90 & 0.45 & 0.84 & 17m:01s & 1 & 0.46 & 0.92 & 46m:23s & -12m:21s & -42\% & +0.10 & +0.01 & +0.08 \\
GS01 & 3 & 0.50 & 1 & No & GF & 13m:28s & 1 & 0.88 & 0.94 & 9m:21s & 0.80 & 0.46 & 0.38 & 22m:49s & -4m:07s & -31\% & -0.20 & -0.42 & -0.56 \\
GS02 & 3 & 0.67 & 1 & Yes & FG & 20m:51s & 1 & 0.70 & 0.77 & 10m:07s & 1 & 0.69 & 0.78 & 30m:58s & -10m:44s & -51\% & 0 & -0.01 & +0.01 \\
GS04 & 2 & 1.5 & 2 & No & GF & 15m:00s & 1 & 0.80 & 0.80 & 12m:16s & 1 & 0.91 & 0.97 & 27m:16s & -2m:44s & -18\% & 0 & +0.11 & +0.17\\
\midrule
Av & 3 & 3.1 & 4.7 & - & GF & 12m:58s & 1 & 0.72 & 0.83 & 9m:22s & 0.93 & 0.61 & 0.67 & 22m:20s & -3m:37s & -28\% & -0.07 & -0.11 & -0.16 \\
Av & 4 & 3.5 & 6.3 & - & FG & 27m:29s & 0.98 & 0.64 & 0.80 & 15m:30s & 1 & 0.73 & 0.89 & 42m:59s & -11m:59s & -44\% & +0.03 & +0.09 & +0.10 \\
\hline
\end{tabular}
\caption{\small Specification task outcomes. Columns: group ID, number of participants, average and maximum experience with regulation (years), collaboration experience, treatment sequence, for P1 and P2: specification completion time, requirements completeness, requirements consistency, requirements correctness, total specification completion time, differences between P1 and P2.}
\label{tab:coordination}
\end{table*}

In the quasi-experiment, 10 of 11 groups, except GS03 in which two force-majeure withdrawals occurred, completed the specification tasks, including fully onsite groups, the online GS01 group, and GP05, with only a legal expert participating online.

Across both periods, the four groups with FG treatment sequence (GP04, GP05, GP07, GS02) had an average total specification time of 42m:59s (see Table~\ref{tab:coordination}). In contrast, in the groups with GF treatment sequence (GP01, GP02, GP03, GP06, GS01, GS04), the corresponding average was 22m:20s, showing a difference of 20m:39s. This pattern was also observed at the level of groups: the longest total time of any GF group (27:20) was shorter than the shortest FG total (30:58). Notably, the average total time for groups with GF treatment sequence (22m:20s) was lower than the average specification time in Period 1 (P1) alone among the groups with FG sequence (27m:29s).
Guided coordination averaged 12:58 when done first (GF) and 15:30 when done second (FG). Free-form coordination executed first in FG sequence took on average 27m:29s, and only 9m:22s when executed after guided coordination in GF sequence groups. 

\begin{colora}
    These findings provide indicative evidence for Hypothesis 1 that the average total specification time is  shorter in groups that used templates guided coordination in Period 1 (GF sequence) than the average total time in the groups that applied free-form coordination in Period 1 (FG sequence), with a difference of 20m:39s between the averages.
\end{colora}

Specification time decreased from Period 1 to Period 2 in all groups, except GP06, where P20.6L and P23.6A were critical of the idea of coordination. The average time reduction for groups with the GF treatment sequence was 3m:37s (28\%), and for groups with the FG sequence, 11m:59s (44\%), corresponding to a difference of 8m:22s (16 percentage points). Among groups with GF sequence, the largest reductions occurred in GP01 (with minimal organizational guidance) (52\% or 9m:42s), and GP03 (36\% or 3m:57s). In contrast, the smallest reductions in groups with FG sequence occurred in GP05 (21\% or 7m:21s) (with hybrid participation) and GP07 (42\% or 12m:21s). In absolute terms, three of four FG groups exceeded the largest time reduction among the GF groups in GP01, however, in relative terms, only GP04 did. For comparison, if GP01, which initially received only minimal organizational guidance, is excluded, three of four FG groups also exceeded the relative time reduction observed in GP03.

\begin{colora}
\textit{RQ3}: Feature analysis results indicate that AM4RRE was perceived as useful for coordination in regulatory requirements specification (\textit{CC05}), and ensuring specification transparency (\textit{CC11}), potentially useful for capturing legal (\textit{CC06}) or other role-specific knowledge (\textit{CC07}), supporting compliance implementation in software design (\textit{CC13}), and other downstream SE processes (\textit{CC14}).

Quasi-experiment results complement the positive feature analysis results and suggest that applying AM4RRE templates before free-form coordination (GF sequence) was associated with shorter total specification time than in groups that applied free-form coordination first (FG sequence). These results are also consistent with \textit{Hypothesis 2}--the average reduction in specification time from Period 1 to Period 2 was greater in groups with the FG treatment sequence than in groups with the GF sequence, by approximately 8m:22s, or 16 percentage points.
\end{colora}
\newpage

\subsection{RQ4: To what extent can AM4RRE support the quality of regulatory requirements?}\label{sec:rq4}
To answer RQ4, we report feature analysis results for criteria CC01--CC04 (Table~\ref{tab:evaluation}) and the quality of specifications produced in the quasi-experiment (Table~\ref{tab:coordination}).

For \textit{regulatory requirements completeness (CC01)}, 16 of 37 participants found AM4RRE useful, and the average evaluation was 1.24.
For support of \textit{regulatory requirements consistency (CC02)}, AM4RRE averaged 1.24, and 17 of 37 participants found it useful.
For support of \textit{consistency of regulatory and other requirements (CC03)}, the average evaluation was 1.05, and 20 of 37 participants found it potentially useful. Finally, for supporting \textit{regulatory requirements correctness (CC04)}, AM4RRE received an average rating of 1.38, and 18 of 37 participants evaluated AM4RRE as useful. Each of CC01-CC04 was evaluated negatively only once.

Next, we report the results of requirements specifications quality analysis. Internal completeness was maximal (1.00) across all the groups, except for three free-form coordination periods in GP07 (0.90), GP01 (0.80), GS01 (0.80).
In six groups with GF treatment sequence, specifications produced in P1 guided by AM4RRE templates had an average consistency of 0.72 and an average correctness of 0.83. In P2, under free-form coordination, consistency was 0.11 (15\%) lower (0.61), and correctness was 0.16 (19\%) lower (0.67). In five of the six GF groups, AM4RRE templates were associated with higher consistency and correctness, except in GS04, where the free-form coordination period showed higher consistency and correctness by 0.11 (14\%) and 0.17 (21\%), respectively. In four groups with FG treatment sequence, P1 had an average consistency of 0.64 and an average correctness of 0.80. In guided P2, average consistency increased by 0.09 (14\%) to 0.73, and correctness increased by 0.09 (11\%) to 0.89. In three of the four FG groups both were higher in the guided period. In GS02 consistency was 0.01 lower.

Average completeness under guided coordination was higher in the groups with GF treatment sequence (0.07) in comparison to FG groups (0.03), and consistency was higher by similar margins in both GF (0.11) and FG (0.09) sequence groups. The correctness advantage of guided coordination was larger in GF groups (0.16) than in FG groups (0.10). Persistent quality defects were related to using legal subject names (e.g., data subject) instead of corresponding stakeholder names (e.g., EU adult users), despite considering managerial specifications. None of the groups addressed compliance concerns from the RE case descriptions, possibly because they mainly focused on legal and software architecture specifications and needed to specify only two requirements.

\begin{colora}
\textit{RQ4}: According to the feature analysis results, participants found AM4RRE useful for supporting correctness (CC04), completeness (CC01), and consistency of regulatory requirements (CC02), and potentially useful for supporting consistency between regulatory and other requirements (CC03).

The quasi-experiment partially supports \textit{Hypothesis 3}. Regardless of treatment sequence, guided coordination was associated with higher requirement completeness in three groups where it varied, higher consistency in 8 of 10 groups, and higher correctness in 9 of 10 groups.
\end{colora}

\subsection{RQ5: To what extent can AM4RRE support the regulatory RE process?}\label{sec:rq5}
We report feature analysis results for criteria on regulatory RE process support (\textit{CC05}-\textit{CC12}, see Section~\ref{sec:rq3} for results according to \textit{CC05-CC07}, \textit{CC11}).

For \textit{compensating for the absence of legal knowledge and guiding non-experts in legal specifications (CC08)}, AM4RRE averaged 0.69, with 17 of 35 participants rating it as potentially useful. P23.6A assigned a ``not useful'' evaluation, and P19.5A, P20.6L, and S2.1R gave a ``potentially not useful'' rating. The average evaluation by participants in legal roles was relatively low--0.85. Participants did not make specific comments about CC08; however, we speculate that the low rating reflected broader concerns regarding the potential loss of legal nuance (see Section~\ref{sec:rq3}). Other comments suggested that AM4RRE can give insight into other viewpoints and their knowledge, including a legal viewpoint (see the quotes next).

\begin{colorb}
    \small
    \textit{``Now, I better understand what our legal people do.''} - P1.1L in an oral comment.
\end{colorb}

\begin{colorb}
    \small
    \textit{``It helps with seeing things from different perspectives.''} - P13.4R in final evaluation.
\end{colorb}

For \textit{support of the tracing in regulatory RE (CC09)}, participants averaged 0.97, with 18 of 37 participants considering AM4RRE potentially useful. Two participants assigned a rating below zero: P23.6A and S4.2L. Notably, participants in RE roles assigned the highest average evaluation of 1.29.

Regarding AM4RRE's  \textit{support of consistency in the regulatory RE process (CC10)}, participants on average rated it 1.05, and 28 of 37 considered AM4RRE ``potentially useful''. S4.2L was the only participant who assigned a negative evaluation.

For \textit{supporting interpretation of regulations (CC12)}, participants gave an average rating of 1.19; 15 participants considered it ``potentially useful'' and 15 ``useful''. The only participant who assigned a rating below zero for this criterion was P16.5L.
In a follow-up discussion, participants described template-guided coordination as very ``mechanistic'' consideration of different interpretation options and regulatory requirements. They note that in practice, this process is guided by legal expertise, and interpretation results can be captured in a form allowing systematic reuse (see the quotation below).

\begin{colorb}
    \small
    \textit{``The conceptual design of the framework seems reasonable; in the practical exercise, it felt a little like ``brute-forcing'' the requirements, as in establishing all relevant potential ones to capture them. Like many other frameworks, this will likely become more effective with a repository of elements that can be reused.''} - P16.5L in AM4RRE evaluation.
\end{colorb}

\begin{colora}
\textit{RQ5}: Feature analysis results indicated that AM4RRE is potentially useful for supporting traceability (\textit{CC09}), consistency (\textit{CC10}), and regulations interpretation (\textit{CC12}) in regulatory RE. To a lesser degree, AM4RRE was evaluated as potentially useful for compensating for the absence of legal knowledge (\textit{CC08}).
\end{colora}

\subsection{RQ6: To what extent can AM4RRE support achieving the goals of different viewpoints?}\label{sec:rq6}
We report feature analysis results for \textit{CC13}-\textit{CC18} on support for viewpoint-specific goals. Details for CC13-CC14 are available in Section~\ref{sec:rq3}).

The average evaluation of AM4RRE across CC13-CC18 criteria was 1.19. For \textit{enabling decision-making about regulatory compliance (CC15)}, AM4RRE was on average evaluated as 1.22, with 17 of 37 participants considering AM4RRE ``useful''.
For \textit{documenting the required information (CC16)}, the average evaluation was 1.32, with 18 of 37 participants considering AM4RRE ``useful''. Participants in legal roles gave the highest average evaluation (1.62), perhaps because they rely most on such documentation.
For \textit{facilitation of compliance governance (CC17)}, the average evaluation was 1.29, with 17 of 35 participants considering AM4RRE ``useful''. Notably, participants in managerial roles, for whom this goal is most relevant, gave the highest average (1.50). For \textit{addressing business concerns (CC18)}, the average evaluation was 1.11, with 15 of 37 participants considering AM4RRE ``useful'' and managerial roles most related to business goals also assigning the highest average evaluation--1.33. \textit{CC16} and \textit{CC17} were evaluated as ``not useful'' only by P23.6A. \textit{CC15} and \textit{CC18} were evaluated as ``potentially not useful'' only by P23.6A, S1.1L, and S2.1R.

\begin{colora}
\textit{RQ6:} According to feature analysis results, participants most often evaluated AM4RRE as useful (+2) for documenting the information for compliance (\textit{CC16}), decision-making in regulatory compliance (\textit{CC15}), facilitating governance of regulatory compliance (\textit{CC17}), and addressing business concerns during regulatory compliance implementation (\textit{CC18}).
\end{colora}

\section{Discussion and future work}\label{sec:discussion}
Next, we discuss the core results and suggest directions for future work.

\paragraph{Applicability of AM4RRE for CbD according to CRA and GDPR (RQ1)}
AM4RRE was applied in regulatory RE for CbD under the GDPR and the CRA, although only one of its five models (the artifact model) was tailored. AM4RRE templates supported coordination in both PbD and SbD cases, though AM4RRE evaluations were generally lower for SbD. The largest PbD-SbD differences, in both template and overall evaluations, concerned completeness (CC01, TC05), consistency (CC02, TC07), capturing legal knowledge (CC06) along with regulation interpretation (CC12) in the AM4RRE evaluation, and explicating role knowledge (TC02) in template evaluation. Despite the smaller sample, SbD evaluation participants reported more known regulatory RE methods for CbD, which were also more technology-focused. These findings suggest regulation-specific differences, consistent with legal practice demanding specialized expertise across fields of law. Future research could examine such specificity and ways to tailor regulatory RE approaches, for example by tailoring all AM4RRE models.

\paragraph{Practical applicability of AM4RRE components and templates (RQ2)}
The results indicate that our conceptualization of regulatory RE for CbD as specification and coordination across viewpoints is compatible with existing practice. Regulatory RE can thus be seen as a complex RE situation in which fine-grained requirements may not be immediately specifiable~\cite{katina2014system}, but can be specified after coordinating and considering regulations' impact on existing specifications. Characterizing similar complex RE situations is an important topic for future work.

AM4RRE was applied and positively perceived both by participants who had not worked together before and by colleagues from one organization. We speculate that participants' ability to complete specification templates with minimal guidance suggests that AM4RRE's artifact orientation may support integration with existing viewpoint-specific processes without requiring significant redesign. Using specification instances in place of absent participants also suggests potential for flexible and asynchronous coordination. AM4RRE was also applied in onsite, online, and hybrid coordination formats. Together, these findings suggest that AM4RRE may accommodate variability in regulatory RE processes and ways of working. Future research could compare artifact- and activity-oriented RE approaches in addressing this variability, including different coordination formats (e.g., fully asynchronous).

We explored goal and context models to explicate different viewpoints. Although previous studies used goal models to represent regulations, to our knowledge, their use to guide and coordinate roles in regulatory RE has not previously been explored. Future research could investigate how well goal models capture the regulatory RE context and directly support requirements specification.

Our results suggest that organizational context (e.g., division of responsibilities) and goals (e.g., compliance governance) influence the applicability of regulatory RE approaches beyond organizational size alone. Feedback from P23.6A and P20.6L suggests that coordination may not be perceived as important even in large enterprises with dedicated legal experts. We also argue that enterprises where a single SE role handles both legal and SE specifications may be motivated to separate and coordinate these specifications to improve regulatory RE outcomes (e.g., verifiability). Future evaluations of AM4RRE could assess its applicability across organizational contexts.

Although participants suggested several improvements to AM4RRE-based templates, the results indicate their applicability to support coordination. We argue that difficulties with the \textit{TObj} template arose from the need to combine legal knowledge for identifying legal concepts with SE knowledge for allocating them between requirements and system levels of abstraction. This highlights the need to explore how regulatory RE should address the close intersection of legal and SE knowledge.

Since the evaluation did not cover the full effort required for AM4RRE tailoring (only some tasks of legal roles resembled regulation-specific tailoring), future research should evaluate the costs and benefits of AM4RRE in a real organizational context. Such evaluations could investigate whether AM4RRE introduces additional overhead, as suggested by P20.6L and P23.6A, and whether this overhead is offset by its benefits. Other research directions include tailoring AM4RRE to multiple regulations simultaneously and integrating it into existing SE processes and artifacts.

\paragraph{AM4RRE support for coordination in regulatory RE for CbD (RQ3)}
Feature analysis indicates positive perceived usefulness across both template and AM4RRE criteria: capturing and explicating role-specific knowledge through templates (TC01, TC02) and AM4RRE (CC06, CC07), and specification transparency (TC03, CC11). Participants in legal roles rated knowledge capture positively (1.00) and interpretation support—presumably requiring their involvement—higher (1.15).

In the quasi-experiment, AM4RRE templates were associated with shorter average specification time in groups that executed guided coordination first and higher requirements quality in most of the groups, with some exceptions. We compare these findings with prior studies, discuss group differences, potential confounding factors, and learning effects (how familiarity with the approach or case affects subsequent tasks~\cite{vegas2015crossover,ko2015practical}). It is important to note that the treatment comprised only completed AM4RRE templates preceded by introductory tasks. Before specification tasks, all groups were exposed to AM4RRE context, role, and goal models,viewpoint-specific templates, and the regulatory requirements specification template. Fully isolating AM4RRE as the treatment could therefore yield different results.

To our knowledge, no prior regulatory RE study evaluated structured engineering-legal coordination with both roles present. Studies in privacy threat modeling~\cite{sion2020dpmf,boltz2026enabling} partly cover legal and SE viewpoints for identifying threats in existing design and data flows, but do not address broader RE concerns related to software context and requirements. Boltz et al.\cite{boltz2026enabling} emphasize legal interpretation and use the same legal concepts as AM4RRE (legal object, legal subject), but do not evaluate their approach with threat modeling practitioners. Engiel et al.~\cite{engiel2017tool} tested an approach applicable to engineering-legal RE with software engineers only, eliciting more requirements than ad hoc practice in equal time. Similarly, in our study, GP02 and GP04 specified more requirements with AM4RRE templates than with free-form coordination, although this measure was not predefined. Alhirabi et al.~\cite{alhirabi2023parrot} evaluated their PbD tool separately with software engineers and legal experts: 71\% found it useful for implementing six PbD principles, and 46\% found it easy to use and efficient. For AM4RRE, 89\% perceived it positively for compliance in software design; template evaluations for ease of information specification were 55\% (legal), 50\% (managerial), 77\% (RE), and 85\% (software architecture). Although direct comparison of study results is not possible, our study contributes to existing state of research by evaluating AM4RRE with legal and engineering practitioners working together.

When considering \textit{coordination format} as a potential confounding factor, GP05, with hybrid participation, had the longest specification time but one of the largest increases in consistency and correctness under guided coordination. Online GS01 had the third-longest specification time among GF groups, but had substantially higher consistency and correctness in guided P1 than free-form P2. These findings suggest that it is important to explore if AM4RRE may be particularly useful in hybrid or remote coordination. The results of GP05 and GS04 suggest a possible association between the specification time and requirements quality. GS04 had the second-longest total time among GF groups and the smallest overall time reduction, but it was the only GF group to achieve higher consistency and correctness in P2.

Quality gains in GP05 and GS04 may reflect stronger learning effects linked to greater \textit{average experience with regulation}, while GS01's lower average may explain its lack of evident learning effects in P2. Experience may also explain why two experienced participants in GP02 achieved higher consistency and correctness than the three less experienced participants in GP03 in a similar time.

Among groups with the same treatment sequence, those with \textit{more participants} tended to show greater time reductions (e.g., GP04 and GS02 or GP02 and GP03), but no clear quality improvements. GP07, with five participants, achieved correctness comparable to GP02, with two participants, but with lower consistency. Since smaller groups used specification instances for absent roles, these findings suggest completed templates may partly compensate for fewer participants.

Shorter P2 specification times suggested a potential \textit{learning effect} across both sequences in all groups except GP06, where P20.6L and P23.6A were critical of coordination. Since distinguishing learning from treatment effects requires more controlled conditions, we examined only indicative associations. We compared average time reductions between groups with FG and GF treatment sequences with AM4RRE template use. However, in groups with the GF treatment sequence, guided coordination may have induced stronger learning than free-form coordination in FG groups, contributing to subsequent time reductions. Average free-form times support this possibility: 27m:29s in P1 (FG) and 9m:22s in P2 (GF). Guided coordination averaged 12m:58s in P1 (GF) versus 15m:30s in P2 (FG). Longer guided coordination in P2 may reflect participant differences, fatigue, or switching effort observed in some groups. The almost threefold difference in average free-form specification times may reflect learning associated with prior AM4RRE template use. Still, this comparison does not clearly isolate learning from period effects or between-group differences.

\paragraph{AM4RRE support for requirements quality (RQ4)}
Feature analysis showed that participants evaluated AM4RRE highest for requirements correctness (1.38), followed by completeness and consistency between regulatory requirements (both 1.24), and consistency between regulatory and other requirements (1.05). These perceptions align with the quasi-experiment in which correctness in guided coordination in GF groups had the greatest average advantage (0.16).

In the quasi-experiment, requirements quality showed a different pattern of potential learning effects from specification time. Across treatment sequences, guided coordination was associated with higher completeness in the three groups where it varied, higher consistency in eight of ten groups (except GS02 and GS04), and higher correctness in nine (except GS04). Marginal changes under guided P2 in GS02 (consistency -0.01; correctness +0.01) may relate to its shortest total specification time among FG groups and third-largest time reduction overall. Under guided coordination, average completeness was unchanged across periods, while consistency and correctness were 0.01 and 0.06 higher in P2 (FG), suggesting that guided coordination may better support applying knowledge acquired in P1. Under free-form coordination, P1 (FG) had higher means: completeness by 0.04, consistency by 0.03, and correctness by 0.13. These differences may reflect participant characteristics, difficulty switching from guided to free-form coordination, or overreliance on acquired knowledge.

We observed other manifestations of possible learning effects: participants in SE roles sometimes relied on memorized information. Legal experts often used general regulatory knowledge instead of the case descriptions or their specifications. These behaviors may reflect learning during the evaluation, prior expertise, or both. RE methods may help practitioners internalize knowledge within and beyond their roles, reducing explicit method use. This may have unintended consequences, such as role interference, as when a GP05 participant in a managerial role provided architecture-related information. We suggest that regulatory RE approaches should adapt as practitioners’ familiarity with regulations or the project grows.

Participants used different \textit{coordination forms}, such as displaying completed templates or conveying contents verbally without showing them. We observed that coordination forms were related to specification syntax--GP01 participants reported that complex template syntax hindered coordination. Together with observations on the legal object template, this highlights the importance of syntactic differences in interdisciplinary work, and potentially within software engineering teams.

One important finding is that coordination can support regulatory requirements quality but does not guarantee it. Across all groups and treatments, specifications used legal concepts (e.g., data subject) instead of corresponding SE concepts derived through legal interpretation. Engineers may understand these specifications, especially when collaborating with legal experts, but leaving interpretations implicit limits usefulness for SE and risks misinterpretation when coordination context is lost~\cite{kosenkov2025privacy}. This issue may affect regulatory and other multi-viewpoint RE, motivating dedicated quality assurance and further research.

\paragraph{AM4RRE contribution to regulatory RE (RQ5)}
Support for traceability was on average evaluated around +1 both for templates (TC06) and for AM4RRE (CC09). Interestingly, participants in RE roles evaluated CC09 considerably higher (1.29) than other roles (0.83-0.92), while evaluating AM4RRE lower for coordination (0.86 vs. 1.46-1.55). S2.1R linked their negative evaluation to the ``delayed'' and ``insufficient'' interaction between RE and legal roles, which could be relevant for other participants in RE roles. Participants in legal roles evaluated AM4RRE for compensating for absent legal knowledge (\textit{CC08}) as relatively low (0.85), describing coordination as ``brute-forcing'' requirements (P16.5L), mechanistic or overloaded (P20.6L), and specifications as risking ``los[ing] meaning and nuance'' (P24.7L). Participants in legal and occasionally other roles sometimes raised provisions or regulations beyond the cases in discussions (e.g., US privacy regulation raised by P2.1M). These observations suggest that legal experts draw on broad legal knowledge and context, which must be captured to preserve their interpretations and maintain backward traceability to regulations.

The comment by S3.1A (see Section~\ref{sec:rq3}) suggests that AM4RRE has potential to integrate into different SE process models, support both independent work on viewpoint-specific specifications and targeted collaboration across coordination formats. This may offer an alternative to less structured coordination as in Agile SE.

\paragraph{AM4RRE support for goal achievement (RQ6)}
Average usefulness evaluation for CC13–CC18 goal support was positive but varied across roles. Most ratings for other viewpoints’ second-level goals were also positive across goal models. Future research could explore goal support across roles (e.g., decision-making) to align regulatory RE research with practice, identify each goal’s primary stakeholders, and prioritize goals relevant across stakeholders. Relatively low AM4RRE evaluation for compliance support across SE processes (0.95) suggests a need for additional SE viewpoints spanning the software development life cycle. Different evaluations for compliance support in software design between RE (0.71) and software architecture (1.55) may reflect differences in expectations or participation.

\paragraph{The role of RE in engineering-legal coordination}
Our findings suggest that RE roles were insufficiently involved in the quasi-experiment in two respects. First, no group addressed compliance from the RE viewpoint (case description or templates) in regulatory requirements specification. Although managerial and RE roles were involved in the coordination process and their specifications or case descriptions were occasionally reviewed, regulatory requirements primarily reflected software architecture and legal specifications or case descriptions. Possible explanations for this include study design (only two requirements per period were required), insufficient AM4RRE guidance, RE engagement patterns, and prioritizing regulatory architecture constraints over alignment with existing requirements. We argue that these observations reflect existing challenges--few studies address alignment between regulatory and non-regulatory requirements, and regulatory constraints are considered easy to overlook~\cite{kiyavitskaya2007extracting,breaux2008analyzing}. We attribute RE participants’ still positive average evaluation to perceived coordination potential. Future work could examine whether addressing existing non-regulatory requirements should be integrated with CbD implementation or handled separately.
Second, we speculate that RE roles were passive or perceived as secondary. Participants were free to choose who organized and documented coordination, although the goal and role models suggested RE leadership. Nevertheless, none of the seven groups with RE participants selected an RE role to record specifications--four chose the legal expert (GP04-05, GS01, GS02) and three--the software architect (GP03, GP06-07). This observation resembles the finding that regulatory compliance research focusing on downstream SE processes often does not account for RE~\cite{kosenkov2025systematic}. We therefore suggest that RE research and practice should better formulate and establish the RE role in supporting compliance implementation.

Our findings highlight the importance of RE guidance for ensuring the quality of regulatory requirements specifications. In four of the six GF groups, specification times decreased alongside consistency and correctness. We speculate that learning effects may have encouraged reliance on acquired knowledge and faster specification while quality declined. Together with the use of legal instead of system-specific SE concepts, these findings suggest the need for dedicated RE guidance and quality assurance.

Beyond providing indicative evidence of AM4RRE feasibility and perceived usefulness, this study highlights the multifaceted role of RE in supporting regulatory compliance, particularly compliance by design. Regulatory RE approaches should support explication of relevant viewpoints. Modeling goals and challenges may help to uncover task and information dependencies among viewpoints. RE further needs to provide viewpoint coordination mechanisms, such as shared artifacts, templates, and tools. RE can support requirements specification that remains traceable to legal specifications for legal experts and to SE specifications that are understandable and actionable for software architects and other SE roles. In this dynamic process, AM4RRE was already perceived as supporting these needs by structuring viewpoint specifications and making information interdependencies explicit. We plan to explore AM4RRE operationalization with tool support in real organizational settings. While AM4RRE emphasizes pragmatic information exchange through direct role involvement, it also provides a basis for exploring feasible automation. AI-based tools may capture viewpoints and context, but they may not give users enough agency or understanding and may still require expert involvement~\cite{singhal2026code,sleimi2021automated}.

\paragraph*{Key Takeaways}
Next, we summarize the important findings. Key \textbf{research takeaways} are:
\begin{itemize}[leftmargin=28px,nosep]
\item RE can support downstream compliance implementation by communicating required knowledge and information between legal and SE viewpoints;
\item regulatory RE demands mechanisms for interpreting applicable regulatory norms in the context of the existing software and SE specifications;
\item dynamic aspects of RE processes, including coordination among roles, should be considered in regulatory RE and potentially other RE settings;
\item regulatory RE approaches should reflect practice and integrate with SE processes and the wider organizational context;
\item modeling goals and challenges can provide mechanisms to contextualize regulatory RE processes and capture different viewpoints;
\item research should further explore engineering and tailoring of RE approaches for varying contexts and roles directly involved in RE;
\item regulatory RE research could further concretize the mediating role of RE between legal and SE viewpoints by exploring how it supports both.
\end{itemize}

Our results suggest the following \textbf{key takeaways for practitioners}:
\begin{itemize}[leftmargin=28px,nosep]
\item it is important to approach regulatory RE as a lever for supporting compliance implementation in subsequent SE processes;
\item regulatory RE requires special attention to coordination of regulatory and non-regulatory requirements and existing design;
\item with or without a dedicated legal role, maintaining separate legal specifications can help capture legal rationale and ensure backward traceability;
\item it is important to consider different coordination formats (e.g., hybrid) and forms (e.g., intermediated by specifications) to support regulatory RE better;
\item regulatory RE activities should be guided by intended benefits, such as shorter specification times and improved requirements quality;
\item in interdisciplinary settings, requirements quality should be observed, and resulting specifications should be understandable without experts' involvement;
\item approaches and tools for regulatory RE should support seamless integration with existing SE processes and tools.
\end{itemize}

\section{Threats to validity}\label{sec:threats}
Our conclusions rely primarily on feature analysis, complemented by quasi-experimental results; next, we discuss validity threats for both.

For \textbf{construct validity} in feature analysis, we developed evaluation criteria using common requirements quality attributes, goal- and process-related concerns identified in recent empirical studies~\cite{kosenkov2025privacy,kosenkov2026towards}. We analyzed feature analysis results using a dedicated coordination criterion (CC05) and criteria covering its aspects (see Section~\ref{sec:rq3}). Participants rated each criterion after the corresponding activity. Participants evaluated templates after completing them and AM4RRE after completing the specification tasks in the quasi-experiment. Without an established comparable systematic method, participants drew on their practical and workshop experience. All participants had exposure to the tasks of the role they represented, enabling comparison. The first author developed and operationalized AM4RRE, introduced it, and led all sessions, which may have influenced evaluations. To limit this influence, participants evaluated templates and AM4RRE individually in writing, and the session lead did not review results until session end. Additionally, handouts allowed free-form comments, and participants could express opinions during sessions. We explained criteria for feature analysis during sessions and summarized them in template evaluation forms. Sessions were not recorded; oral comments illustrating feature-analysis results derive from the first author’s field notes and follow-ups with some participants.

For the quasi-experiment, we designed the tasks so that no single role could specify the requirements alone; at a minimum, the legal expert and the software architect had to interact. During guided coordination, participants were encouraged to use only completed templates, but were allowed to use case descriptions if required. During free-form coordination, participants were not allowed to use templates. Consistency of requirements specified under free-form coordination was assessed against the case descriptions and under guided coordination--against completed templates.

For \textbf{internal validity}, we followed guidelines for method evaluation~\cite{kitchenham1996desmet} and quasi-experiments~\cite{wohlin2012experimentation,vegas2015crossover,ko2015practical,kampenes2009systematic} as far as our settings allowed. Phased planning and execution supported systematic evaluation and refinement of methods. We also report the applied methodology in detail.

To address \textbf{selection bias}, we distributed a standard invitation through multiple channels. The invitations named viewpoint coordination as the focus of evaluation, although we encouraged participation regardless of interest in such coordination. Fifteen participants were recruited from the same company (C6), allowing us to include individuals motivated by organizational factors.

We formed groups by availability and did not randomize treatment sequences. To keep sessions viable, we avoided the FG sequence for groups likely to be delayed. We assigned the FG treatment sequence to GP05 because of high GDPR experience, and to GS02 because of partial prior collaboration and high experience in CRA and cybersecurity. As a result, FG groups were larger on average (four vs. three participants) and more often had all roles present. Comparisons of total time between sequences thus reflect group composition and treatment order. To introduce AM4RRE templates, we used untimed introductory tasks (CG01) immediately before guided coordination without a free-form counterpart. To address the potential validity threat due to this, we designed introductory tasks to overlap with the timed tasks minimally and excluded the architecture overview and legal object templates.

All groups received standardized handouts, with minor PbD/SbD differences and the same regulatory requirements template — the guided version provided placeholders, and the free-form version provided it as an example. Guidance on guided coordination was added after GP01. Although this could affect quasi-experimental results, results without GP01 still point to the same trends and associations.

For \textbf{conclusion validity}, two referees additionally assessed requirements correctness and consistency. The correctness referee was blinded to coordination type. The consistency referee was not blinded, because the assessment required different reference materials by condition.

We excluded missing specification elements rather than scoring them as zero, consistently in all calculations for feature analysis and quasi-experiment. This affects the guided periods of GP06, GP07, and GS02 and the free-form periods of GP01, GP07, and GS01, where participants discussed but did not record requirement elements.

To support \textbf{reliability of treatment implementation}, the session lead refrained from intervening during feature analysis and the quasi-experiment. When substituting for one or two missing roles, he introduced only case or specification samples and pointed to the required information on request. Given the small sample and limited control, we report descriptive results and associations rather than causal relationships. We also report data on potential confounding factors: regulatory experience, knowledge of regulatory RE methods for CbD, and collaboration experience.

Within the feature analysis, 
template and AM4RRE evaluations mainly converged (e.g., TC03 and CC11), although evaluations differed between roles. Because feature analysis and the quasi-experiment shared participants, we treat agreement as corroboration rather than independent confirmation. Measured requirements quality corroborates perceived quality support (CC01--CC04); time results lack a feature-analysis counterpart and remain exploratory.

For \textbf{external validity}, we recruited practitioners with varied positions and with GDPR or CRA experience. For SbD sessions, we waived the one-year experience requirement because the main CRA obligations apply only from December 2027. Several participants represented roles differing from their job titles, but most had relevant backgrounds, experience, or knowledge of the corresponding role (e.g., S7.3L studied law at university). Fifteen of the 40 participants came from C6, including all members of GP02--GP04 and GP07, so their responses may reflect shared organizational practices or opinions. Most participants worked in large companies, limiting generalization for small and medium organizations. Results are further bounded by two EU regulations, fictitious cases, and tasks requiring two requirements per period, unlike real project settings involving more requirements and longer coordination. Comments by P20.6L and P23.6A suggest that coordination needs may depend more on compliance priorities and organizational practices than on company size. AM4RRE may therefore be relevant where organizations require both legal and engineering involvement in implementing compliance by design.
\\

\section{Conclusion}\label{sec:conclusion}
To address growing regulatory demands and support compliance by design, software engineering practitioners need adaptable, practice-oriented regulatory RE methods. Such methods should support knowledge-intensive regulatory analysis and compliance implementation throughout software engineering processes. Our findings suggest that direct coordination between legal and engineering viewpoints is a pragmatic and promising basis for specifying regulatory requirements for software design.

The Artifact Model for Regulatory Requirements Engineering (AM4RRE) presented in this study operationalizes engineering-legal coordination through viewpoint templates and explicit relations between specification content. Evaluation with 37 practitioners using fictitious GDPR and CRA cases provided indicative evidence of feasibility and perceived usefulness. Practitioner feedback suggested that AM4RRE is compatible with their existing regulatory requirements engineering practice. 

Feature analysis indicated positive evaluations of ten of eleven viewpoint-specific templates; the exception was the one requiring simultaneous application of legal and engineering knowledge. These templates were perceived as useful for capturing, explicating, and communicating role-specific knowledge. They also guided coordination in the quasi-experiment. AM4RRE as a whole received positive average usefulness ratings for viewpoint coordination, requirements quality, regulatory RE processes, and viewpoint-specific goals.

These perceptions partially aligned with the quasi-experiment results. Across both treatment sequences, guided coordination was associated with higher completeness in three groups where it varied, higher consistency in eight of ten groups, and higher correctness in nine of ten groups. However, the use of legal instead of system-specific SE concepts highlights a continuing need for RE guidance and quality assurance. Groups using AM4RRE-based templates for guided coordination before free-form coordination had shorter total specification times and almost three times shorter average free-form specification time than groups that used free-form specification first. The groups using the templates afterward showed greater average time reductions between the two periods. Although these associations do not establish causal effects, the results on specification quality are consistent with the evaluations assigned in feature analysis.

We argue that approaching regulatory requirements engineering as a coordination challenge among multiple viewpoints reflects the complexity of practice. However, our observations suggest that practitioners can overlook the contribution of requirements engineering in this process. We suggest that requirements engineering methods must be further refined to meet the practical demands of coordination and support the goals of the involved viewpoints. This can equip requirements engineering practitioners with the knowledge and tools to explicitly and meaningfully contribute to resolving complex software engineering challenges, such as processing regulations and implementing regulatory compliance.
\\

\textbf{Data availability statement}. The open dataset for this publication is available on \href{https://zenodo.org/records/23726817}{Zenodo (DOI: 10.5281/zenodo.23726817)}.

\setlength{\bibsep}{4pt}
\bibliography{bibliography}

\end{document}